%% file: PRO-21-001_temp.tex
\pdfoutput=1
\documentclass[11pt,twoside,a4paper,cmspaper,final,collab]{cms-tdr}
\def\svnVersion{17664f0}\def\svnDate{2023/09/08}\def\cmsCernNoTag{CERN-EP-2022-139}\def\cmsCernDate{\today}\def\cmsMessage{Published in the Journal of Instrumentation as \href{http://dx.doi.org/10.1088/1748-0221/18/09/P09009}{\doi{10.1088/1748-0221/18/09/P09009}.}}
\begin{document}\cmsNoteHeader{PRO-21-001}

\newcommand{\murad}{\unit{\ensuremath{\mu\text{rad}}\xspace}}
\newlength\cmsTabSkip\setlength{\cmsTabSkip}{1ex}

\renewcommand{\cmsCollabName}{The CMS and TOTEM Collaborations}
\renewcommand{\cmsNUMBER}{PRO-21-001}
\renewcommand{\cmslogo}{\includegraphics[height=2.33cm]{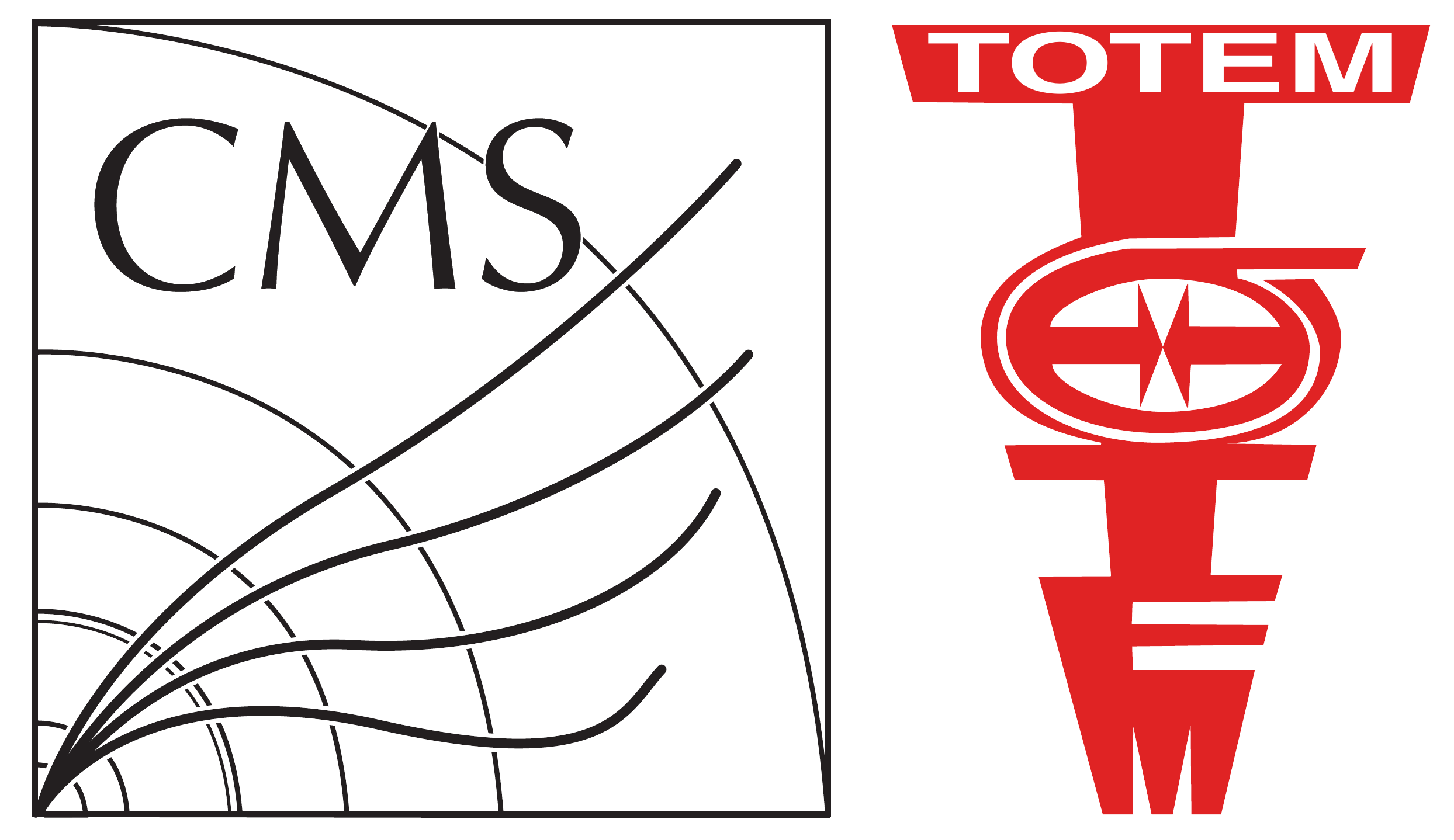}}
\renewcommand{\cmsTag}{CMS-\cmsNUMBER\\&TOTEM-2022-001\\}
\renewcommand{\cmsPubBlock}{\begin{tabular}[t]{@{}r@{}l}&CMS \cmsSTYLE\
\cmsNUMBER\\&TOTEM-2019-001\\\end{tabular}}
\renewcommand{\appMsg}{See appendices A and B for lists of collaboration members.}
\renewcommand{\cmsCopyright}{\copyright\,\the\year\ CERN for the benefit of the CMS and TOTEM Collaborations.}

\cmsNoteHeader{PRO-21-001}
\title{Proton reconstruction with the CMS-TOTEM Precision Proton Spectrometer}

\author*[cern]{A. Cern Person}

\date{\today}

\abstract{
 The Precision Proton Spectrometer (PPS) of the CMS and TOTEM experiments collected 107.7\fbinv in proton-proton (pp) collisions at the LHC at 13\TeV (Run 2). This paper describes the key features of the PPS alignment and optics calibrations, the proton reconstruction procedure, as well as the detector efficiency and the performance of the PPS simulation. The reconstruction and simulation are validated using a sample of (semi)exclusive dilepton events. The performance of PPS has proven the feasibility of continuously operating a near-beam proton spectrometer at a high luminosity hadron collider.
}

\hypersetup{%
pdfauthor={CMS Collaboration},%
pdftitle={Proton reconstruction with the CMS-TOTEM Precision Proton Spectrometer},%
pdfsubject={CMS},%
pdfkeywords={CMS,PPS,protons}}

\maketitle 

\tableofcontents
\clearpage

\section{Introduction}

The Precision Proton Spectrometer (PPS) detector system has been installed and integrated into the CMS experiment~\cite{Chatrchyan:2008zzk} during Run 2 of the LHC with 13\TeV proton-proton collisions. It is a joint project of the CMS and TOTEM~\cite{Anelli:2008zza} Collaborations and measures protons scattered at very small angles at high instantaneous luminosity~\cite{Albrow:2014lrm}.
The scattered protons that remain inside the beam pipe, displaced from the central beam orbit, can be measured by detectors placed inside movable beam pipe insertions, called Roman pots (RP), which approach the beam within a few\mm. The PPS detectors have collected data corresponding to an integrated luminosity of 107.7\fbinv during the LHC Run 2, which occurred between 2016 and 2018. 

The physics motivation behind PPS is the study of central exclusive production (CEP), \ie the process $\Pp\Pp \to \Pp^{(\ast)} + \PX + \Pp^{(\ast)}$ mediated by color-singlet exchanges (\eg photons, Pomerons, Z bosons), by detecting at least one of the outgoing protons.
In CEP, one or both protons may dissociate into a low-mass state ($\Pp^{\ast}$); dissociated protons do not produce a signal in PPS.
The $\PX$ system is produced at central rapidities, and its kinematics can be fully reconstructed from the 4-momenta of the protons, thereby giving access to standard model (SM), or beyond SM (BSM) final states that are otherwise difficult to observe in the CMS central detectors because of the large pileup (multiple interactions per bunch crossing) at high luminosities.
CEP provides unique sensitivity to SM processes in events with Pomeron and/or photon exchange, and BSM physics, \eg via searches for anomalous quartic gauge couplings, axion-like particles, and new resonances~\cite{Akiba:2016ofq,CMS:2021ncv,Chapon:2009hh,Fichet:2014uka,Baldenegro:2018hng}.

This paper is organized as follows. The CMS detector and PPS are described in Section~\ref{sec:detector}.
The LHC optics and the concept of proton transport is presented in Section~\ref{sec:optics_intro}, followed in Section~\ref{sec:datasets} by a description of the data sets used.
Sections~\ref{sec:alignment} and~\ref{sec:optics} describe the detector alignment procedure and the LHC optics calibration.
Section~\ref{sec:reconstruction} details the proton reconstruction with the PPS detectors.
Sections~\ref{sec:aperture} and~\ref{sec:simulation} document the study of LHC aperture limitations and the simulation of the proton transport and PPS detectors, and Section~\ref{sec:uncertainties} describes the uncertainties affecting the proton reconstruction.
A validation of the reconstruction using a (semi)exclusive dimuon sample is presented in Section~\ref{sec:dilepton}.
The measurement of the proton reconstruction efficiency is discussed in Section~\ref{sec:efficiency}.
Section~\ref{sec:timing} describes a study of the performance of the proton vertex matching criteria from time-of-arrival measurements.
Finally, a summary is presented in Section~\ref{sec:summary}.

\section{The CMS detector and PPS}
\label{sec:detector}
The central feature of the CMS apparatus is a superconducting solenoid of 6\unit{m} internal diameter, providing a magnetic field of 3.8\unit{T}. Within the solenoid volume are a silicon pixel and strip tracker, a lead tungstate crystal electromagnetic calorimeter, and a brass and scintillator hadron calorimeter, each composed of a barrel and two endcap sections. Forward calorimeters extend the pseudorapidity coverage provided by the barrel and endcap detectors. Muons are measured in gas-ionization detectors embedded in the steel flux-return yoke outside the solenoid. 

Events of interest are selected using a two-tiered trigger system. The first level (L1), composed of custom hardware processors, uses information from the calorimeters and muon detectors to select events at a rate of around 100\unit{kHz} within a fixed latency of about 4\mus~\cite{Sirunyan:2020zal}. The second level, known as the high-level trigger (HLT), consists of a farm of processors running a version of the full event reconstruction software optimized for fast processing, and reduces the event rate to around 1\unit{kHz} before data storage~\cite{Khachatryan:2016bia}.

A more detailed description of the CMS detector, together with a definition of the coordinate system used and the relevant kinematic variables, is reported in Ref.~\cite{Chatrchyan:2008zzk}.

\subsection*{The PPS detectors}
\label{sec:pps-detectors}

\begin{figure}
 \includegraphics[height=4.1cm]{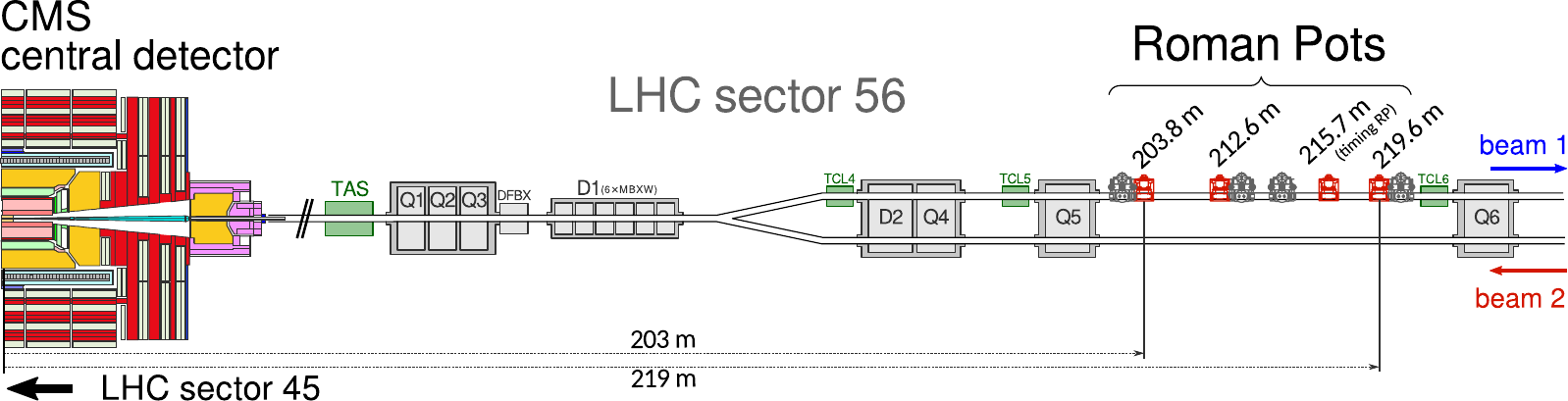}

 \caption{
  Schematic layout of the beam line between the interaction point and the RP locations in LHC sector 56, corresponding to the negative $z$ direction in the CMS coordinate system and the outgoing proton in the clockwise beam direction. 
  The accelerator magnets are indicated in grey and the collimator system elements in green. 
  The horizontal RPs, which constitute PPS, are marked in red.
  The vertical RPs are indicated in dark grey; they are part of the TOTEM experiment.
  The vertical RPs are not used during high luminosity data taking; nevertheless, they provide PPS with a reference measurement for the calibration and alignment of the detectors.}
   \label{fig:layout}
\end{figure}

Figure~\ref{fig:layout} shows the layout of the RP system installed at around 200--220\unit{m} from the CMS interaction point (LHC interaction point 5 (IP5)), along the beam line in the LHC sector between the interaction points 5 and 6, referred to as sector 56. A symmetric set of detectors is installed in LHC sector 45. Some RPs approach the beam vertically from the top and bottom, some horizontally. During standard machine operation, scattered protons undergo a large displacement in the horizontal direction and a small vertical displacement at the RP positions. The horizontal RPs are hence used. The vertical RPs are used in special configurations of the machine and in low luminosity proton-proton fills for the calibration and alignment of the detectors.
 
Each detector arm consists of two RPs instrumented with silicon \textit{tracking} detectors that measure the transverse displacement of the protons with respect to the beam, and one RP with \textit{timing} detectors to measure their time-of-flight. The tracking RP closer to the IP5 is referred to as ``near'', the other as ``far''. 
Silicon strip sensors with a reduced insensitive region on the edge facing the beam were initially used~\cite{totemstrips}. Each RP housed 10 silicon strip sensor planes, half at a $+45^{\circ}$ angle and half at a $-45^{\circ}$ angle with respect to the bottom of the RP. These sensors could not sustain a large radiation dose and could not identify multiple tracks in the same event. For this reason they have been gradually replaced by new 3D silicon pixel sensors: one RP (in each arm) during the 2017 data-taking run and all tracking RPs in 2018 were instrumented with 3D pixel sensors. Each such RP hosts six 3D pixel sensor planes~\cite{Albrow:2014lrm}. A summary of the RP configurations used in 2016-2018 is provided in Table~\ref{tab:rp-config}.

The difference between the proton arrival times in the detectors on both sides of the IP5 is used to reject background events with protons from pileup interactions, or beam-halo particles. Timing detectors were operational in 2017 and 2018, with four detector planes hosted in a single RP.
They consisted of single- and double-sided single crystal chemical vapor deposition (scCVD) diamond sensor planes~\cite{Antchev:2017pjj}; during 2017 data taking one of the four planes consisted of ultra-fast silicon sensors~\cite{Sola:2017zty} instead of diamond ones. 

\begin{table}[ht]
\begin{center}
\topcaption{RP configurations in different years. The numbers represent the RP distances from the IP5, the sensor technology is indicated in parentheses. The RP layout was always symmetric about the IP5. There were always two tracking RPs per arm; the one closer to the IP5 is denoted as ``near'', the other as ``far''. In 2016, no timing RPs were used.}
\begin{tabular}{llll}
\hline
Year & Near tracking RP & Far tracking RP & Timing RP \\
\hline
2016 & 203.8\unit{m} (strips)   & 212.6\unit{m} (strips)  & \NA \\
2017 & 212.6\unit{m} (strips)   & 219.6\unit{m} (pixels)  & 215.7\unit{m}\\
2018 & 212.6\unit{m} (pixels)   & 219.6\unit{m} (pixels)  & 215.7\unit{m}\\
\hline
\end{tabular}
\label{tab:rp-config}
\end{center}
\end{table}

\section{LHC optics and proton transport}
\label{sec:optics_intro}

PPS is a proton spectrometer that uses the LHC accelerator magnets between the interaction point (IP) and the RPs.
Scattered protons are detected in the RPs after having traversed a segment of the LHC lattice containing 29 main and corrector magnets~\cite{Bruning:782076}.

Since the protons that reach the PPS detectors travel more than 200 m inside the vacuum pipe of the LHC and very close to the LHC beams,
we use the technique normally employed to model beams inside an accelerator. The trajectory of the protons in the vicinity of the central orbit~\cite{Wiedemann:1993jy,Wilson:2001hw} can be described as follows. 
The proton kinematics $\mathbf{d}$ at a distance $l$ from the IP (\eg at the RPs) is related to the proton kinematics at the IP, $\mathbf{d}^{\ast}$, via the transport equation:

\begin{equation}
\label{eq:transport_vector_equation}
  \mathbf{d}(l) = T(l,\xi) \cdot  \mathbf{d}^{\ast} .       
\end{equation}
Superscript $^*$ in general is used in the following to denote the value of the given parameter at the interaction point, $z=0$.
The proton kinematics is described by $\mathbf{d}=(x, \theta_x, y, \theta_y, \xi)^{T}$, where $(x,y)$ and $(\theta_x, \theta_y)$ indicate the transverse position and angles; $\xi$ denotes the fractional momentum loss
\begin{equation}
\xi = (p_\text{nom} - p) / p_\text{nom} ,
\label{eq:introduction-xi}    
\end{equation}
where $p_{\text{nom}}$ and $p$ are the nominal beam momentum and the scattered proton momentum, respectively~\cite{Antchev:2014voa, Nemes:2256433}.

In exclusive reactions the momentum losses of the two scattered protons, $\xi_{1}$ and $\xi_{2}$, can be used to assess the mass of the centrally produced state
\begin{equation}
\label{eq:mx_formula}
m_{X}=E_\text{cm}\sqrt{ \xi_{1} \xi_{2}} ,
\end{equation}
and its rapidity 
\begin{equation}
\label{eq:rapidity_formula}    
y=\frac{1}{2}\ln\frac{\xi_{1}}{\xi_{2}} ,
\end{equation}
where $E_\text{cm}$ stands for the proton-proton centre-of-mass energy (13\TeV in LHC Run 2).

The transport matrix is defined as:
    \begin{eqnarray}
    T(s,\xi)=\left(
        \begin{array}{ccccc} 
            v_x         & L_x     & m_{13}   & m_{14}  & D_x  \\
            \frac{\rd  v_x}{\rd  l}        & \frac{\rd  L_x}{\rd  l}    & m_{23}   & m_{24}  & \frac{\rd  D_x}{\rd  l} \\
            m_{31}      & m_{32}  & v_y      & L_y     & D_y  \\
            m_{41}      & m_{42}  & \frac{\rd  v_y}{\rd  l}     & \frac{\rd  L_y}{\rd  l}    & \frac{\rd  D_y}{\rd  l} \\ 
            0           & 0       & 0        & 0       & 1
        \end{array}  
    \right),
    \label{eq:transport_matrix}
    \end{eqnarray}
where the most important quantity for the proton spectrometer is $D_{x}$, the horizontal dispersion; the other matrix elements are the so-called optical functions ($v_{x}, L_{x}, m_{i,j}$ and their 
vertical counterparts)~\cite{Willeke:1988zu}. The definition of the relevant optical functions and their determination are described in Section~\ref{sec:optics}.
The optical functions depend on LHC parameters like the betatron function value $\beta^{*}$ at the IP5 and the crossing angle. Throughout this document, we refer to the half crossing angle,  \ie half the angle between
the beams at their crossing point.

\begin{figure}
\centerline{
	\includegraphics{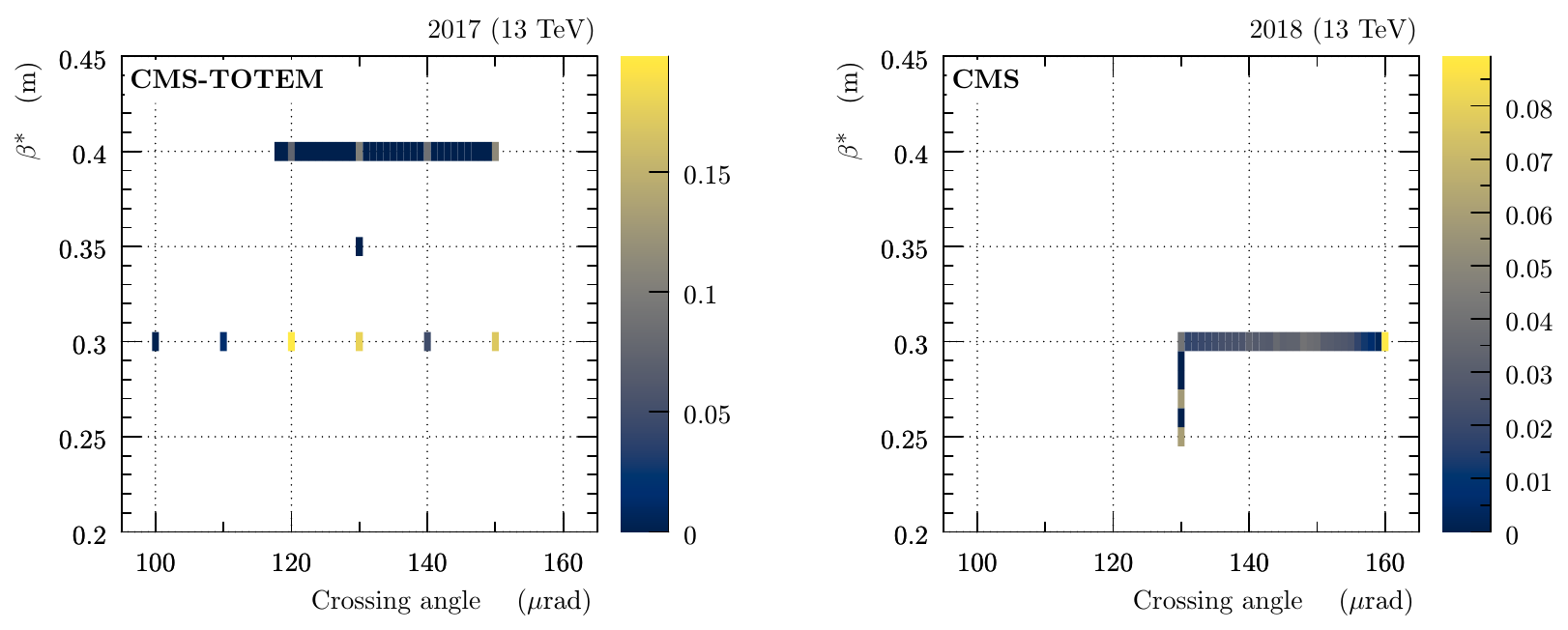}
}
\caption{
Frequency distributions of $\beta^{\ast}$ vs.~crossing angle configurations as extracted from data. {Left}: year 2017. {Right}: year 2018.
}
\label{fig:intro-xangle-beta-st}
\end{figure}

Figure~\ref{fig:intro-xangle-beta-st} shows the distributions of $\beta^{\ast}$ vs.~crossing angle for different data taking periods as extracted from data certified for analysis. In 2017, most of the data were recorded at four discrete values of the crossing angle: 150, 140, 130 and 120\murad. The highest value was used at the beginning of the fills, then the crossing angle was reduced as the instantaneous luminosity dropped. The value of $\beta^{\ast}$ was set to 0.4\unit{m} (0.3\unit{m}) in periods before (after) Technical Stop 2 (TS2). In 2018, the crossing angle was changed continuously from 160\murad at the beginning of the fill down to 130\murad. At this point, $\beta^{\ast}$ was changed in two discrete steps, from 0.3 to 0.27 and finally to 0.25\unit{m}. In 2016 (not shown in the figure) $\beta^{\ast} = 0.4\unit{m}$ was used together with the crossing angle values of 185\murad and 140\murad for the pre-TS2 and post-TS2 periods, respectively.

\section{Data sets}
\label{sec:datasets}

Two types of data are used for the calibration and alignment of the PPS detectors: data taken in high-intensity LHC ``physics'' fills and data taken in special ``alignment'' fills. The low beam intensity is an essential feature of the alignment fills, which provide additional data for alignment and optics calibration. The various beam intensities are typically achieved by injecting various numbers of bunches in the LHC, since the number of protons per bunch is typically the same, up to $1.2\times 10^{11}$. The RP distances from the LHC beams are typically expressed in multiples of ``beam sigmas'', the RMS values of the beam transverse profile. The values of the beam sigma are the same for the alignment and physics fills: $\sigma_\text{beam} \approx 0.1\unit{mm}$ horizontally and $\sigma_\text{beam} \approx 0.4\unit{mm}$ vertically.

The physics fills are standard LHC fills. There are up to 2500 bunches per beam, yielding an instantaneous luminosity of about $10^{34}\unit{cm}^{-2}\unit{s}^{-1}$. The average number of inelastic proton interactions at the IP (pileup) is typically between 15 and 55. Only horizontal RPs are inserted in these fills, to a distance of $15\,\sigma_\text{beam}$.

\begin{figure}
\centering
\includegraphics[height=4.5cm]{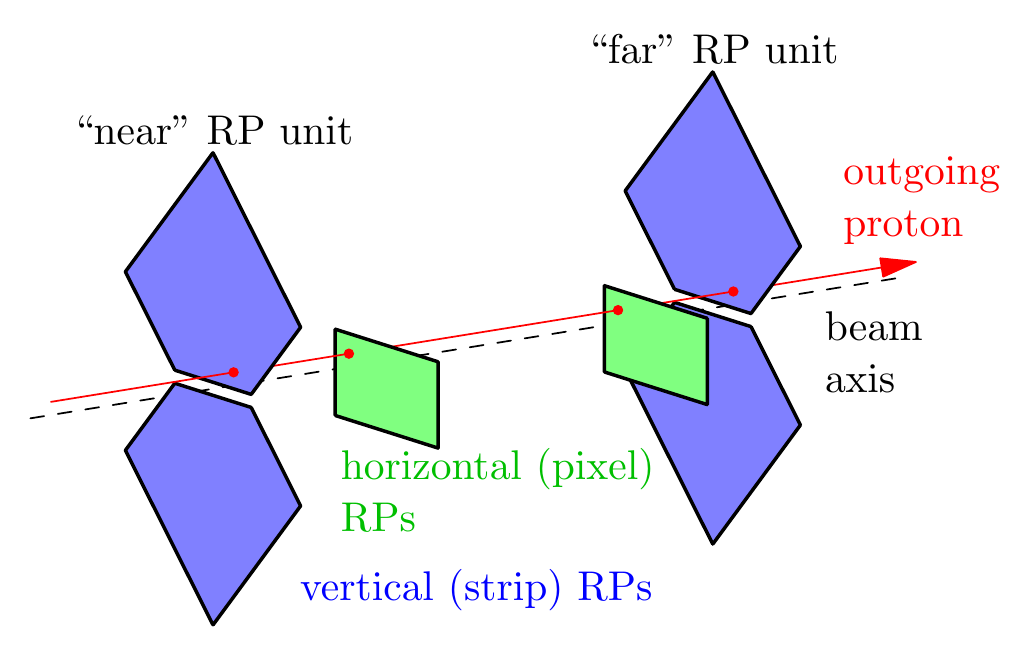}
\caption{Illustration of a proton crossing both the vertical (blue) and the horizontal (green) RPs (overlapping configuration).}
\label{fig:alig-overlap}
\end{figure}

The alignment fills use the same LHC optics as the physics fills, but much lower beam intensity---typically only two bunches are injected per each beam. This gives instantaneous luminosities of the order of $10^{30}\unit{cm}^{-2}\unit{s}^{-1}$ and average pileup about 20. The primary purpose of these fills is to establish the RP position with respect to the LHC collimators using a procedure analogous to the LHC collimator alignment~\cite{totem_peformance_2013}. This is a precondition for systematic RP insertion close to the high-intensity LHC beams. Because of the low intensity, the safety rules allow insertion of both horizontal and vertical RPs very close to the beam: at $6.5\,\sigma_\text{beam}$ horizontally and at $5\,\sigma_\text{beam}$ vertically. At these distances, the horizontal and vertical detectors overlap, as shown in Fig.~\ref{fig:alig-overlap}, which allows the relative alignment of the RPs in each arm. With the use of the vertical RPs, it is possible to detect elastically scattered protons that are used for horizontal RP alignment with respect to the beam. The alignment procedure is detailed in Section~\ref{sec:alignment}. In the alignment fills the very small separation of the horizontal RPs from the beam allows the recording of additional data essential for optics calibration (cf.~Section \ref{sec:optics}). Typically there are two alignment fills per year of LHC operation.

In Run 2, PPS was operated from 2016 to 2018. The PPS data sets are divided in data-taking periods. The PPS performance is often sensitive to the LHC settings (optics, collimators, etc.), which often vary with time; they are changed during LHC technical stops (TSs). For instance, the LHC optics was modified during the second technical stop (TS2) in 2016 and $\beta^{\ast}$ was changed after TS2 in 2017. The technical stops are also opportunities for changing the position of the detectors in the RPs. For example, in TS1 and TS2 in 2018, the tracking RPs were shifted vertically to better distribute the radiation dose accumulated by the pixel sensors. The sensor inefficiency due to radiation damage is discussed in Section~\ref{sec:efficiency}. Table~\ref{tab:datasets-lumi} summarizes the PPS periods with significantly different LHC/RP settings and the corresponding integrated luminosities~\cite{lum1,lum2,lum3}.

\begin{table}[ht]
\begin{center}
\topcaption{List of the PPS periods with distinct LHC and/or RP settings. The third column from left indicates the time ranges where PPS recorded data. $L_\text{int}$ corresponds to the integrated luminosity recorded during runs certified for use in physics analysis.}
\begin{tabular}{llcc}
\hline
Year & Period & LHC fill number (date) range(s) & $L_\text{int}(\fbinv)$ \\
\hline
2016 & pre-TS2  & 4974 (31 May) to 5052 (29 Jun), 5261 (29 Aug) to 5288 (9 Sep) & 9.8 \\
     & post-TS2 & 5393 (9 Oct) to 5451 (26 Oct)                                 & 5.0 \\
\\[\cmsTabSkip]
2017 & pre-TS2  & 5839 (16 Jun) to 6193 (12 Sep)                                & 15.0 \\
     & post-TS2 & 6239 (24 Sep) to 6371 (10 Nov)                                & 22.2 \\
\\[\cmsTabSkip]
2018 & pre-TS1  & 6615 (26 Apr) to 6778 (12 Jun)                                & 18.5 \\
     & TS1-TS2  & 6854 (27 Jun) to 7145 (10 Sep)                                & 26.8 \\
     & post-TS2 & 7213 (24 Sep) to 7334 (24 Oct)                                & 10.4 \\
\hline
Total&          & & 107.7 \\
\hline
\end{tabular}
\label{tab:datasets-lumi}
\end{center}
\end{table}

\section{Alignment}
\label{sec:alignment}

The alignment of the RPs is a multi-level procedure including aligning the sensor planes within each RP as well as aligning the RPs with respect to the LHC beam. This is one of the inputs for the proton reconstruction (discussed in detail in Section \ref{sec:reconstruction}).

Although conceptually similar, the alignment of RPs is different from that of other CMS subdetectors, because the RPs are moveable devices. At the beginning of each LHC fill they are stored in a safe position away from the beam. Only when the LHC reaches stable conditions are they moved close to the beam. Since the fill-to-fill beam position reproducibility has a limited accuracy, it is desirable to determine the alignment parameters for every fill.

The alignment procedure involves multiple steps. A special ``alignment'' calibration fill determines the absolute position of the RPs with respect to the beam (Section \ref{sec:al-alignment fill}). This calibration then serves as a reference for the alignment of every ``physics'' fill with standard conditions (Section \ref{sec:al-physics fills}). Once the tracking RPs are aligned with respect to the beam, the timing RPs are aligned with respect to the tracking RPs (Section \ref{sec:al-timing}).

\subsection{Alignment fill}
\label{sec:al-alignment fill}

An alignment fill is a special fill, which allows to obtain data essential for calibration, not available in standard physics fills (more details are given in Section \ref{sec:datasets}).

\begin{figure}
\centerline{
	\includegraphics{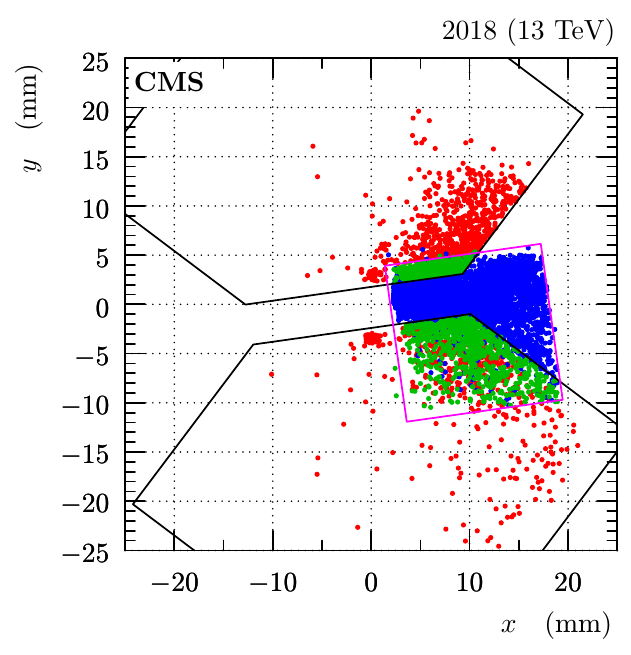}
	\hfil
	\includegraphics{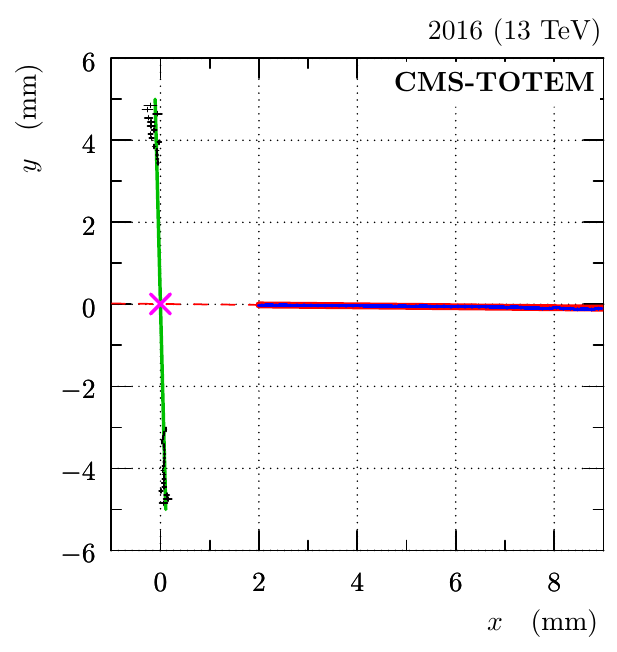}
}
\caption{
{Left}: relative alignment between vertical and horizontal RPs (April 2018). The plot shows track impact points in a scoring plane perpendicular to the beam. The points in red represent tracks only reconstructed from vertical RPs, in blue only from horizontal RPs and in green from both vertical and horizontal RPs. The size and position of the RP sensors is schematically indicated by the black (vertical strip RPs) and magenta (horizontal pixel RPs) contours.
{Right}: determination of the beam position with respect to the RPs (September 2016). Black: profile (mean $x$ as a function of y) of elastic track impact points observed in vertical RPs; green: fit and interpolation. Blue: horizontal profile of minimum bias tracks found in the horizontal RP; red: fit and extrapolation. Magenta cross: the determined beam position. The error bars represent statistical uncertainties.
}
\label{fig:alig-calib-fill}
\end{figure}

The relative alignment among the sensor planes in all the RPs and among all the RPs in one arm is determined by minimizing residuals between hits and fitted tracks \cite{kaspar-phd-thesis}. This is an iterative procedure, since a priori it is not possible to distinguish between misalignments and outliers (unrelated hits due to noise, etc.). Therefore, the iteration starts with a large tolerance, $\mathcal{O}$($100\mum$), that allows for misalignments, and as it proceeds the tolerance is decreased to $\mathcal{O}$($10\mum$) as outliers are discarded. An illustration is shown in Fig. \ref{fig:alig-calib-fill}, left, emphasizing the essential role of the overlap of the vertical and horizontal RPs. The typical uncertainty of the relative RP alignment is few micrometres. By construction, the relative alignment is not sensitive to misalignment modes that do not generate residuals, \eg~a global shift of the full RP system. These modes are addressed in the next step.

The vertical RPs can detect protons from elastic scattering,  \ie~a process with only two protons in the final state, each having $\xi \equiv 0$ as a consequence of momentum conservation. Because the two protons emerge from the same vertex in opposite directions, elastic events are relatively easy to tag (cf.~Section 5.2.1~in Ref.~\cite{Antchev:2015zza}). Because of the azimuthal symmetry of the elastic scattering at the IP and the properties of the LHC optics, the elastic protons arrive at the RPs with impact points in the transverse plane elliptically distributed around the beam. Although only the tails of the elastic hit distributions are within the acceptance (protons with sufficiently large vertical scattering angle, $\abs{\theta^*_y}$), the distributions can be used to extract the beam position with respect to the RPs. This is illustrated in Fig. \ref{fig:alig-calib-fill}, right: the profile of the elastic hit distribution (black) is interpolated between the top and bottom RP (green), which provides information on the horizontal alignment and potential rotations in the $xy$ plane. This is combined with the information from a minimum bias sample, in which most protons detected in the horizontal RPs are due to pileup. The profile from the minimum bias sample (blue) is extrapolated linearly (red) to find the intersection (magenta cross) with the green line. The intersection indicates the beam position with respect to the RPs, with a typical uncertainty of about 10\mum.

\subsection{Physics fills}
\label{sec:al-physics fills}

\begin{figure}
\centerline{
	\includegraphics{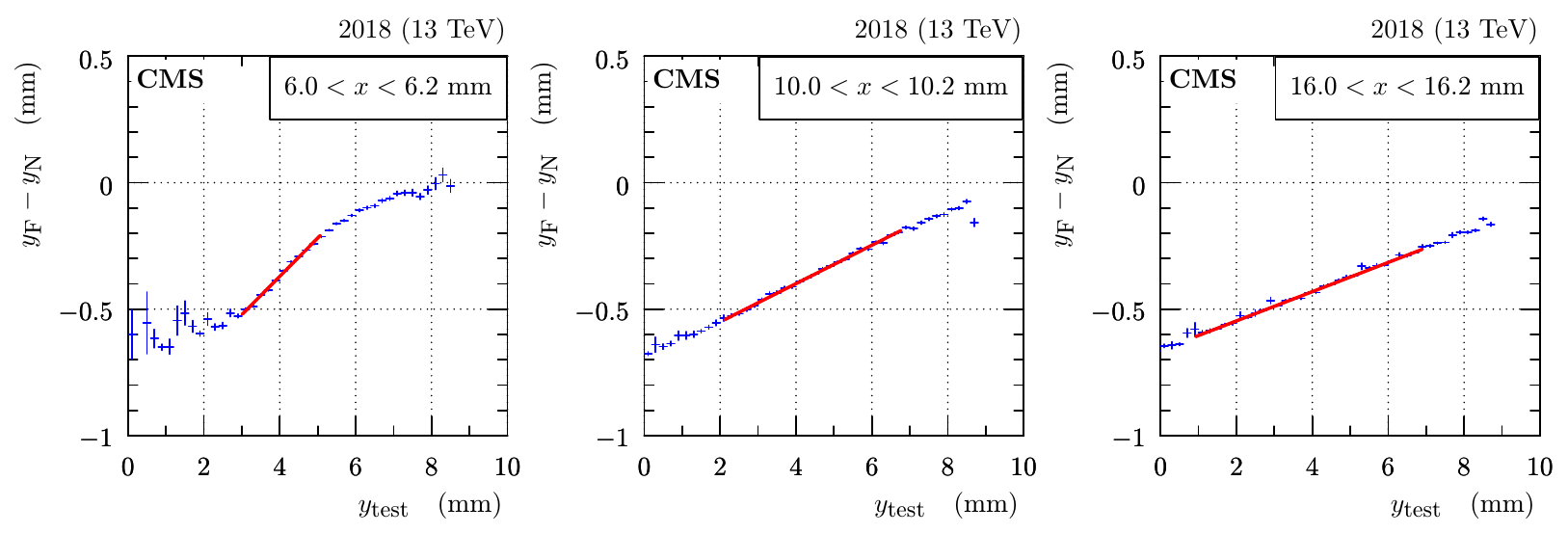}
}
\caption{
Illustration of $y_\mathrm{F} - y_\mathrm{N}$ dependence on $y$ (fill 7139, 2018, near RP in sector 56). The three plots correspond to three $x$ selections as indicated in the legends. Blue: profile histogram of the dependence, red: linear fit to the central part. The error bars represent statistical uncertainties.
}
\label{fig:alig-phys-fill-slope}
\end{figure}

\begin{figure}
\centerline{
	\includegraphics{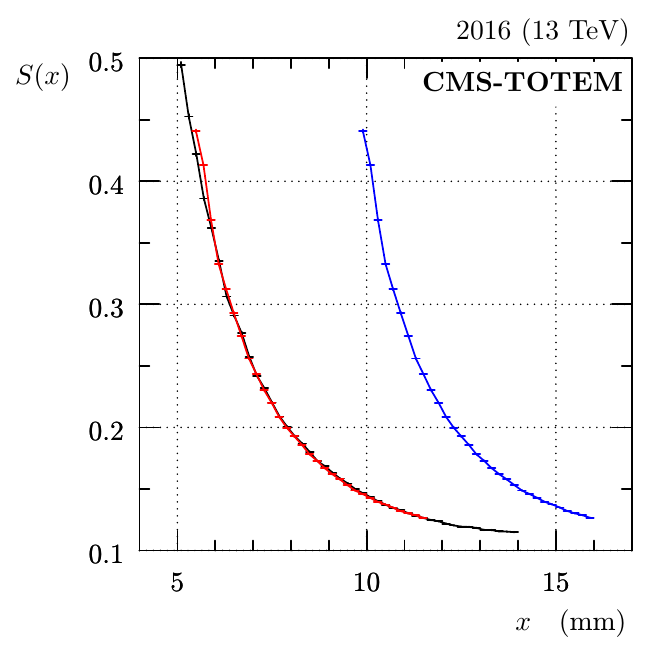}
	\hfil
	\includegraphics{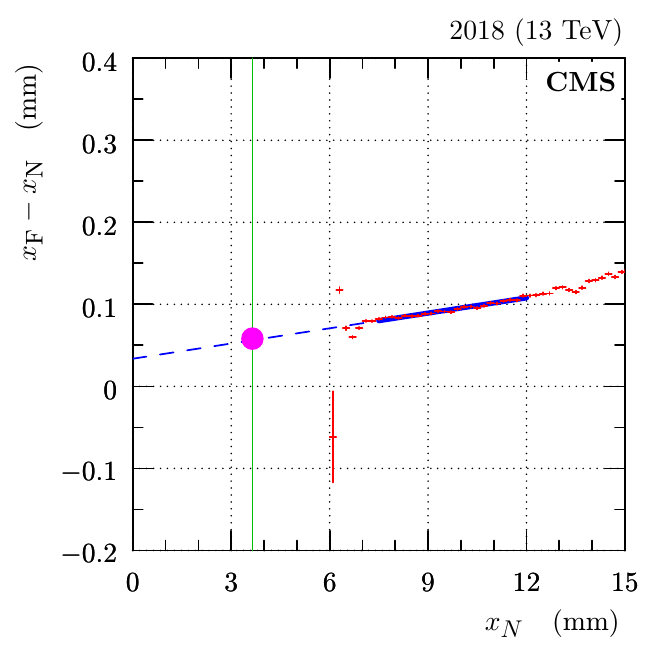}
}
\caption{{Left}: illustration of the absolute horizontal alignment (fill 5424, 2016 post-TS2, far RP in sector 45). Black: data from the reference alignment fill, blue: data from a physics fill before the alignment and red: data from the physics fill, aligned to match with the black reference. The error bars represent the bin sizes (horizontally) and statistical uncertainties (vertically).
{Right}: illustration of horizontal near-far relative alignment (fill 7052, 2018 and sector 45). Red: mean value of $x_\mathrm{F} - x_\mathrm{N}$ as function of $x_\mathrm{N}$. Blue: fit and extrapolation to the horizontal beam position (vertical green line, \eg~from the left plot). The value of the relative near-far alignment correction is indicated by the magenta dot. The error bars represent the bin sizes (horizontally) and statistical uncertainties (vertically).}
\label{fig:alig-phys-fill-x}
\end{figure}

For each high-luminosity LHC fill (``physics'' fill), the horizontal RP alignment is obtained by matching observations from the fill to those from the reference ``alignment'' fill, cf.~Section \ref{sec:al-alignment fill}. Various matching metrics have been used, and some of the first choices are discussed in~Ref.~\cite{Kaspar:2256296}). Eventually the procedure converged to:
\begin{equation}
S(x) = \hbox{slope of profile } (y_\mathrm{F} - y_\mathrm{N}) \hbox{ vs. } y_\text{test},
\label{eq:al-match-metric}
\end{equation}
where $y_\mathrm{N}$ and $y_\mathrm{F}$ stand for the vertical track positions in the near and far RP, respectively. Similarly, $y_\text{test}$ refers to the vertical track position in the RP being aligned. The shape of the profile is illustrated in Fig. \ref{fig:alig-phys-fill-slope}, where the value of $S$ corresponds to the slope of the red line. The $x$ dependence of the $S$ function is generated by the LHC optics, cf.~Section \ref{sec:optics}: $y$ is mostly given by the vertical effective length, $L_y(\xi)$, and $\xi$ is largely correlated with $x$ because of the large horizontal dispersion. The optics has been verified to be stable in time and therefore $S(x)$ is suitable for matching observations between different fills. Furthermore, the function from Eq.~(\ref{eq:al-match-metric}) is convenient because of its slope character: vertical misalignments (shifts in $y$) cause no bias and unavailable parts of the phase space (\eg~because of localized radiation damage) do not have any detrimental impact since the slope can still be determined from the available part. The matching procedure is illustrated in Fig. \ref{fig:alig-phys-fill-x}, left: the $S(x)$ curve from the test fill (blue) is shifted left and right until the best match with the $S(x)$ curve from the reference fill (aligned with the method from Section \ref{sec:al-alignment fill}) is found. The shift between the blue and red curves is then used as the alignment correction.

The relative alignment between the RPs within the same arm is then refined with a dedicated method with a better sensitivity --- good calibration of the relative alignment is essential for some of the proton reconstruction techniques. The relative near-far alignment method is based on comparing horizontal track positions in the near and far RPs, $x_\mathrm{N}$ and $x_\mathrm{F}$, respectively. The procedure is illustrated in Fig. \ref{fig:alig-phys-fill-x}, right: the profile $x_\mathrm{F} - x_\mathrm{N}$ vs.~$x_\mathrm{N}$ (red) is extrapolated (blue dashed) to the value of $x_N$ corresponding to the beam position (green). The extrapolated value of $x_\mathrm{F} - x_\mathrm{N}$ (magenta dot) then gives the relative-alignment correction. In general, the $x_\mathrm{F} - x_\mathrm{N}$ difference can be generated either by misalignments (independent of the horizontal position) or by the optics (roughly proportional to horizontal displacement from the beam). The extrapolation to the beam position, where the displacement from beam is $\approx$0, thus suppresses the optics contribution and keeps the misalignment component only.

\begin{figure}
\centerline{
	\includegraphics{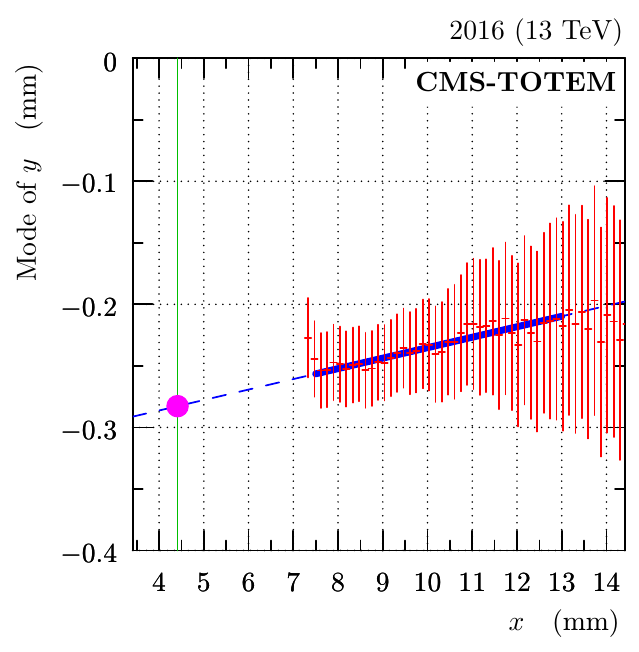}
}
\caption{Illustration of the vertical alignment (fill 5424, 2016 post-TS2, far RP in sector 45). Red: mode (most frequent value) of $y$ as a function of $x$, Blue: fit and extrapolation to the horizontal beam position (indicated by the vertical green line and extracted from Fig. \ref{fig:alig-phys-fill-x}, left). The value of the vertical alignment correction is indicated by the magenta dot. The error bars represent the systematic uncertainties.}
\label{fig:alig-phys-fill-y}
\end{figure}

The vertical alignment is obtained by extrapolating (blue) the observed vertical profile (red) to the horizontal beam position (green), as shown in Fig. \ref{fig:alig-phys-fill-y} where the alignment correction is marked with the magenta dot. The extrapolation to the beam position suppresses the optics contributions and keeps the misalignment component only. The mode (most frequent value) of $y$, contrary to the mean of $y$, is a local estimator not considering the tails of the $y$ distribution, which can be truncated because of the limited sensor size or other acceptance related effects. This vertical alignment method is sufficiently sensitive to provide both absolute per-RP and relative near-far alignment.

\begin{figure}
\centerline{\includegraphics[width=\textwidth]{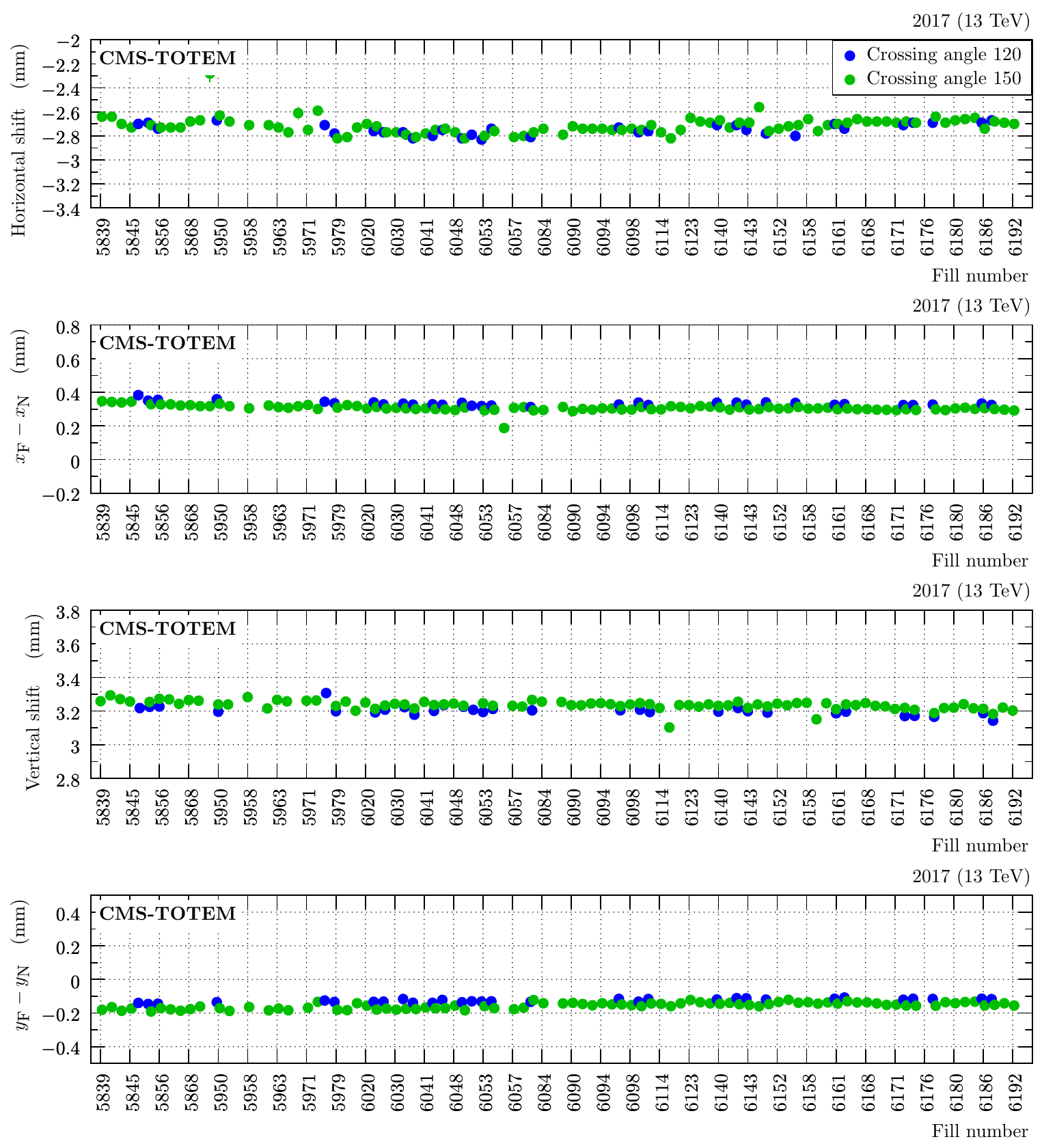}}
\caption{
Example of per-fill alignment results (2017 post-TS2, sector 56 or near RP in sector 56). Horizontal axis contains representative LHC fills where PPS was active. The colors indicate two values of crossing angle, 120 $\mu$rad (blue) and 150 $\mu$rad (green). Rows from top to bottom: absolute horizontal alignment, near-far relative horizontal alignment, absolute vertical alignment and near-far relative vertical alignment. The error bars (mostly invisible) represent a combination of statistical and systematic uncertainties.
}
\label{fig:alig-phys-fill-res}
\end{figure}

Figure \ref{fig:alig-phys-fill-res} shows a summary of per-fill alignment results for one alignment period. It also illustrates one of the many systematic validations performed; compatible results are expected from data sets obtained with different values of the crossing angle, $\beta^*$, or different central-detector triggers (the vast majority of the protons reaching the RPs are due to pileup unrelated to the triggering event).

Figure \ref{fig:alig-phys-fill-res} also confirms the expectation of fill independence of the alignment results. A fit of the results is used to remove occasional outliers, improve fill-to-fill stability and increase the overall accuracy of the alignment. In Run 2, there were two alignment periods where significant time variation was observed for some RPs. A notable example is 2016 pre-TS2 (additional details are discussed in Section 3.5 of Ref.~\cite{Kaspar:2256296}) where a package of sensors was initially wrongly inserted into a RP and over time the package slowly drifted to its nominal position due to the spring included in the RP assembly. Even in these cases, the variation was slow enough that fits could be applied to suppress the excessive fill-to-fill fluctuations and thus improve the results.

The alignment uncertainties are presented in Table \ref{tab:alig-unc}. They are estimated from fill-to-fill result fluctuations in cases where identical results are expected.

\begin{table}[ht]
\centering
\topcaption{Summary of per-fill alignment uncertainties.}
\begin{tabular}{ccc}
\hline
Projection & Absolute & Relative (near-far) \\
\hline
Horizontal & 150\mum & 10\mum \\
Vertical   & 100\mum & 10\mum \\
\hline
\end{tabular}
\label{tab:alig-unc}
\end{table}

\subsection{Timing RPs}
\label{sec:al-timing}

\begin{figure}
\centerline{
	\includegraphics{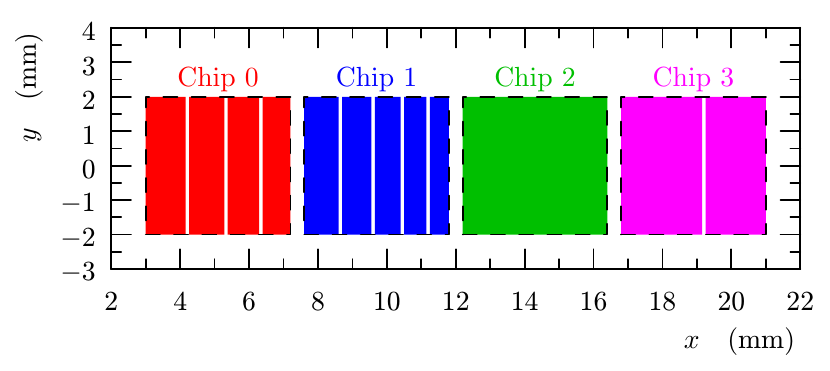}
}
\caption{
Example of timing detector segmentation in one plane (plane 1 in 2018 configuration). The beam is at $x = 0\mm$. Chip boundaries are drawn as dashed black rectangles. Pads are visualized as thick vertical strips, their colors indicate the chip relation.
}
\label{fig:alig-timing}
\end{figure}

The timing RPs consist of four sensor layers, called ``planes'', perpendicular to the LHC beam. As shown in Fig. \ref{fig:alig-timing}, each plane is composed of four physical pieces of diamond substrate, called ``chips''. Each chip has a structure of readout electrodes in the form of thick vertical strips, called ``pads''. This structure constitutes the horizontal segmentation of the timing detector and, in general, is different for each plane and chip.

The timing sensors are aligned with respect to the tracking RPs to associate local tracks using timing and tracking RPs (cf.~Fig. \ref{fig:reco-timing}). Since the timing RPs have only horizontal segmentation, only $x$ alignment is performed. The alignment is performed individually for each plane and pad as well as for each LHC fill.

\begin{figure}
\centerline{
	\includegraphics{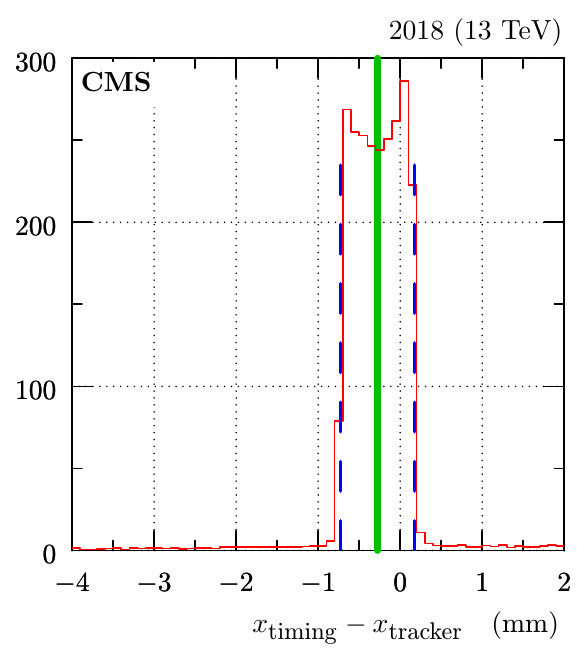}
}
\caption{Illustration of the timing-RP alignment method (fill 7137, 2018, sector 56, plane 1 and pad 9). 
The red histogram shows the difference between the horizontal track position in the timing sensor, $x_\text{timing}$,
and the track interpolated from the tracking RPs, $x_\text{tracker}$. The vertical blue dashed lines indicate the identified pad boundaries, the green line the pad center.}
\label{fig:alig-timing2}
\end{figure}

As illustrated in Fig. \ref{fig:alig-timing2}, the alignment method is based on a histogram of horizontal residuals between the hit position in the timing sensor and the track interpolated from the upstream and downstream tracking RPs. The histogram of these residuals (red) reveals the ``shape'' of the pad, the pad edges (dashed blue) as well as the pad centre (green). The alignment correction is given by the offset of the green line from zero. Estimated correction uncertainty is $100\mum$, driven by the uncertainties of the extracted pad edge positions.

\begin{figure}
\centering
\includegraphics{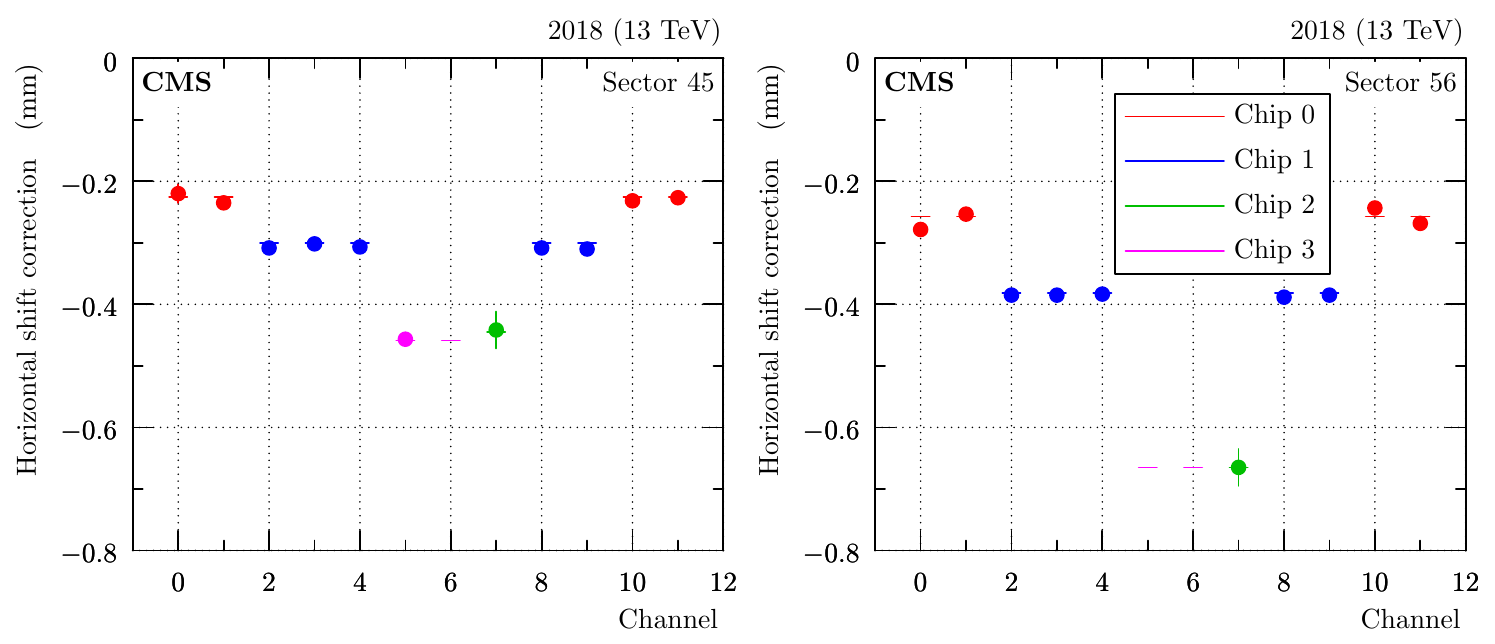}
\caption{
An example of alignment corrections in a single timing RP sensor plane (fill 7137, 2018, plane 1). Two different markers are used: the dots represent per-channel measurements, while the short horizontal lines represent per-chip averages. The same color is used for channels/pads placed on the same diamond chip, following the scheme in Fig. \ref{fig:alig-timing}. For chip 3 (most far from the beam) sometimes the track statistics is insufficient for alignment determination. In such cases the magenta thick dot is missing. The error bars represent a combination of statistical and systematic uncertainties reflecting the sharpness of the pad boundaries shown in Fig.~\ref{fig:alig-timing2}. {Left}: sector 45, {right}: sector 56.
}
\label{fig:alig-timing-res}
\end{figure}

A typical example of alignment corrections is shown in Fig. \ref{fig:alig-timing-res}. As expected, we find compatible results for the pads on the same physical chip, cf.~Fig. \ref{fig:alig-timing}. The average per-chip correction is indicated by the short horizontal line. The result pattern can be explained by the mechanical process of gluing the chips on the board --- the chips cannot mechanically overlap, only additional gaps can be introduced. This leads to a cumulative misalignment monotonically increasing (in absolute value) with the chip number, as revealed by the results. Chip 3, the most far from the beam, often gets an insufficient number of tracks (because of the LHC collimators, cf.~Section \ref{sec:aperture}) and the correction from chip 2 is used in this case.

\clearpage
\newpage

\section{Optics model and calibration}
\label{sec:optics}

\subsection{Introduction}
\label{sec:optics_introduction}
In Run 2, the LHC optics settings and conditions were modified every year. The key concepts and the tools to constrain the main optical functions using collision data for 2016 have been described in Refs.~\cite{Antchev:2014voa, Nemes:2256433}. During physics runs, the luminosity of the LHC beams decreases naturally due to bunch intensity decay.
Luminosity can be regained for the experiments by adjusting the crossing angle and betatron amplitude to increase the so-called luminosity geometry factor. To achieve this goal in 2017 the levelling of the crossing angle and of the betatron amplitude $\beta^{*}$ was introduced. In 2018 the levelling of both parameters became continuous~\cite{Fartoukh:2293518}.

The modelling of this varying optics and its calibration required a generalization of the well-established 2016 methods; the higher number of events permitted, and also required, a more careful dispersion calibration. The vertical position of the beams crossing point, $y^{*}$, also changed with respect to 2016. In the last two years of Run 2, the optics had a sizable vertical dispersion $D_{y}$, which is an important optical function for the reconstruction. An optics uncertainty model based on collision data is also presented.
The optics calibration methods of Run 2 are briefly discussed from the viewpoint of the HL-LHC in Ref.~\cite{CMS:2021ncv}.

\subsubsection{Proton transport at the LHC}
\label{subsec:optics_introduction}

The transport matrix and the optical functions have already been introduced in Section~\ref{sec:optics_intro}. In the following, the meaning of the transport 
matrix elements is explained, with emphasis on the connection between the $\beta$ amplitude and the optical functions used in the reconstruction. Specifically, the horizontal and vertical magnifications
\begin{equation}
v_{x,y}=\sqrt{\beta_{x,y}/\beta^*}\cos\Delta\mu_{x,y},
\label{eq:vxy}
\end{equation}
and the effective lengths
\begin{equation}
 L_{x,y}=\sqrt{\beta_{x,y}\beta^*}\sin\Delta\mu_{x,y},
\label{eq:Lxy}
\end{equation}
are functions of the betatron amplitudes $\beta_{x,y}$, their value $\beta^{*}$ at IP5 and the relative phase advance
\begin{equation}
\Delta\mu_{x,y}=\int^\text{RP}_\text{IP}\frac{\rd  l}{\beta_{x,y}}.
\end{equation}

The beam size can be calculated from the beam emittance $\varepsilon$ of the LHC and from the betatron amplitude
\begin{align}
    \sigma(x)=\sqrt{\beta_{x}\varepsilon} \approx 13\mum,
    \label{eq:beam_size}
\end{align}
using a representative $\beta_{x}=0.3$\,m value, where $\varepsilon$ is computed from the normalised emittance $\varepsilon_{N}=(\beta_\mathrm{L}\  \gamma_\mathrm{L})\varepsilon=3.75$\mum~rad. Here $\beta_\mathrm{L}=v/c$; $v$ is the velocity of the beam particles, $c$ is the speed of light and $\gamma_\mathrm{L}=(1-\beta_\mathrm{L}^{2})^{-\frac{1}{2}}$ is the Lorentz factor. The subscript ``L'' is used in $\beta_\mathrm{L}$ and $\gamma_\mathrm{L}$ to avoid confusion. The Liouville theorem dictates that
\begin{align}
    \pi\ \sigma(x)\ \sigma(x')=\pi\varepsilon,
    \label{eq:liouville}
\end{align}
where $\sigma(x')$ is the beam divergence,  \ie the angular spreading, of the LHC beams; the symbol $x'$ stands for $\rd{x}/\rd{l}$~\cite{Wilson:2001hw}. Therefore, from Eq.~(\ref{eq:liouville}) it follows that  $\sigma(x')=\sqrt{\beta_{x}^{-1}\varepsilon}\approx 40\murad$ for the representative $\beta_{x}=0.3$\unit{m} value, which gives the limit on the resolution of the scattering angle $\theta_{x,y}^{*}$ of PPS~\cite{Bruning:782076}.

As already mentioned, in 2017, the necessity to improve the lifetime of the beams led to the change or ``levelling" of both the betatron amplitude, $\beta^{*}$, at IP5  in discrete steps and the horizontal crossing angle. In 2018 both parameters were modified continuously (cf. Table~\ref{tab:beam_parameters}). For comparison at IP1 (ATLAS) the crossing angle bump was in the vertical plane during Run 2 to avoid long range beam-beam interactions~\cite{Herr:604005}. The levelling is based on the so-called Achromatic Telescopic Squeezing (ATS) optics~\cite{Fartoukh:2293518}; one of its features is that the optical functions Eq.~(\ref{eq:vxy}) and Eq.~(\ref{eq:Lxy}) remain constant despite the change in $\beta^{*}$. Therefore, the $\beta^{*}$ levelling is a transparent operation from the viewpoint of the reconstruction. The horizontal dispersion $D_{x}$ determines the proton trajectory in the horizontal plane and depends on the crossing angle levelling at IP5; therefore $D_{x}$ is calibrated separately for each reference crossing angle.

            \begin{table}[ht]
            \begin{center}
            \topcaption{Summary of main beam parameter values, crossing angle and $\beta^{*}$, during the Run 2 period per year. In 2017 the values changed in discrete steps, whereas in 2018 there was a continuous change within the interval.}
                \begin{tabular}{  c  c  c }
            \hline
             Year & Half horizontal crossing angle ($\mu$rad) & $\beta^{*}$ (m)\\\hline
             2016 & 140,185 & 0.4\\
             2017 & 120,130,140,150 & 0.3,~0.4\\
             2018 & [130,~160]   & [0.25,~0.4]\\\hline
            \end{tabular}
            \label{tab:beam_parameters}
            \end{center}
        \end{table}

The transport equation Eq.~(\ref{eq:transport_vector_equation}) can be explicitly written at the RPs in the form
        \begin{equation}
            \label{eq:transport_equations}
            \begin{aligned}
            x &= x_{0} + D_{x}\cdot  \xi +  L_{x}(\xi) \cdot  \theta^{*}_{x} + v_{x}(\xi) \cdot  x^{*}, \\
            y &= y_{0} + D_{y}\cdot  \xi +  L_{y}(\xi) \cdot  \theta^{*}_{y} + v_{y}(\xi) \cdot  y^{*}, 
            \end{aligned}
       \end{equation}
that describes the connection between the proton kinematics at the IP5 and at the RPs, where $x_{0}$ and $y_{0}$ are the horizontal and vertical beam position, respectively. The
horizontal dispersion $D_{x}$ is a function of $\xi$, therefore it is useful to define a function 
that provides the horizontal position of a proton with momentum loss $\xi$ directly
    \begin{equation}
        x_\rd (\xi)=D_{x}(\xi) \cdot  \xi
        \label{definition_of_horizontal_dispersion},
    \end{equation}
and one can define similarly $y_\rd (\xi)$~\cite{Anelli:2008zza,Grote:2003ct}.

The coupling terms $m_{ij}$ in the transport matrix Eq.~(\ref{eq:transport_matrix}) connect the horizontal and vertical scattering planes. At the LHC, like for most accelerators, these terms are set to zero nominally $m_{13},\ldots,m_{42}\approx 0$ for collision optics. They receive perturbative-level corrections because of skew quadrupole corrector magnets. The effect of the coupling on the reconstruction of the proton kinematics was negligible for all years.

The optics calibration assumes the beam-based alignment of the detectors, after which the beams appear at $x_{0}=y_{0}=0$, cf. Eq.~(\ref{eq:transport_equations})~\cite{Kaspar:2256296}. The horizontal position of the protons, $x(\xi)$,
is a nonlinear function of $\xi$, which can be approximated for low $\xi$ values
            \begin{equation}
                x \approx \left.D_{x}(\xi)\right|_{\xi=0}\cdot \xi,
                \label{eq:shift_due_to_dispersion} 
            \end{equation}
where the resolution in $x$ is limited by the spreading because of the scattering angle term $L_{x}\cdot \theta_{x}^{*}$ and by the contribution of the vertex $x^{*}$, cf. Eq.~(\ref{eq:transport_equations}).

\subsection{Calibration of the LHC optics}
\label{sec:optics_calibration}

The horizontal dispersion $D_{x}$ is the most important optics quantity, because it allows one to convert the $x$-coordinate measurements at the RPs into the fractional proton momentum loss $\xi$.
The determination of $D_{x}$ from the measured proton tracks is briefly reviewed in the next section (cf. also Ref.~\cite{Antchev:2014voa}). The 2017 and 2018 optics calibration procedure goes a step further and also exploits (semi)-exclusive $\mu\mu$ production; the exclusivity of the process plays a key role in the calibration, as illustrated in Section~\ref{sec:dilepton}.

In the last step of the calibration procedure, the vertical dispersion $D_{y}$ is determined from minimum bias RP data. The calibration of the dispersion functions is followed by the calibration of the remaining optical functions in the transport matrix Eq.~(\ref{eq:transport_matrix}), namely the horizontal, $L_{x}(\xi)$, and vertical, $L_{y}(\xi)$, effective lengths, and the corresponding magnification functions; other optical functions are less relevant for the proton reconstruction.

The above optics calibration steps rely on the nominal transport model, which is taken from LHC databases. The transport matrix is defined by the machine settings $\mathcal{M}$, which are obtained from several data sources. The proper version of the
LHC magnet lattice description, known as ``sequence", is used each year. The nominal magnet strength file for a given beam optics is always updated using measured data: the currents of the magnets power converters $I_{\text PC}$ are first retrieved using
TIMBER~\cite{Billen:1000757}, an application to extract data from heterogeneous databases containing information about the whole LHC infrastructure. The currents $I_{\text PC}$ are converted to magnet strengths with the LHC software architecture (LSA)~\cite{Roderick:1215575}, which uses the conversion curves from the field description for the LHC (FIDEL)~\cite{Aquilina:1742065}.

\subsubsection{The $L_{y}=0$ method}
\label{sec:Dx_calibration}

This procedure uses the minimum bias data recorded during the special low-luminosity runs mentioned in Section~\ref{sec:al-alignment fill}. The method has been applied for each year within Run 2; for 2017 and 2018, a separate calibration was carried out for each crossing angle. The procedure assumes the calibration of the vertical effective length $L_{y}$ for low-$\xi$ values, below $\xi\approx4$\%, using the elastic candidate events measured in the vertical RPs; this additional step is reported in detail in Refs.~\cite{Antchev:2014voa,Nemes:2131667}.

\begin{figure}[h!]
    \centering
        \includegraphics[width=0.7\textwidth]{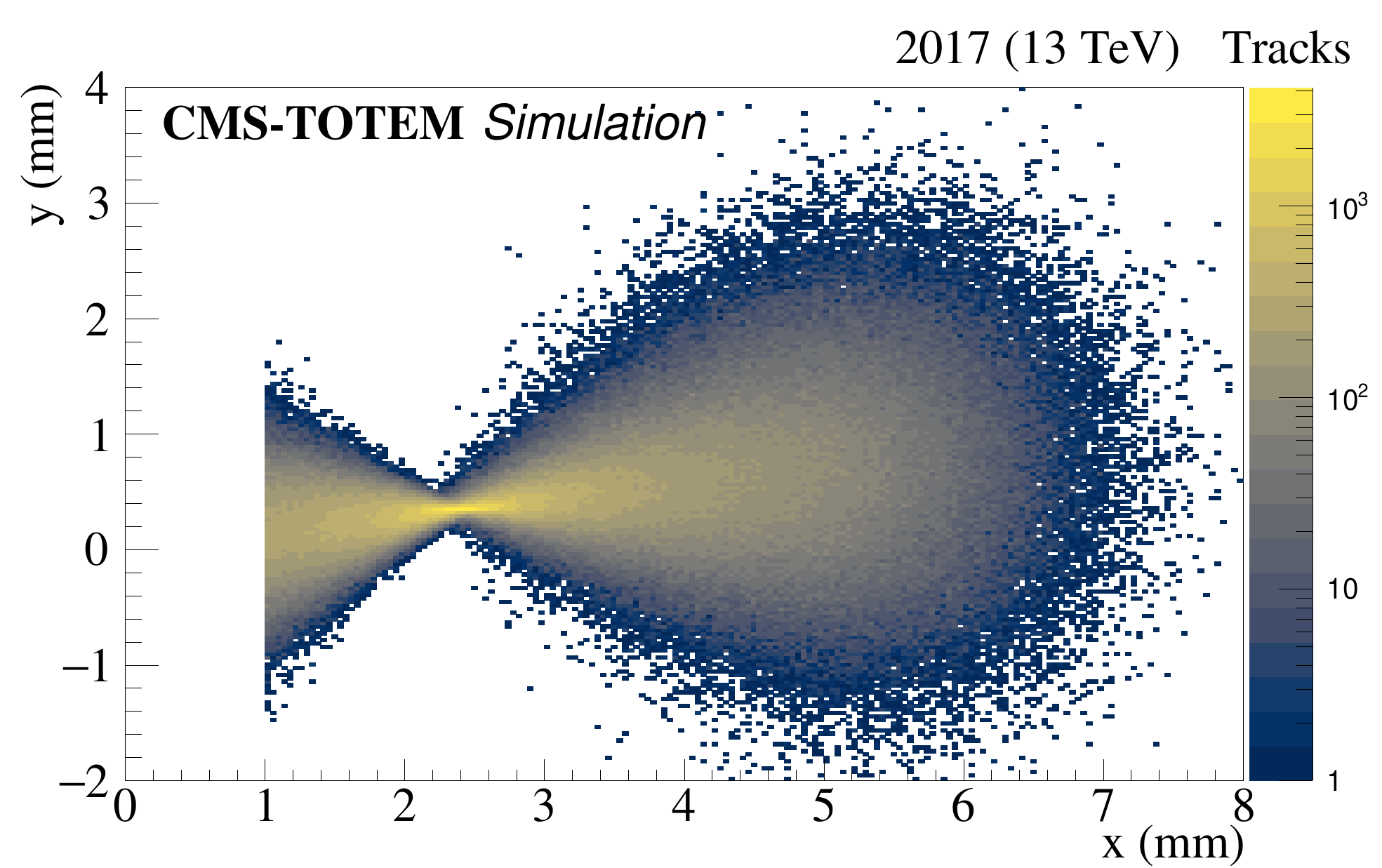}
    \caption{The $(x,y)$ distribution of simulated proton tracks in the near RP in sector 56 using MAD-X. It illustrates the ``pinch'' or focal point at $x = x_\mathrm{F}$ where the vertical effective length vanishes: $L_{y}(\xi_\mathrm{F}) = 0$, given the relation $y \approx {L_{y}(\xi_\mathrm{F})}\cdot \theta_{y}^{*}$. The simulation takes into account that the small vertical dispersion moves particles upward according to $\Delta y=D_{y}\ \xi$  with increasing $x$, and $\xi$.}
    \label{fig:Ly_procedure}
    \end{figure}

The LHC optics are calculated with the methodical accelerator design (MAD-X) program, a general purpose beam optics and lattice software~\cite{Grote:2003ct}. The vertical effective length $L_{y}(\xi)$ is a function of the proton momentum loss $\xi$, and can be calculated with MAD-X at each RP location with good accuracy. The calibration is based on the observation that $L_{y}(\xi)$ is positive at $\xi=0$,
monotonically decreases with increasing $\xi$ reaching large negative $L_{y}$ values and it vanishes at about
$\xi\approx~4\%$. According to Eq.~(\ref{eq:transport_equations}) at this $\xi_\mathrm{F}$ value every proton is transported to the same vertical coordinate $y = 0$ regardless of the vertical scattering angle $\theta_{y}^{*}$ (the vertex contribution is neglected). At the same time these protons appear at the horizontal location $x_\mathrm{F}\approx D_{x} \cdot \xi_\mathrm{F}$. Consequently, the $(x,y)$ distribution of the protons has to exhibit a ``pinch", or focal point, at this horizontal location $x_\mathrm{F}$, cf. Fig.~\ref{fig:Ly_procedure}.

\begin{figure}
    \centering
        \includegraphics[width=0.7\textwidth]{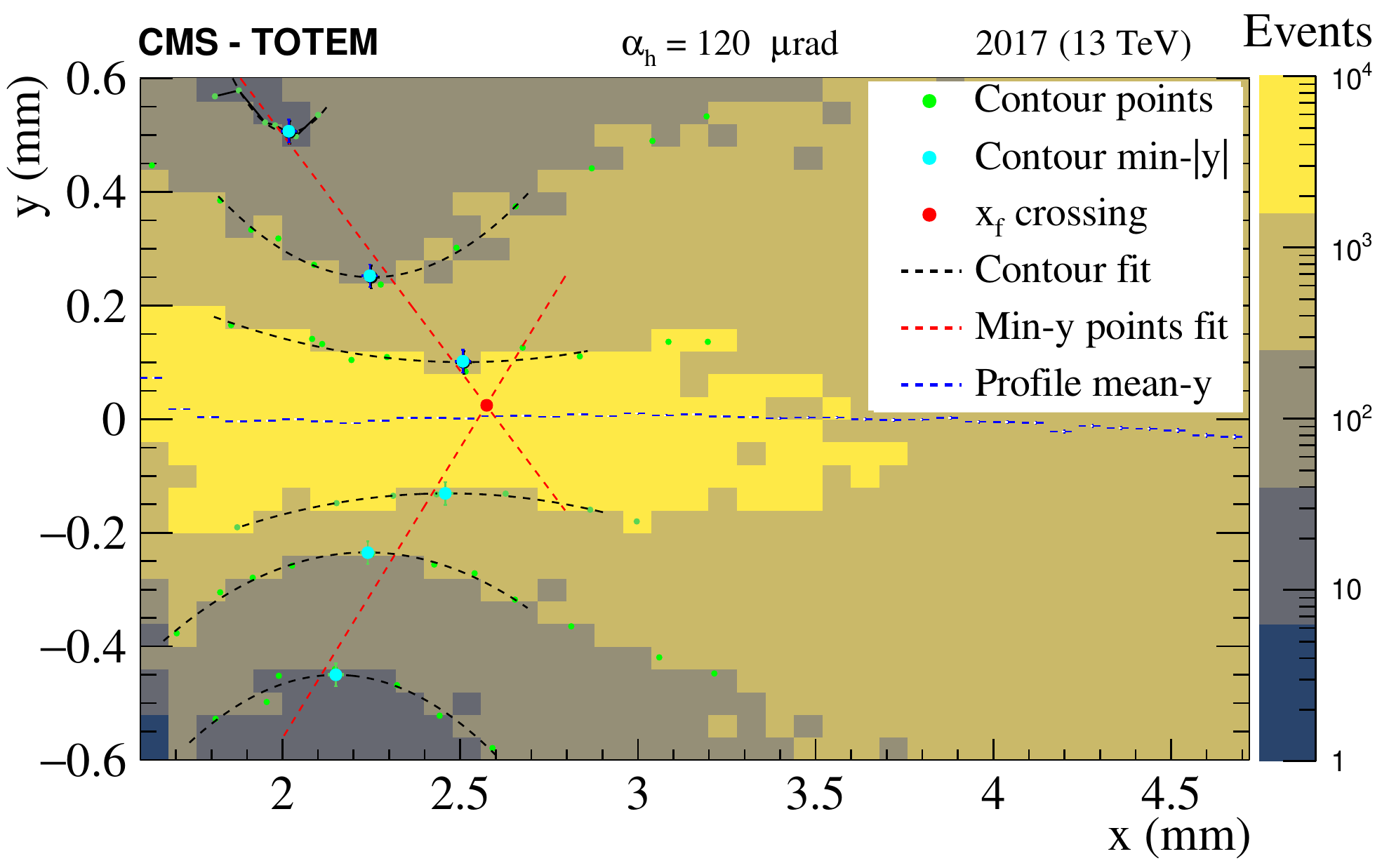}    
    \caption{The $(x,y)$ distribution of the proton impact points in the near RP detector in sector 45 for 2017 using minimum bias data, along with parabolic fits of the contours around the ``pinch''. The minima of the parabolas are fitted with a straight line. The intersection of the two lines is marked with a red dot, and indicates the estimated focal point coordinate $x_\mathrm{F}$. To make the contour curves and extrapolation symmetric, the mean of the histogram was aligned to 0 to remove the $y$ offset created by the vertical dispersion $\Delta y=D_{y}\ \xi$. The vertical error bar on the contour minima, blue points, represents the statistical uncertainty of the fit.}
    \label{fig:Ly_procedure_data}
    \end{figure}

The LHC optics transport is the same for all protons, thus the focal point can be observed and measured with the horizontal RP detectors using large statistics minimum bias data, cf. Fig.~\ref{fig:Ly_procedure_data}. The figure shows the $(x,y)$ distribution of the proton impact points in the RP detectors for 2017 for a representative half crossing angle $\alpha_{h}=120\,\mu$rad. The plot shows the parabolic fit of the contour curves around the ``pinch'' point. The minima of the parabolic curves are fitted with a linear function and the fits are extrapolated. The intersection of the linear fits is marked with a red dot, and indicates the estimate of the focal point position $x_\mathrm{F}$. The fit of the contour lines and the extrapolation are used in order to estimate the bias coming from the scattering angle $\theta_{x}$ and extrapolate to the point where the bias vanishes. The measurement is repeated with the distribution obtained after a selection on the scattering angle $\theta_{x}$ to reduce the horizontal spreading around the focal point; in this case the parabolic fits are not needed.

The dispersion is estimated as
\begin{align}
    \left.{D_{x}(\xi)}\right|_{\xi=\xi_\mathrm{F}} = \frac{x_\mathrm{F}}{\xi_\mathrm{F}}.
\end{align}
The measured $D_{x}$ values are used to calibrate the LHC optics model, as described in the next sections. The uncertainty of the $L_{y}=0$ method includes the uncertainty of the contour fits, their minimum and their linear extrapolation;
the systematic uncertainty due to remaining bias is estimated with a Monte Carlo simulation.

  \subsubsection{Calibration using the (semi)-exclusive $\mu\mu$ process}
  
In 2016 PPS collected its first (semi)-exclusive dilepton sample~\cite{Cms:2018het}, $\Pp\Pp \to \Pp^{(\ast)}\ell+  \ell\Pp^{(\ast)}$, where a pair of leptons ($\ell = \Pe$, \PGm) is reconstructed in the
central CMS apparatus, one of the protons is detected in PPS, and the
second proton either remains intact or is excited and then dissociates
into a low-mass state, indicated by the symbol $\Pp^{(\ast)}$, and escapes undetected.
Section 11 focuses on the $\mu\mu$ measurement, whereas the implications on
the optics calibration are presented here.

The (semi)-exclusivity implies a high-purity data set: in these events, the central $\mu\mu$ system carries the momentum lost by the two forward protons. Therefore, the difference of the fractional momentum loss reconstructed from PPS and from the
central CMS detectors can be determined; the correction to $D_{x}$ is computed such that this difference vanishes. The improved calibration result for $D_{x}$ remains within the uncertainty of the $L_{y}=0$ method and the final $D_{x}$ result is the weighted average of the two measurements.
The uncertainty of $D_{x}$ is the combined uncertainty of the $L_{y}=0$ and the (semi)-exclusive $\mu\mu$ methods. The evolution of the dispersion $D_{x}(\xi)$ (or $x_{d}(\xi)$ cf.~Eq.~(\ref{definition_of_horizontal_dispersion})) with $\xi$ can be also validated using the $\mu\mu$ results. The $D_{x}$ results are shown in Table~\ref{tab:dispersion_results} with a conservative 8\% uncertainty in $D_{x}$, which applies to $x_\rd $ as well.

The dispersion asymmetry between the two arms was observed in 2016 and persisted in 2017 and 2018 as well; it is attributed to crossing angle asymmetry and quadrupole magnet misalignment within their nominal tolerance.

            \begin{table}[ht]
            \begin{center}
            \topcaption{Measured horizontal dispersion values $D_{x}$ in the near RP at low $\xi$ between 2\% and 4\% (the exact $\xi_\mathrm{F}$ value depends on the detector and the year). The resulting $D_{x}$ value is the weighted average of the $L_{y}=0$ and (semi)-exclusive $\mu\mu$ results. The quoted 8\% uncertainty in $D_{x}$ applies to the $x_{d}$ function as well.}    
                \begin{tabular}{  c  c  c  c }
            \hline
             Year & Half crossing angle ($\mu$rad) & Sector 45 (cm) & Sector 56 (cm) \\\hline
             2016 & 185 & $-9.7\pm0.4$  & $-6.7\pm0.4$ \\
             2017 & 120 & $-10.4\pm0.8$ & $-7.9\pm0.6$ \\
             2018 & 120 & $-11.3\pm0.9$ & $-8.7\pm0.7$ \\\hline
            \end{tabular}
            \label{tab:dispersion_results}
            \end{center}
        \end{table}

\subsubsection{Optics matching}
\label{sec:optics_matching}

The purpose of the optics fitting (or ``matching'') is the calibration of the LHC optics model using the measured dispersion values and other measured constraints. The calibration procedure consists of a $\chi^{2}$ minimization with MINUIT, where the initial optics model of the fit is taken from the LHC databases, as mentioned in Section~\ref{sec:optics_calibration}~\cite{James:1975dr}.

The first step is to constrain the quadrupole field model using the elastic candidates from the alignment fills, described in Ref.~\cite{Antchev:2014voa}. In the second step the measured dispersion values from Table~\ref{tab:dispersion_results} are used as inputs to the $\chi^{2}$ function, with additional constraints reflecting the LHC optics uncertainties:
\begin{equation}
    \label{chi2_constructions}
    \chi^2 = \chi_\text{design}^2 + \chi_\text{measured}^2.
\end{equation}

The following measurements from both LHC beams contribute to $\chi_{measured}^2$:
        \begin{itemize}
            \item the readings of three beam position monitors (BPMs) (at $l=22$\unit{m}, 58\unit{m}, 199\unit{m}), with an uncertainty $\sigma_{x,\text{absolute}}\approx 0.43$\mm;
            \item the beam position at RP 210, near, vertical, with an uncertainty $\sigma_{x}=0.5$\mm;
            \item the two measured dispersion values $D_{x}$ (1 per arm) with their measured uncertainty, cf. Table~\ref{tab:dispersion_results}.
        \end{itemize}

To match, or fit, the dispersion values and the LHC optics model, the relevant LHC machine parameters are varied during the minimization.  The matching procedure exploits the fact that a quadrupole magnet misaligned by a $\delta x$ offset gives a correction to the dipole field, whereas the quadrupole fields remain
unchanged. The following machine parameters have
    to be matched for the two LHC beams separately to obtain the orbit model for the proton reconstruction:
            \begin{itemize}
            \item horizontal (half) crossing angle $\alpha_{h}$;
            \item quadrupole positions ($\sigma_{x}=0.5$\mm, 6 parameters);
            \item kicker strength ($\sigma_{k}\approx3$\%, 3 parameters).
        \end{itemize}

\begin{figure}[ht]
\centering
\includegraphics[width=0.8\textwidth]{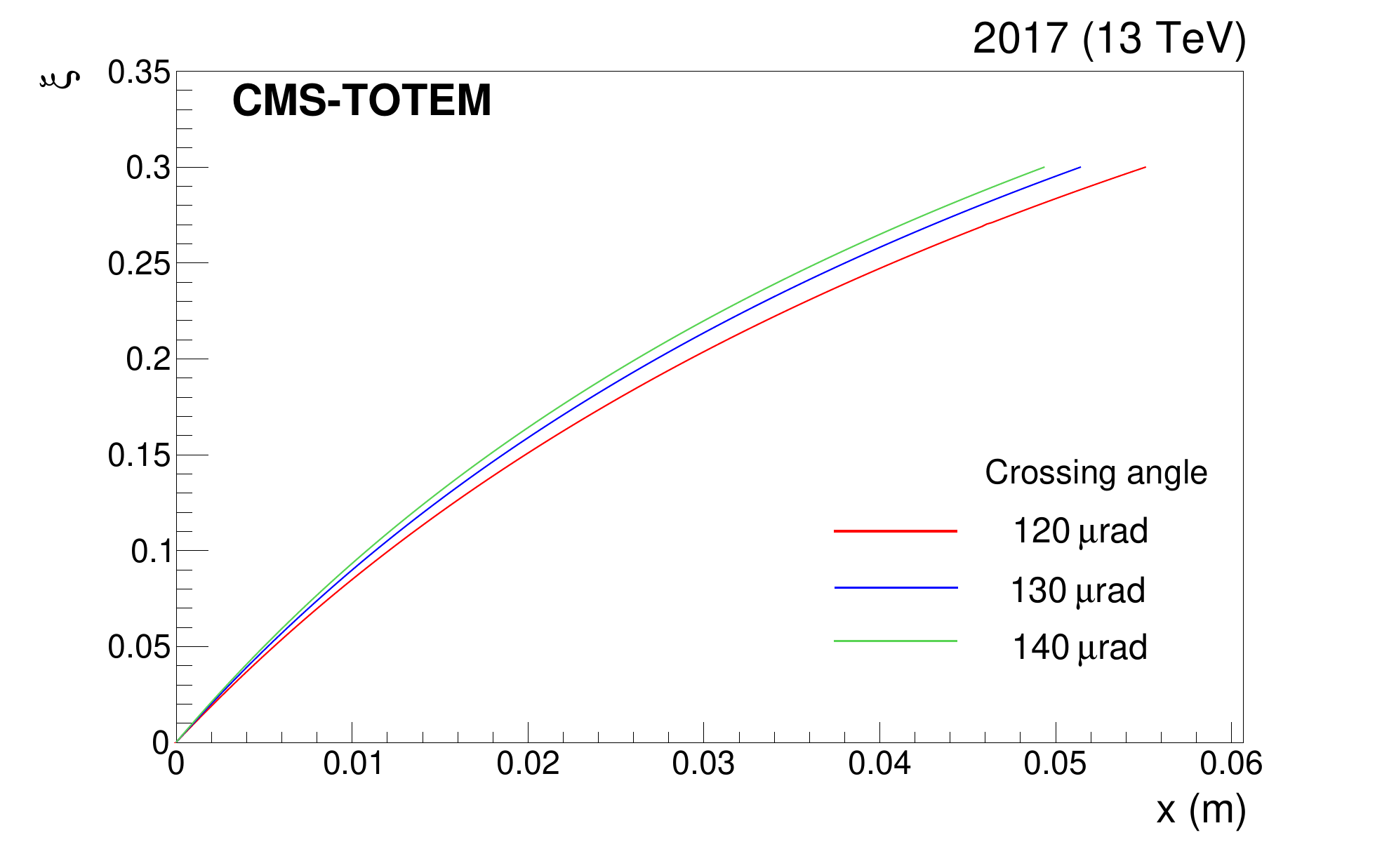}
\caption{The momentum loss of the protons $\xi$ as a function of $x$ in the near RP of sector 45. The dispersion function is $\xi$ dependent itself and the figure shows directly the nonlinear $x(\xi)=D({\xi})\cdot \xi$ function. The $\xi(x)$ function depends on the crossing angle as well; the figure shows the dependence for three reference angles, so the function can be interpolated to arbitrary intermediate angles.}    
\label{fig:x_to_xi_curves}
\end{figure}

    With this procedure a good confidence level was achieved for the lattice model of the two LHC beams. The matched MAD-X optics model is used to extend the measured dispersion values from Table~\ref{tab:dispersion_results} to higher $\xi$ values. An
example of the fitted result is shown in Fig.~\ref{fig:x_to_xi_curves}.\footnote{This matching procedure has been reviewed by the beam department (BE)  experts of the LHC.} 

The optics model MAD-X shows that the different interpretations of the dispersion asymmetry between sector 45 and 56 (crossing angle rotation, quadrupole misalignment, etc.) lead to negligible differences
in the systematic uncertainty, for example in the evolution of $D_{x}$ with $\xi$.

\subsubsection{Calibration of the vertical dispersion $D_{y}$}
\label{sec:calibration_of_the_vertical_dispersion}

In 2016 the vertical dispersion $D_{y}$ was close to zero, whereas in 2017 and 2018 the optics changed and a vertical dispersion $D_{y}\approx -1$\cm was applied. Despite its small value, the vertical dispersion has a strong effect on the $\xi$ dependence of the vertical reconstruction of $\theta_{y}^{*}$ and $y^{*}$ because of the nonlinearity of the other optical functions.

The vertical dispersion $D_{y}$ is estimated from the $(D_{y}/D_{x})$ ratio measured on the $(x,y)$ plane; the value is refined by perturbing it so as to match the measured $\theta_{y}^{*}$ and $y^{*}$ values as well.
\begin{figure}[ht]
    \centering
        \includegraphics[width=0.7\textwidth]{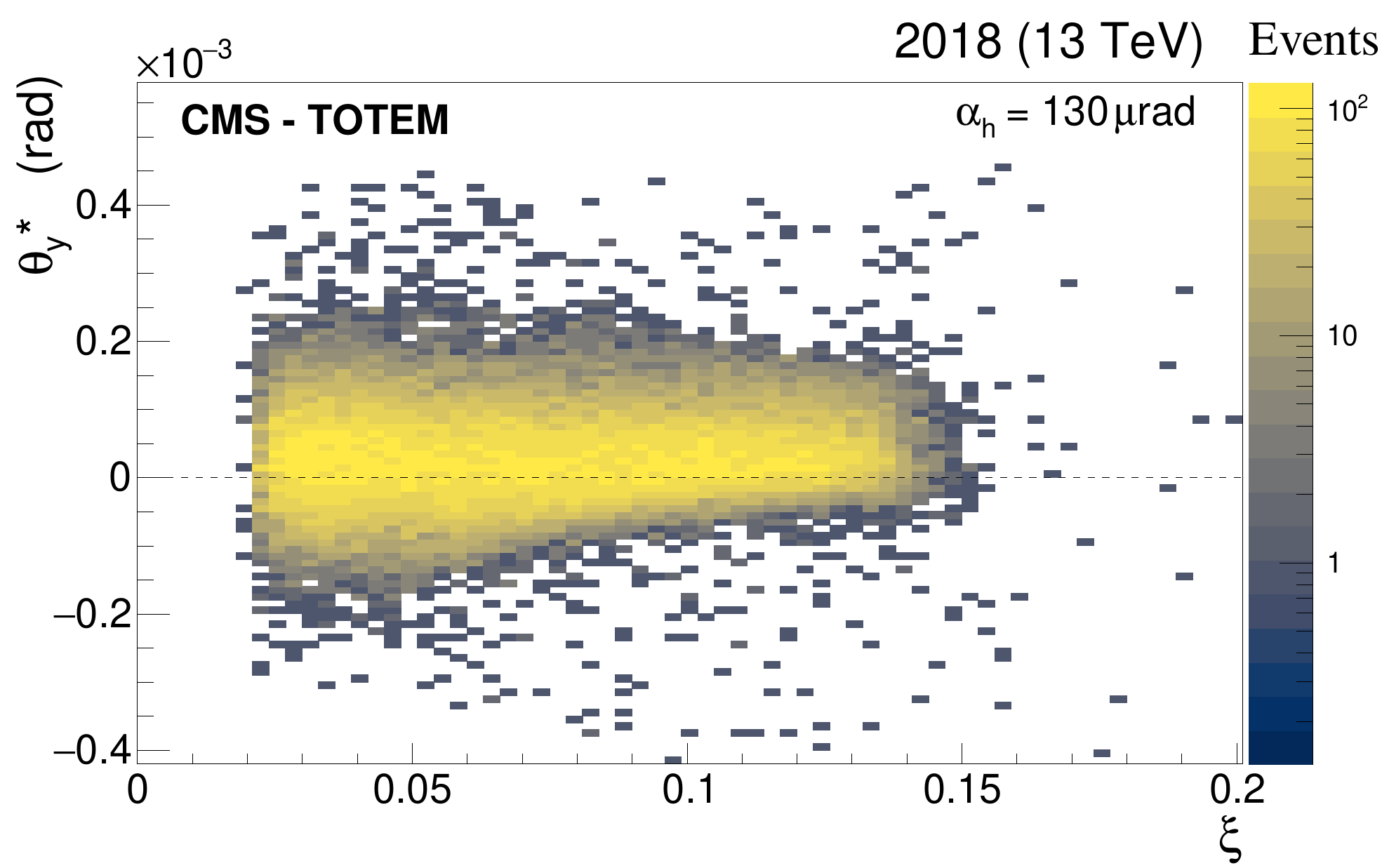}
        \caption{The vertical scattering angle $\theta_{y}^{*}$ as a function of $\xi$ after calibration of the vertical dispersion $D_{y}$ for sector 45 and fill 6923. The mean of the scattering angle distribution is consistent with 0. The distribution is also affected by the vertical acceptance limitations starting from about $\xi\approx 5$\% because of the vertical acceptance limits of the detector cf.~Fig.~\ref{fig:alig-overlap}.}
        \label{fig:Dy_calibration}
        \end{figure}

The measured vertical dispersion values are summarized in Table~\ref{tab:vertical_dispersion_results}. The values are small enough that the crossing angle dependence can be neglected. The vertical dispersion values $D_{y}$ are validated with minimum bias data, cf. Section~\ref{sec:reconstruction} and also Fig.~\ref{fig:Dy_calibration}.

            \begin{table}[ht]
            \begin{center}
            \topcaption{Final measured vertical dispersion values $D_{y}$ in the near RP per year. The uncertainty is derived conservatively from the measured $(D_{y}/D_{x})$ ratio.}				
                \begin{tabular}{  c  c  c }
            \hline
             Year &  Sector 45 (cm) & Sector 56 (cm) \\\hline
             2016 &  $0\pm0.02\stat$  & $0\pm0.02\stat$ \\
             2017 &  $-1.36\pm0.02\stat\pm{0.1\syst}$ & $-1.99\pm0.02\stat\pm{0.16\syst}$ \\
             2018 &  $-1.36\pm0.02\stat\pm{0.1\syst}$ & $-1.87\pm0.02\stat\pm{0.15\syst}$ \\\hline
            \end{tabular}
            \label{tab:vertical_dispersion_results}
            \end{center}
        \end{table}

\subsection{Optics description and uncertainty model}

The LHC optics model, calculated with MAD-X, can be described in several efficient ways for the event reconstruction and physics analysis~\cite{Grote:2003ct}. In the year 2016, the description of the proton transport used orthonormal polynomials to fit the $(x,y)$ coordinates of the protons at the RPs as a function of their input kinematics~\cite{Niewiadomski:1131825}.
 
  Experience with the data and optics modelling showed that the parametrization, or factorization, of Eq.~(\ref{eq:transport_equations}) is sufficient to describe the proton transport between IP5 and the RPs; therefore, since 2016 an expansion using only 1-dimensional $\xi$ dependent optical functions is applied. 
 
    \begin{figure}[ht]
    \centering
        \includegraphics[width=0.85\textwidth]{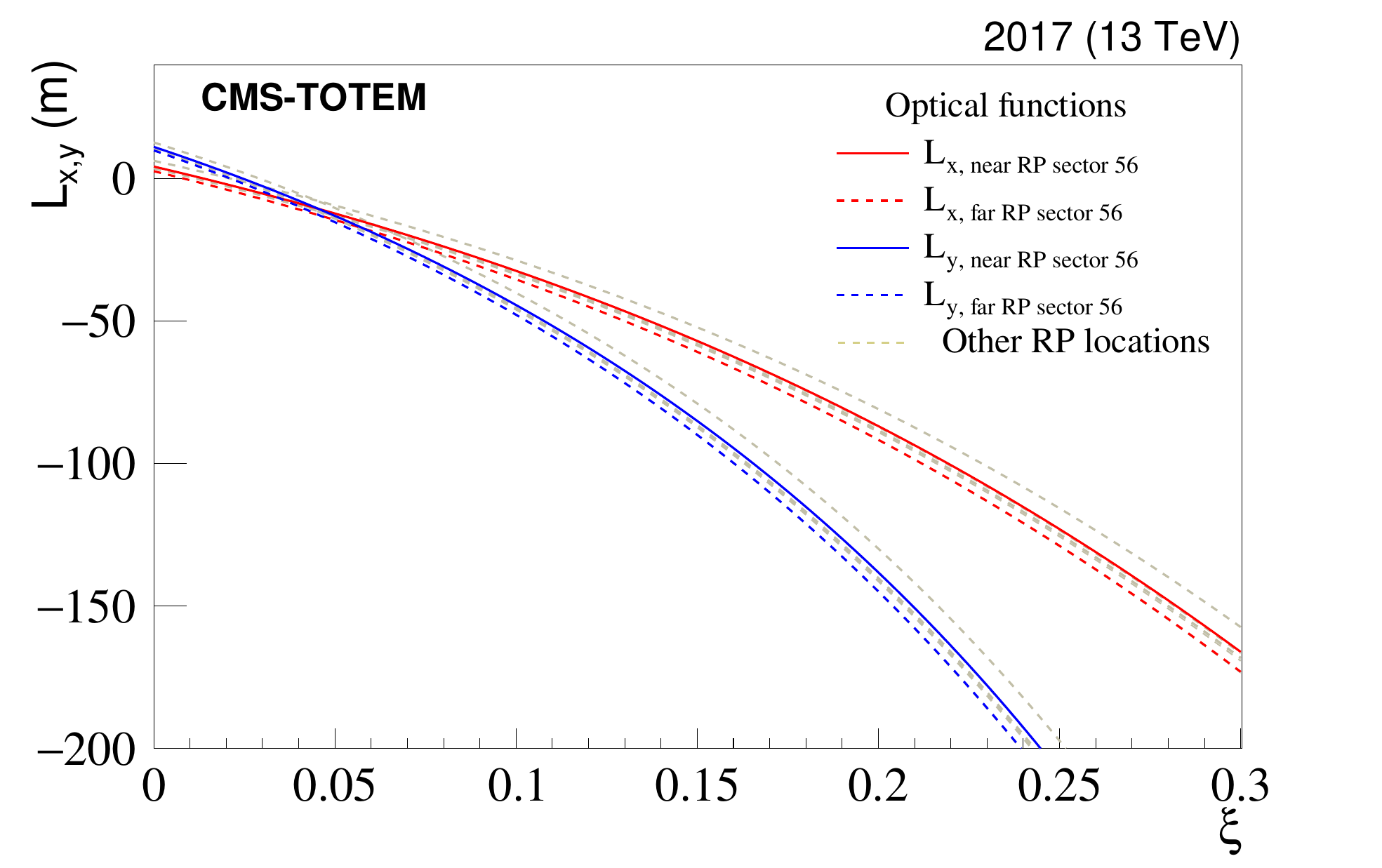}        
    \caption{The horizontal and vertical effective lengths $L_{x}$ and $L_{y}$ transport the scattering angle of the proton at IP5 $\theta_{x}^{*}$ and $\theta_{y}^{*}$ to the position $(x,y)$ at the RPs. The figure shows the $\xi$ dependence of the two functions. The horizontal effective length $L_{x}(\xi)$ decreases faster than the vertical function $L_{y}$; both of them cross zero at low $\xi$, below $\xi=5$\%. The grey dashed lines show the effective lengths for the TOTEM RPs used for calibration.}
    \label{fig:Lx_Ly_vs_xi}
	\end{figure}

As discussed earlier, in 2017 the levelling of the beam crossing angle was introduced. This is straightforward to take into account using the optical function concept with an additional extrapolation function among reference crossing angles, as shown in Eq.~(\ref{eq:interpolation_function}) and Fig.~\ref{fig:x_to_xi_curves}:
    \begin{equation}
    \label{eq:interpolation_function}
        x(\alpha_{h}, \xi) = x_{120}(\xi) + \frac{120 - \alpha_{h}}{120 - 140} \  \left[x_{140}(\xi) - x_{120}(\xi)\right].
    \end{equation}
The linear function is motivated by MAD-X and is compatible with the dispersion measurements within uncertainties. The other optical functions remain constant during the levelling of the crossing angle and, due to the telescopic concept of the ATS optics, they also remain constant during the levelling of $\beta^{*}$. The relevance of the ATS telescopic squeezing from the viewpoint of uncertainty model is discussed in Section~\ref{sec:uncertainty_model}.
    
\subsubsection{Optics uncertainty model}
\label{sec:uncertainty_model}
The uncertainties of the horizontal and vertical dispersions $D_{x}$ and $D_{y}$, and of the function $x_\rd (\xi)$ have already been discussed in Sections~\ref{sec:optics_calibration} and~\ref{sec:calibration_of_the_vertical_dispersion} (cf. also Table~\ref{tab:dispersion_results}). The uncertainties of the remaining relevant optical functions are illustrated in the following.

The levelling of the crossing angle and $\beta^{*}$, mentioned earlier, is based on the ATS optics, which has been conceived to cope with requirements expected for HL-LHC~\cite{Fartoukh:2293518}. The most important feature of the ATS optics, from the viewpoint of the forward spectrometers, is that the magnetic fields around the IP are kept stable during the levelling process. The $\beta^{*}$ at these IPs is changed by varying the magnetic fields at IP2 and IP8~\cite{Fartoukh:2293518}. This stability significantly reduces the uncertainty in the optics model and transport matrix for PPS. It also contributes to the alignment stability, which uses the distribution of ${L_{y}}/[{\rd L_{y}/\rd l}]$, cf. Eq.~(\ref{eq:al-match-metric}) and Eq.~(\ref{eq:transport_equations}).

Despite its stability, the LHC~\cite{Evans:2008zzb} is subject to additional imperfections $\Delta \mathcal{M}$, which alter the transport matrix by $\Delta T$:
\begin{equation}
\label{imperf_mach_eq}
	T\left.(l;\, \mathcal{M}\right) \to T\left.(l;\, \mathcal{M}+\Delta \mathcal{M}\right) = T\left.(l;\, \mathcal{M}\right)+\Delta T .
\end{equation}
The principles of the optics uncertainty model are described in Ref.~\cite{Antchev:2014voa}. A more complex approach is however needed in view of the explicit $\xi$ dependence of the optical functions.

The transport of protons in the vicinity of the central orbit, or any other reference orbit with a certain $\xi$, is mainly determined by the quadrupole fields of the alternating focusing and defocusing magnet (FODO) system of the LHC, whereas the
position of the central orbit itself is determined by the distribution of the dipole fields; this includes the dipole fields created by misaligned quadrupole magnets.
    \begin{figure}[ht]
    \centering
               \includegraphics[width=0.8\textwidth]{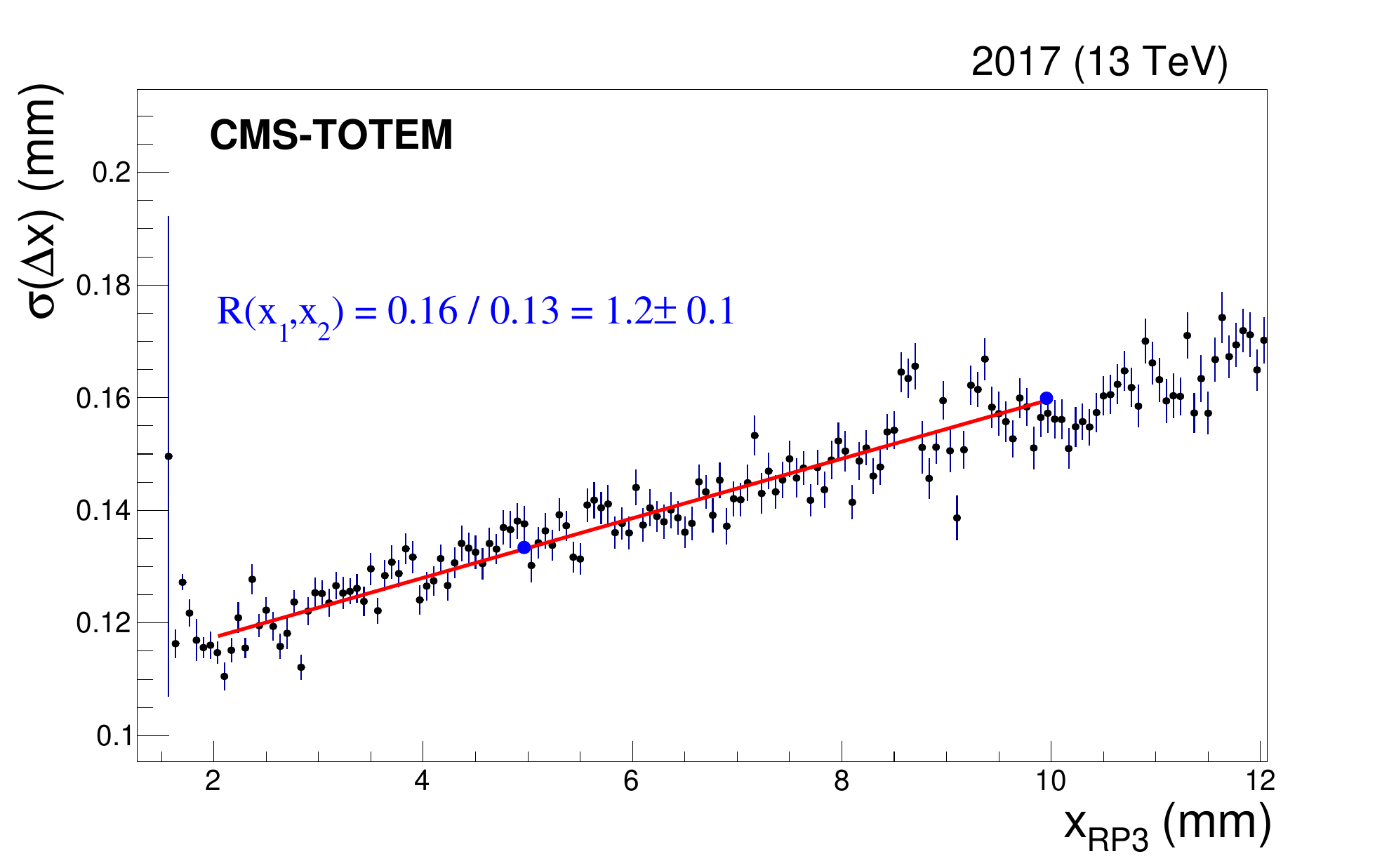}
               \caption{Fit of $\sigma(\Delta x)$ as a function of $x$ for the near RP in sector 45. The fit is used to estimate the uncertainty of the optical function $\rd L_{x}/\rd l$.
					The vertical error bars represent the statistical uncertainties.}
               \label{fig:calib_dLxds2}
        \end{figure}

A typical example is the assessment of the uncertainty of the optical function $\rd L_{x}/\rd l$. The estimation starts with the uncertainty model at low $\xi$; the magnet strengths in MAD-X are perturbed within their nominal uncertainty and the model is refined using the optics constraints from elastic candidates. In the next step the ratio of the optical function is estimated between the low- and high-$\xi$ part using collision data, cf. Fig.~\ref{fig:calib_dLxds2}. The estimation is based on the relation $\Delta x(x_{1}) =\left. {\rd{\mathrm{L}_x}}/\rd{l}\right|_{x=x_{1}}\cdot  \theta_{x}^{*}$ and exploits the fact that the scattering angle distribution of the proton is almost independent of $\xi$, so that
$R(x_{1},x_{2})={\sigma(\Delta x(x_{2}))/\sigma(\Delta x({x_{1}))}}\approx{\left.\rd  L_{x}/\rd  l\right|_{x=x_{2}}}/{\left.\rd  L_{x}/\rd  l\right|_{x=x_{1}}}\,$.

After careful evaluation for this particular function the optics model and the data agree within $\approx$10\%, cf. Fig.~\ref{fig:calib_dLxds2}. The $R(x_{1},x_{2})$ result is translated to $R(\xi_{1},\xi_{2})$ using the dispersion and, together with the low-$\xi$ uncertainty, determines the uncertainty at all $\xi$. A similar procedure leads to the uncertainty of $L_{y}(\xi)$. The LHC optics give strict correlations between the magnifications $v_{x}$, $v_{y}$ and $L_{x}$, $L_{y}$. Therefore, the uncertainty estimation of the effective lengths indirectly provides uncertainties on the magnifications as well.

\subsubsection{Covariances of optical functions}
\begin{figure}[ht]
    \centering
        \includegraphics[width=0.48\textwidth]{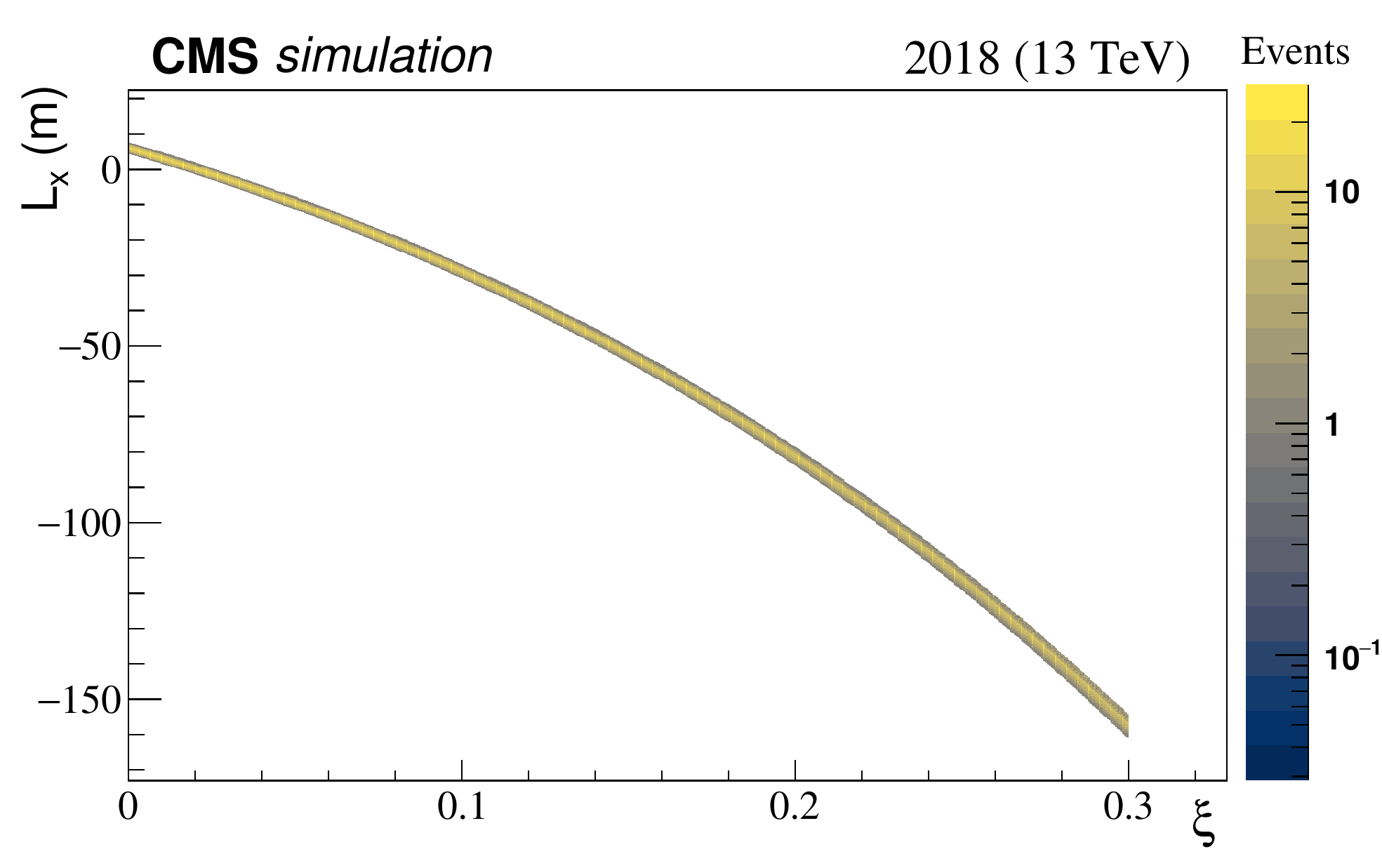}    \includegraphics[width=0.48\textwidth]{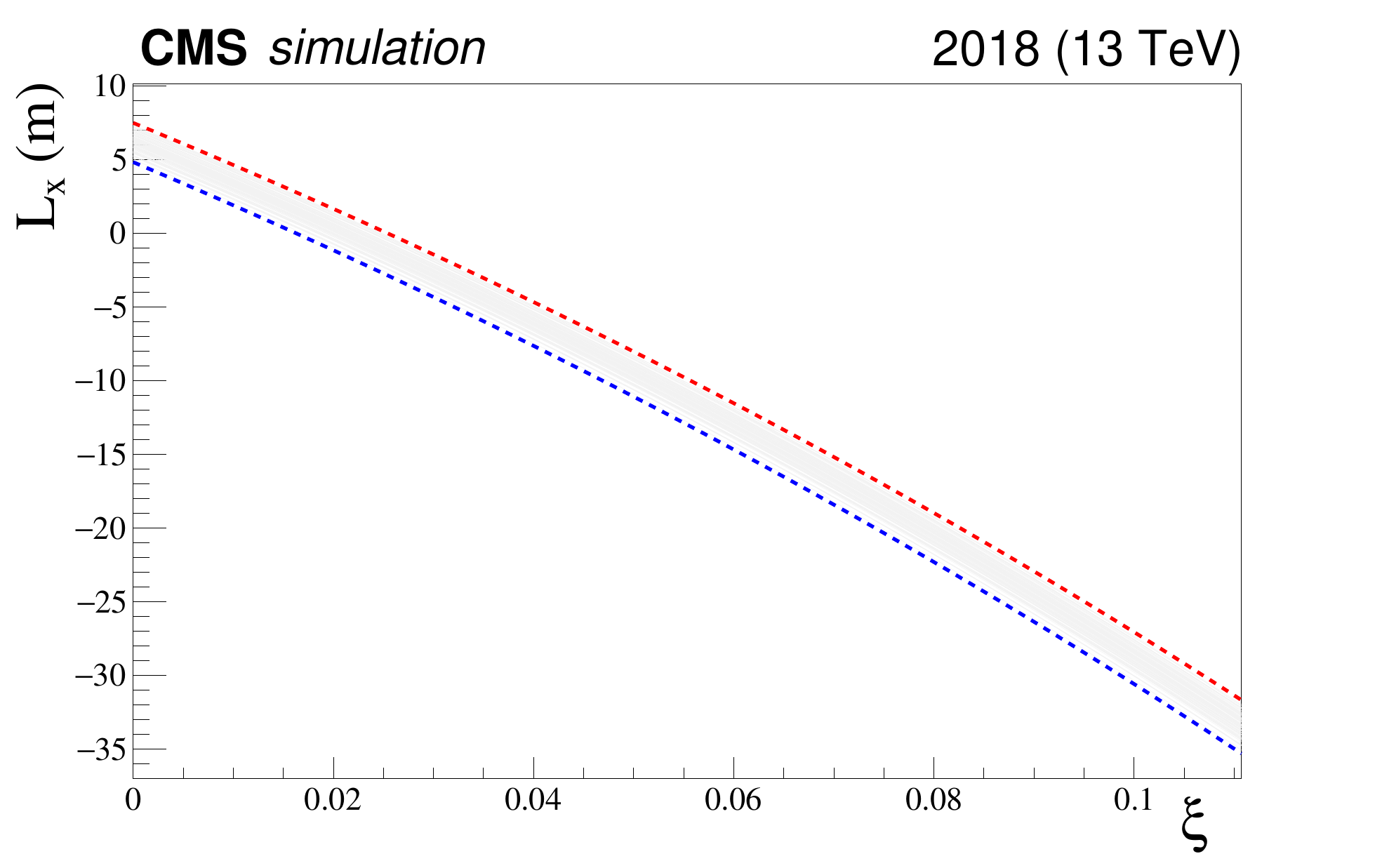}

        \caption{Left: the distribution of the horizontal effective length $L_{x}(\xi)$ values as a consequence of perturbations of the magnetic strength. Right: the correlations of the functions; the red and blue dashed curves represent the two extreme $L_{x}(\xi)$-curves of the Monte Carlo.  The upper and lower envelopes demonstrate that the points of the curve move together at different $\xi$.}
                \label{fig:cov_Lx_fig}
    \end{figure}

To fully estimate the $\xi$ dependence of the uncertainty of the optical functions, the calculation of the covariance matrix between different $\xi$ values for each function is needed.
The magnetic strength $k$ and other relevant beam parameters are perturbed within their nominal uncertainty and the optical functions are calculated for each parameter set. The values of the obtained optical functions $L_{x}$ and the envelope function thus obtained are shown in Fig.~\ref{fig:cov_Lx_fig}. The covariance and correlation matrix for the optical function $L_{x}$ at the fractional proton momentum loss $\xi=3$\%  and  $\xi=10$\% are shown in Table~\ref{tab:cor_Lx}. 

    The correlation matrix, shown in Table~\ref{tab:cor_Lx}, indicates a close to 100\% correlation between the low- and high-$\xi$ regions, which is included in the uncertainty model, cf. also Fig~\ref{fig:cov_Lx_fig}. This means that the variations of the magnetic strength and other beam parameters act in the same way at different $\xi$ values and the
    uncertainty can be described with one parameter. The covariance and correlation matrices are available for all optical functions.

            \begin{table}[ht]
            \begin{center}
            \topcaption{The correlation matrix for $L_{x}$ between different $\xi$ values for the detector RP56-220-fr vertical. }
                \begin{tabular}{  c | c  c  c  c }
                    \multicolumn{1}{c}  {}   &{ $\abs{\xi}=$3\%     }&   $\abs{\xi}=$10\%    \\\cline{2-3}
                    3\%     &  1.0   &   0.996         \\
                    10\%     &   0.996 &   1.0         \\
                \end{tabular}
            \label{tab:cor_Lx}
            \end{center}
        \end{table}

    \begin{figure}[ht]
    \centering    
    \includegraphics[width=0.48\textwidth]{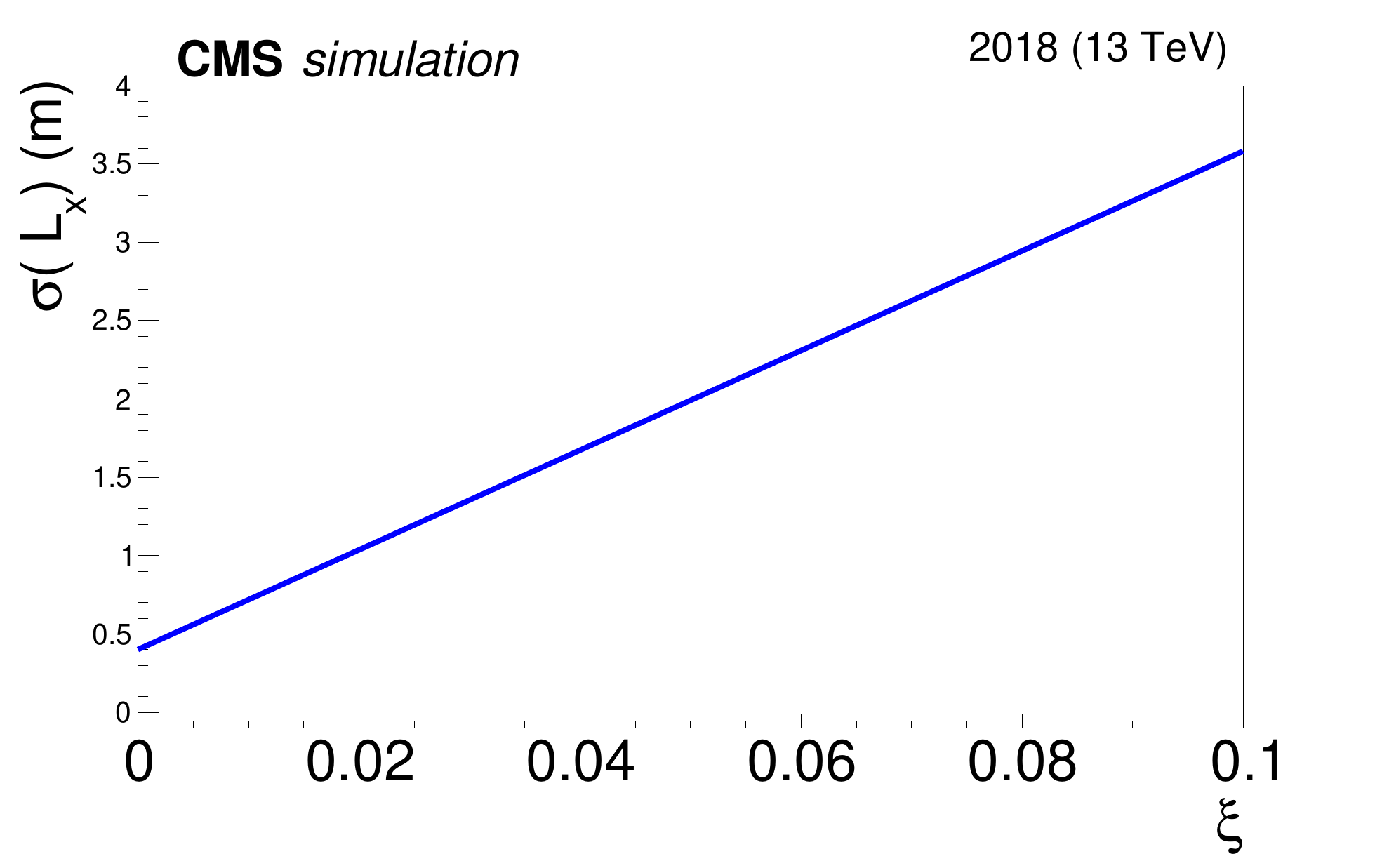}
    \includegraphics[width=0.48\textwidth]{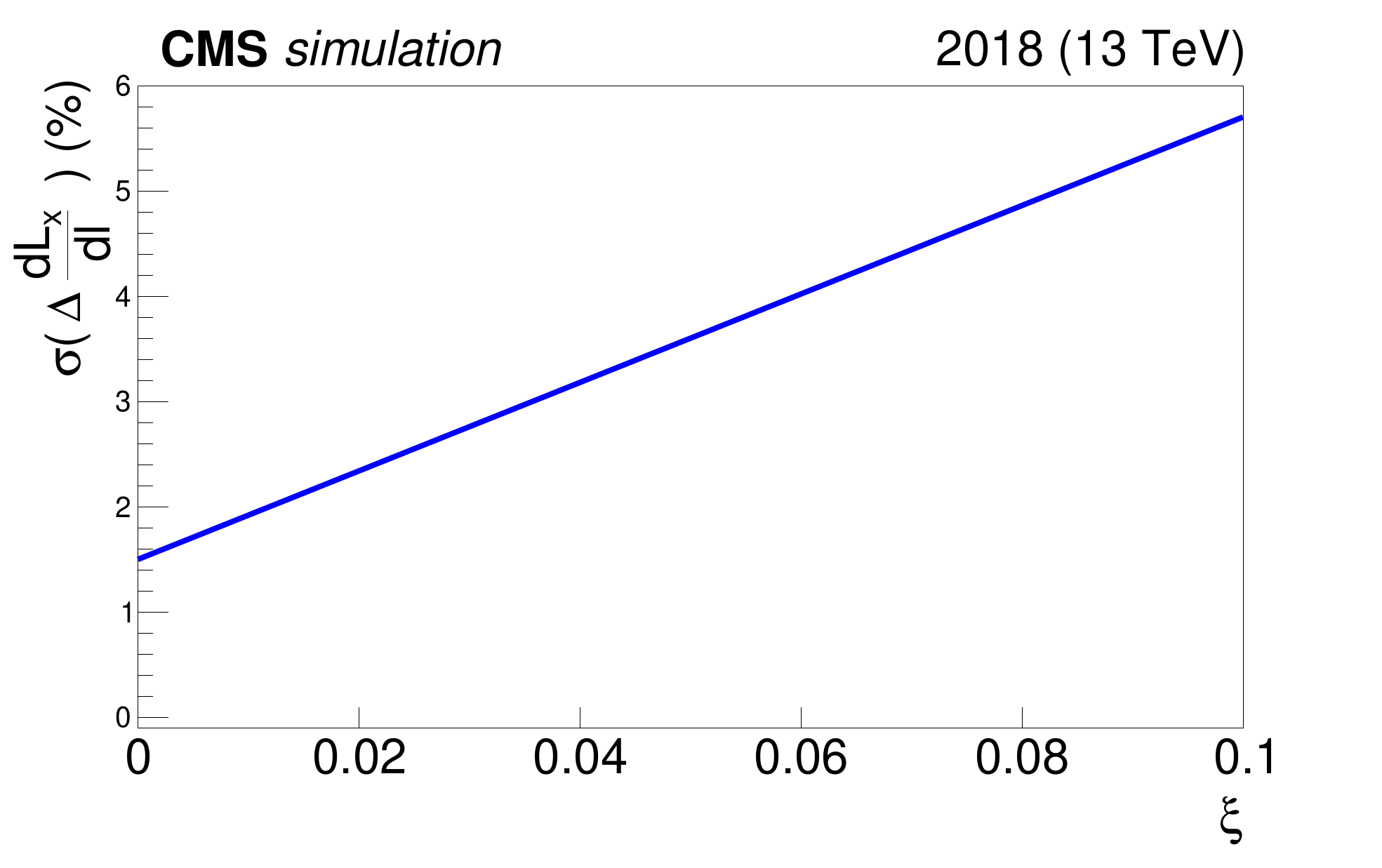}
        \caption{Left: $\xi$ dependent uncertainty function of the horizontal effective length $L_{x}(\xi)$. Right: $\xi$ dependent uncertainty function of the derivative of the horizontal effective length $\rd L_{x}(\xi)/\rd l$.}
    \label{fig:Lx_u}
    \end{figure}

    The optics uncertainty model includes the close to 100\% correlation. This means that the optical function perturbation $\delta o$ can be determined at a given reference $\xi_\text{ref}$ value and can then be scaled with the factor given in Fig.~\ref{fig:Lx_u} to obtain the perturbation at a different $\xi$ value. The optics uncertainty model is included in the PPS proton simulation described in Section~\ref{sec:simulation}.
    
\subsubsection{Inversion of the proton transport equations}
\label{sec:inversion_of_the_transport}

The transport equations Eq.~(\ref{eq:transport_equations}) are linear in $\xi$ and in the horizontal scattering angle $\theta_{x}^{*}$ with coefficient functions like $L_{x}(\xi)$, which are nonlinear. The beam size $\sigma(x)$ from Eq.~(\ref{eq:beam_size}) multiplied by the magnification factor $|v_{x}|\approx 4$ gives $\sigma(x)\  v_{x}\approx 60\,\mu$m in the horizontal plane, a contribution that is negligible when compared with the other two terms. Therefore, Eq.~(\ref{eq:transport_equations}) can be inverted to yield:
        \begin{align}
        \label{eq:reco_xi}
            \xi&=\frac{L_{x,\text{far}}\cdot  x_\text{near} - L_{x,\text{near}}\cdot  x_\text{far}}{D_{x,\text{near}}\cdot  L_{x,\text{far}} -  D_{x,\text{far}}\cdot  L_{x,\text{near}}}, \quad
            \theta^{*}_{x}=\frac{1}{\rd L_{x}/\rd l}\left(\theta_{x}-\frac{\rd  D_{x}}{\rd l}\xi\right),
        \end{align}
where the optical functions, like $L_{x,\text{near}}(\xi)$, are functions of $\xi$. The variable $\xi$ appears on both sides of the first nonlinear equation, whose solution can be found with any iterative method. These formulae are equivalent to those developed and used previously by the TOTEM Collaboration~\cite{Antchev:2015zza}.
Equation~(\ref{eq:reco_xi}) indicates the optical functions whose calibration is most relevant for the reconstruction. The formulae for the vertical reconstruction read: 
	\begin{align}
		\label{eq:reco_y_theta_y_star}
		y^{*}&=\frac{L_{y,\text{far}}\cdot  y_\text{near}' - L_{y,\text{near}}\cdot  y_\text{far}'}{ v_{y,\text{near}}\cdot  L_{y,\text{far}} -  v_{y,\text{far}}\cdot L_{y,\text{near}}}, \quad
		\theta^{*}_{y}=\frac{1}{\rd L_{y}/\rd l}\left(\theta_{y}-\frac{\rd v_{y}}{\rd l}y^{*}\right),
	\end{align}
where $y'=y - D_{y}\  \xi$. The nonlinear Eq.~(\ref{eq:reco_y_theta_y_star}) shows that an otherwise constant offset
in $D_{y}$, or in the vertical alignment would lead to a nonlinear distortion of the reconstructed angle.

\subsubsection{Summary}
In summary, the LHC optics settings and conditions changed every year in Run 2. In this chapter the main concepts and the data-driven tools to constrain the optical functions for 2016, 2017 and 2018
have been presented. The main challenges of Run 2 are the levelling of the instantaneous luminosity by changing the crossing angle and $\beta^{*}$, which requires the careful calibration of the horizontal dispersion
$D_{x}$ and also its change with the crossing angle. The vertical dispersion $D_{y}$ became sizable in 2017 and 2018, and its calibration has been discussed.
An optics uncertainty model based on collision data has been also presented, which includes the covariance matrix of transport elements.

\newpage

\section{Proton reconstruction}
\label{sec:reconstruction}

The proton reconstruction consists in back-propagating the protons from the RPs (where they are measured) to the IP (where the kinematics is determined). The propagation follows the LHC optics discussed in Section \ref{sec:optics}. The input to the propagation consists of the proton tracks detected by the RPs and aligned with respect to the LHC beam (cf.~Section \ref{sec:alignment}). Since the proton tracks at the RPs are linear (no magnetic field), they can be described by four independent parameters (slopes and intercepts along $x$ and $y$). The five proton kinematic variables include: the transverse position of the proton at $z = 0$, $x^*$ and $y^*$, the horizontal and vertical scattering angles, $\theta^*_x$ and $\theta^*_y$, and the fractional momentum loss, $\xi$. Compared to the four parameters measurable by the RPs, the reconstruction problem is underconstrained and a variable must be fixed with external information. Two complementary reconstruction strategies are exploited: ``single-RP'' and ``multi-RP''.

The \textit{single-RP} reconstruction is a simple approach that uses information from single RPs only. Because of the reduced input information, only $\xi$ and $\theta^*_y$ can be estimated:
\begin{equation}
\xi = x_\rd ^{-1}(x), \qquad \theta^*_y = {\frac{y}{L_y(\xi)}},
\label{eq:reco-single-rp}
\end{equation}
where the value of $\xi$ reconstructed from the former equation is inserted into the latter. These equations reflect only the leading terms from the optics decomposition in Eq.~(\ref{eq:transport_equations}). Neglecting the subleading, but still relevant, terms (\eg~the one proportional to $\theta^*_x$) implies a degraded resolution. On the other hand, a notable advantage of this approach is its applicability even when the proton track is not available in the other RP of the arm. Furthermore, this approach has a different (slightly smaller) dependence on the systematic variations with respect to the multi-RP method, cf.~Fig. \ref{fig:unc-systematics}. In this sense the single-RP reconstruction is a very useful check of the calibration. The variables $x^*$, $y^*$ and $\theta^*_x$ cannot be reconstructed with this approach and they are set to zero. For the vertex coordinates this is a reasonable approximation when low $\beta^*$ optics is used (as detailed below).

The \textit{multi-RP} reconstruction exploits the full potential of the spectrometer: it searches for proton kinematics that best match the observations from all RPs and all projections by minimizing the following function:
\begin{equation}
\chi^2 = \sum\limits_{i:\ \mathrm{RPs}}\ \sum\limits_{q:\ x, y} \left[ \frac{d^i_q - (T^i d^*)_q} {\sigma^i_q  }\right]^2 ,
\label{eq:reco-multi-rp}
\end{equation}
where $i$ runs over all the tracking RPs in the arm and $q$ over the two transverse projections. This expression follows the notation of Eq.~(\ref{eq:transport_vector_equation}): the vector $d^i$ represents the (measured) proton position at the $i$th RP, the vector $d^*$ denotes the proton kinematics at the IP and the matrix $T^i$ stands for the proton transport between the IP and the $i$th RP. The quantity $\sigma^i_q$ denotes the position measurement uncertainty at the $i$-th RP in projection $q$. This general formulation allows for using any optics model, $T$, and any number of tracking RPs (greater than 1). A similar approach proved useful already when applied by the TOTEM Collaboration to high $\beta^*$ optics \cite{Niewiadomski:1131825}. Since PPS aims primarily at low $\beta^*$ optics, further optimizations are possible. Low $\beta^*$ optics is characterized by narrow distributions of the interaction vertices in the transverse plane, $\sigma(x^*) \approx \sigma(y^*) = \mathcal{O}(10\mum)$. Consequently, the vertex terms in the optics decomposition of Eq.~(\ref{eq:transport_equations}) give only a small contribution and can be neglected in the reconstruction without any significant loss of accuracy (cf.~Fig. \ref{fig:unc-resol}, right). This, in turn, can resolve the under-determination of the reconstruction discussed earlier. Since there are only 4 measurements available (2 projections times 2 RPs), only 4 proton parameters out of five ($x^*$, $y^*$, $\theta^*_x$, $\theta^*_y$, $\xi$) can be determined. Therefore, by default,  $x^*$ is fixed to 0, which is a reasonable approximation given the LHC optics used by PPS (low $\beta^*$) and the very small $x^*$ RMS in these conditions. In this case, the number of degrees of freedom for the fit is $\text{ndf} = 4 - 4 = 0$ and therefore the fit effectively performs a numerical solution of a set of 4 nonlinear equations. It is equally justified to fix also $y^* \equiv 0$, which results in an alternative fitting model with one less fitted parameter (since $\xi$ is reconstructed from horizontal coordinates) and thus with $\text{ndf} = 4 - 3 = 1$. This option has been tried for validation purposes and yields results compatible with those obtained with the default choice.

The general expression in Eq.~(\ref{eq:reco-multi-rp}) can be decomposed into a set of simpler equations for the conditions relevant to PPS. 
The minimum of $\chi^2$ from Eq.~(\ref{eq:reco-multi-rp}) is described by Eqs.~(\ref{eq:reco_xi}) and (\ref{eq:reco_y_theta_y_star}) when the following conditions are met: (i) if two tracking RPs are used per arm (Run 2 configuration); (ii) if the proton transport can be approximated by the terms explicitly mentioned in Eq.~(\ref{eq:transport_equations}) (a good approximation for 2017 and 2018); (iii) only $x^*$ is assumed to be zero (the case with $\text{ndf} = 0$). Each of these equations gives an explicit expression to determine one of the proton kinematic variables. Only the first equation is nonlinear ($\xi$ on both sides of the equation), whereas the others are linear ($\xi$ is taken from the solution to the first equation). Beyond the usefulness for optics studies as discussed in Section \ref{sec:optics}, this decomposition can speed up the reconstruction software implementation: there is a single nonlinear equation with a single variable that can be solved in different well established ways, \eg~Newton's method. Using this optimisation gives results compatible with the full minimisation according to Eq.~(\ref{eq:reco-multi-rp}).

During Run 2, PPS was operated with two tracking RPs per arm (denoted ``near'' and ``far'', referring to their position with respect to the IP). The input to Eq.~(\ref{eq:reco-multi-rp}) therefore consists of one near and one far RP track, selected such that their combination is consistent with belonging to a proton originating from the IP. The selection is achieved by considering all near-far track combinations and retaining only those fulfilling the so called ``near-far association'' constraints. This selection has a double aim: first, to suppress background, and second, to disentangle multiple forward protons present in the event. The association constraints reflect the expected proton kinematics at the IP (\eg~the RMS of the scattering angles) and the patterns imposed by the LHC optics. For instance, forward protons arrive at the RP detectors at small angles with respect to the LHC beam and therefore $\Delta x$ and $\Delta y$ are expected to be small, of the order of 0.1\mm ($\Delta$ refers to the near-far difference of the track position). Beyond these, selection criteria based on $\Delta\xi$ and $\Delta\theta^*_y$ are also used, based on the single-RP reconstruction of Eq.~(\ref{eq:reco-single-rp}). The constraints have been tuned using both simulation and data, with the aim of optimizing efficiency and purity. The inefficiency (further discussed in Section \ref{sec:efficiency}) can arise either because of overly strict constraints discarding real protons, or overly loose constraints not able to distinguish between two (or multiple) protons in the event. The optimisation of the near-far association constraints is performed for each year. In 2016 and 2017, some of the RPs were equipped with Si strip sensors that reconstruct no more than one track per event. In this case, the association constraints can only suppress background and can thus be relatively loose: typically only the $\Delta\xi$ criterion with a threshold of about 0.01 is applied. In 2018, all tracking RPs were equipped with Si pixel sensors capable of reconstructing multiple tracks. Disentangling individual protons becomes necessary and tighter constraints are needed: typically $\Delta\xi$ (with a threshold of about 0.008), $\Delta x$ and $\Delta y$ criteria are applied.

The quality of the multi-RP reconstruction can be estimated by propagating the reconstructed protons to the RPs and comparing the positions of the measured and the propagated track impact points; the typical difference is smaller than 1\mum (thus at least an order of magnitude better than the spatial resolution of the RPs).

\begin{figure}
\centerline{\includegraphics{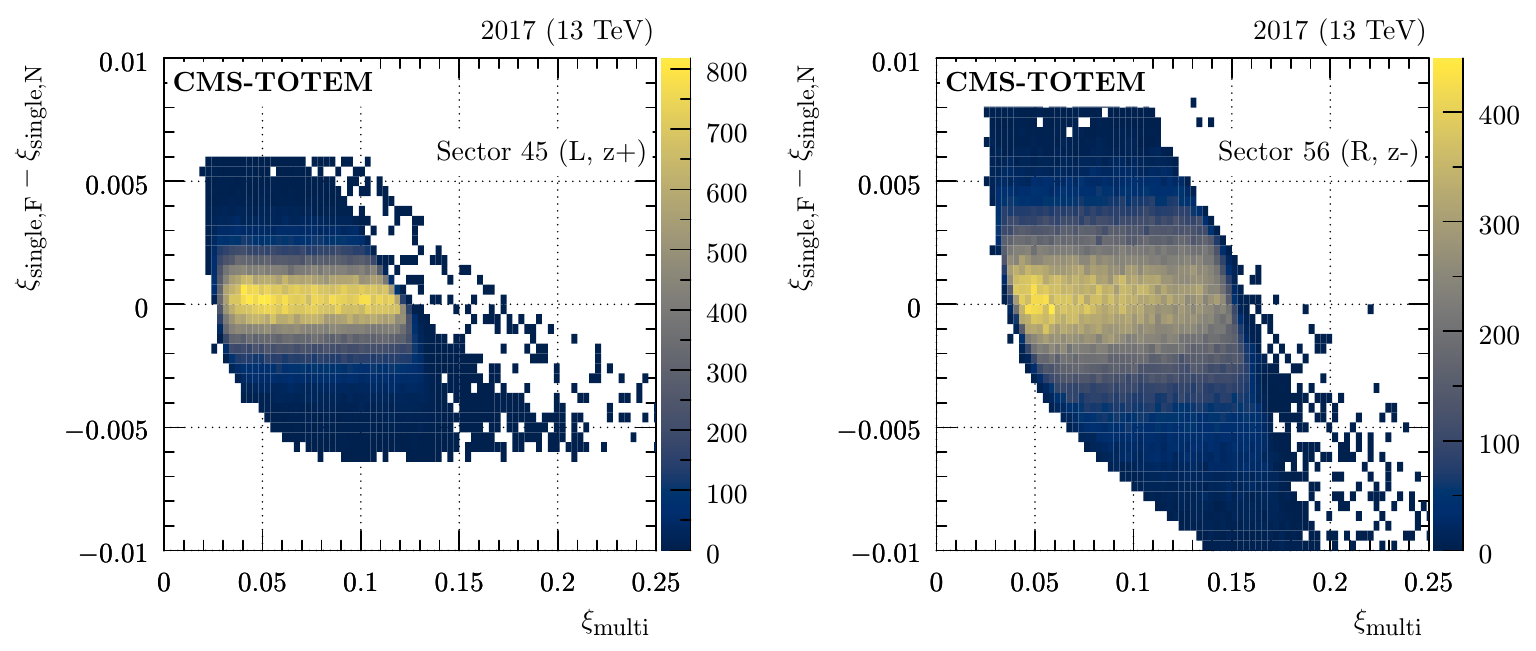}}
\caption{
Comparison of $\xi$ reconstructed with the single-RP method from the near and far RP in each arm, presented as a function of $\xi$ (fill 5849, 2017). The color code represents per-bin event counts. {Left}: sector 45, {right}: sector 56.
}
\label{fig:reco-xi-nf-cmp-ex}
\end{figure}

\begin{figure}
\centerline{\includegraphics{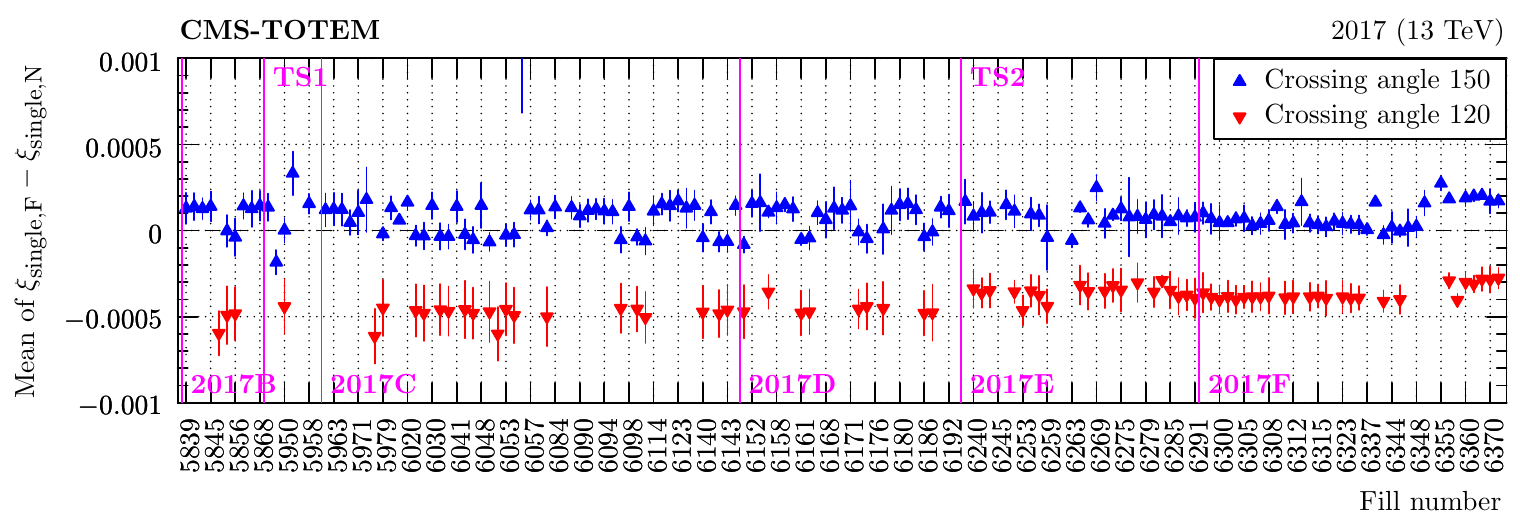}}
\caption{
Mean near-far $\xi$ difference from single-RP reconstruction (in a safe region far from acceptance limitations) as a function of fill number (2017, sector 56). The different colors represent data taken with different values of the crossing angle. The error bars represent the systematic uncertainty estimated as a difference of means evaluated at two different values of $\xi_\text{multi}$.
}
\label{fig:reco-xi-nf-cmp-sum}
\end{figure}

Figure \ref{fig:reco-xi-nf-cmp-ex} compares the results of the single-RP reconstruction of $\xi$ from the near and far RPs. The difference between the left and right plot follows mostly from the optics difference between sectors 45 and 56. The observed part of the phase space (reflected by the discontinuities in the plots) is limited by the distances of the RPs from the beam at low $\xi_\text{multi}$ (where ``multi" stands for reconstructed with the multi-RP method). The LHC aperture limitations (at high $\xi_\text{multi}$, details given in Section \ref{sec:aperture}) and the $\Delta\xi$ association cut (\eg~vertical constraints at about $\pm$0.006 in the left plot). Beyond these acceptance limitations, the difference is distributed symmetrically about 0 and is independent of the reconstructed $\xi$ (multi-RP), as expected if the alignment and the optics calibration are correct. An example of the mean difference for multiple fills is presented in Fig. \ref{fig:reco-xi-nf-cmp-sum}. The mean value is stable in time, as expected. The systematic shift between the blue and red markers (different values of crossing angle) can be attributed to a residual miscalibration and represents a measure of the systematic uncertainty of the reconstruction.

\begin{figure}
\centerline{\includegraphics{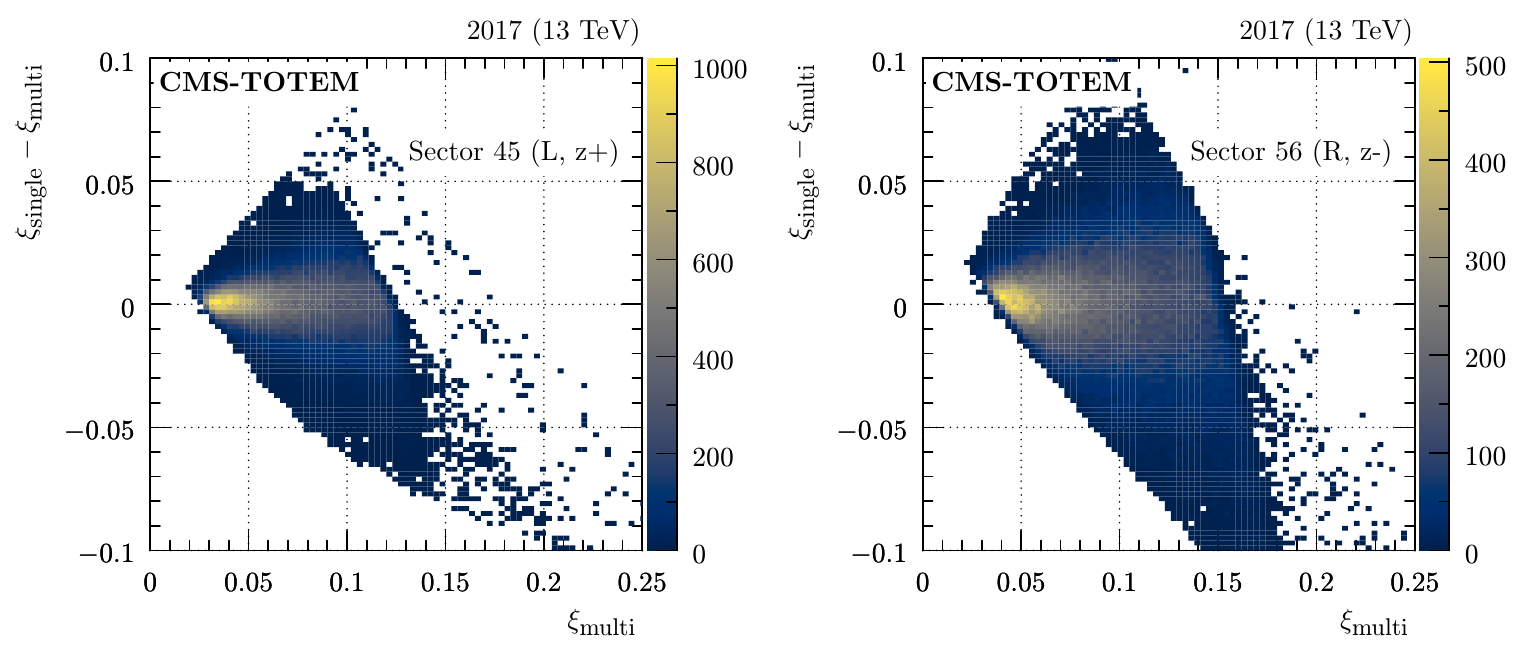}}
\caption{
Comparison of $\xi$ reconstructed with the single-RP and multi-RP methods, presented as a function of $\xi$ (LHC fill 5849, 2017, single-RP reconstruction from the near RPs). The color code represents per-bin event counts. {Left}: sector 45, {right}: sector 56.
}
\label{fig:reco-xi-si-mu-cmp-ex}
\end{figure}

\begin{figure}
\centerline{\includegraphics{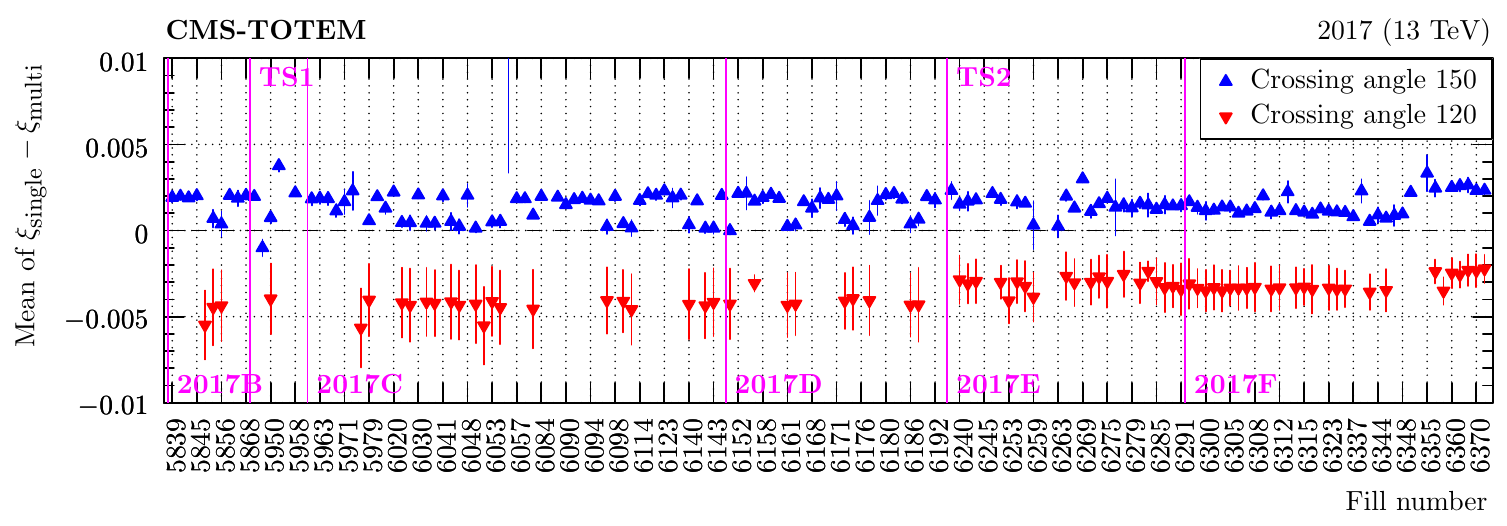}}
\caption{
Mean single-RP vs.~multi-RP $\xi$ difference (in a safe region far from acceptance limitations) as a function of fill number (2017, sector 56, single-RP reconstruction from the near RP). The different colors represent data taken with different values of the crossing angle. The error bars represent the systematic uncertainty estimated as a difference of means evaluated at two different values of $\xi_\text{multi}$.
}
\label{fig:reco-xi-si-mu-cmp-sum}
\end{figure}

Figure \ref{fig:reco-xi-si-mu-cmp-ex} shows a comparison of $\xi$ reconstructed with the single-RP and the multi-RP methods. Within resolution, they are expected to give the same results. As expected, the single-RP reconstruction has a rather low resolution. Apart from acceptance limitations (cf.~Section \ref{sec:aperture}), the single-multi difference is symmetrically distributed about 0 and has a mean independent of $\xi$, again as expected if the alignment and the optics calibration are correct. A summary of the mean single-multi $\xi$ difference for several fills is shown in Fig. \ref{fig:reco-xi-si-mu-cmp-sum}. The mean value is stable with time and close to zero (within the estimated uncertainties, Fig. \ref{fig:unc-summary}). There is a small residual dependence on the crossing angle (colors), which is caused by residual miscalibration and represents a contribution to the systematic uncertainties.

\begin{figure}
\centerline{\includegraphics{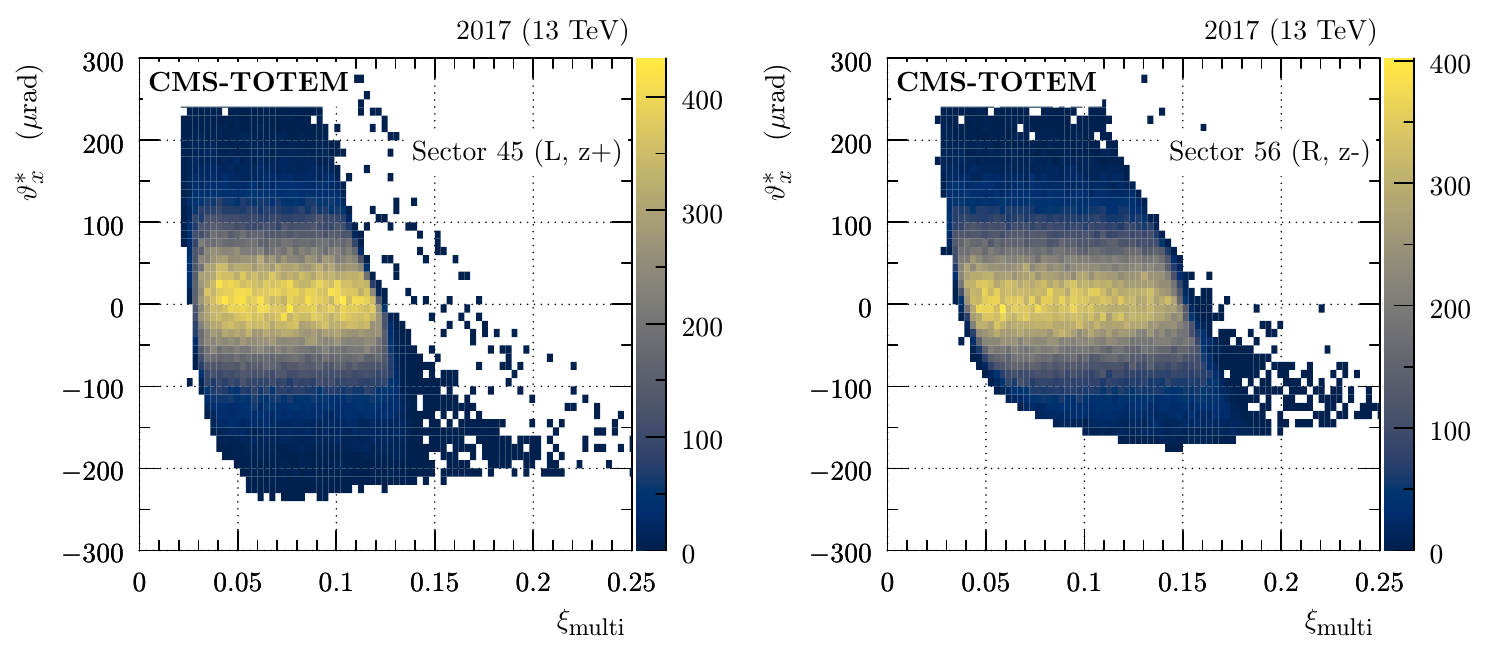}}
\caption{
Histogram of $\theta^*_x$ vs.~$\xi$ as reconstructed with the multi-RP method (fill 5849, 2017). The color code represents per-bin event counts. {Left}: sector 45, {right}: sector 56.
}
\label{fig:reco-thx-ex}
\end{figure}

\begin{figure}
\centerline{\includegraphics{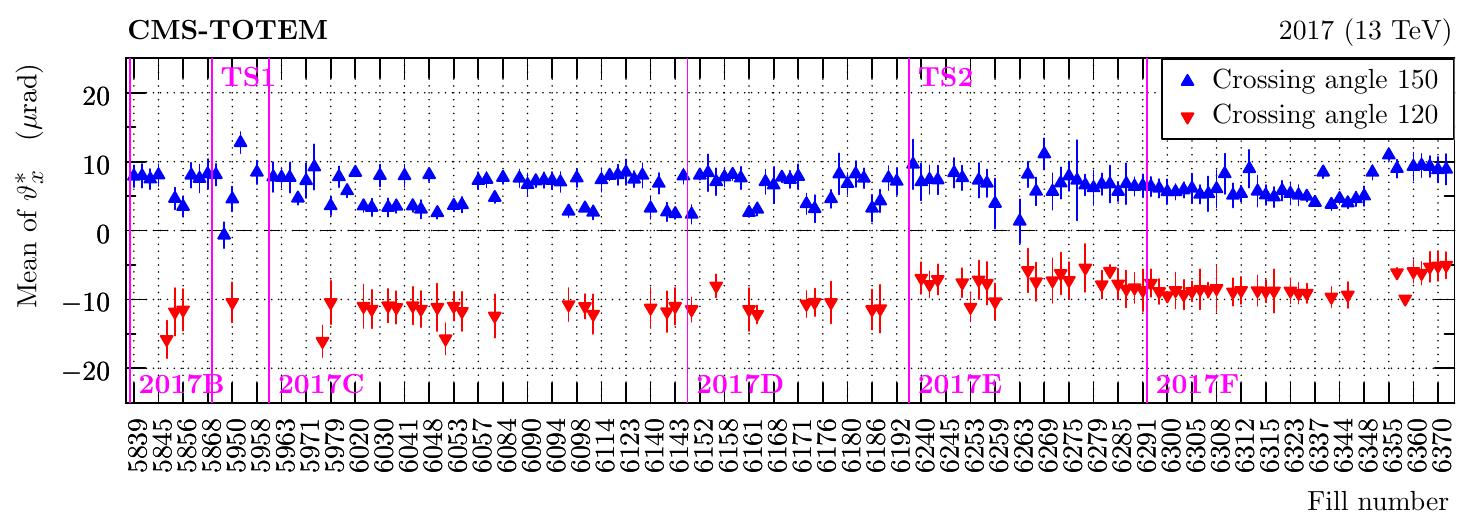}}
\caption{
Mean value of $\theta^*_x$ (in a safe region far from acceptance limitations) as a function of fill number (2017, sector 56). The markers in several colors represent data taken with different values of the crossing angle. The error bars represent the systematic uncertainty estimated as a difference of means evaluated at two different values of $\xi_\text{multi}$.
}
\label{fig:reco-thx-sum}
\end{figure}

Figure \ref{fig:reco-thx-ex} shows an example distribution of the horizontal scattering angle, $\theta^*_x$, vs.~$\xi$ as reconstructed with the multi-RP method. The $\theta^*_x$ distribution is expected to be symmetric about zero. Apart from acceptance limitations (cutoffs at the white-blue boundaries) we observe a result compatible with this expectation. Specifically, the mean value of $\theta^*_x$ does not depend on $\xi$ -- a requirement for well calibrated conditions. Figure \ref{fig:reco-thx-sum} compares mean $\theta^*_x$ from many fills. The mean value is stable over time and close to zero (within approximately $\pm 10\murad$). The small residual dependence on the crossing angle (colors) is again taken as a systematic uncertainty of the reconstruction.

\begin{figure}
\centerline{\includegraphics{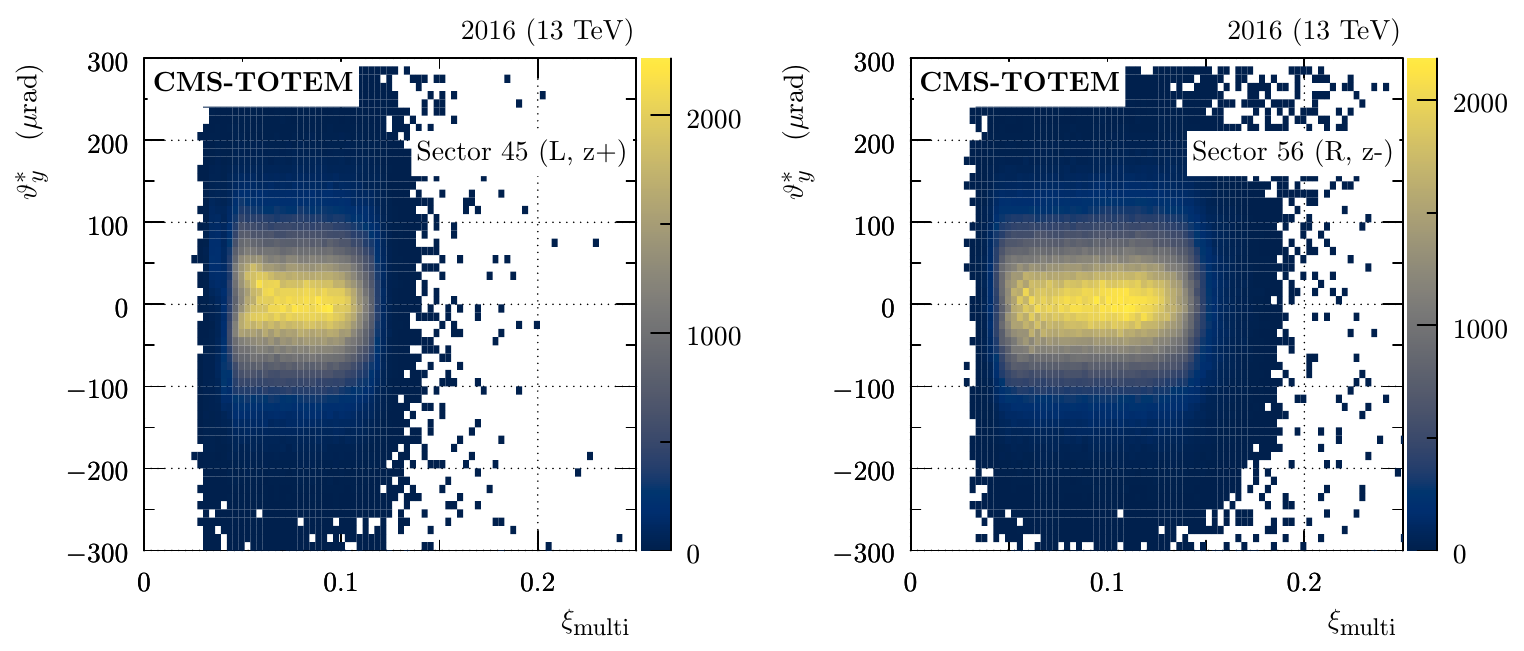}}
\caption{
Histogram of $\theta^*_y$ vs.~$\xi$ as reconstructed with the multi-RP method (fill 5276, 2016). The color code represents per-bin event counts. {Left}: sector 45, {right}: sector 56.
}
\label{fig:reco-thy-ex}
\end{figure}

\begin{figure}
\centerline{\includegraphics{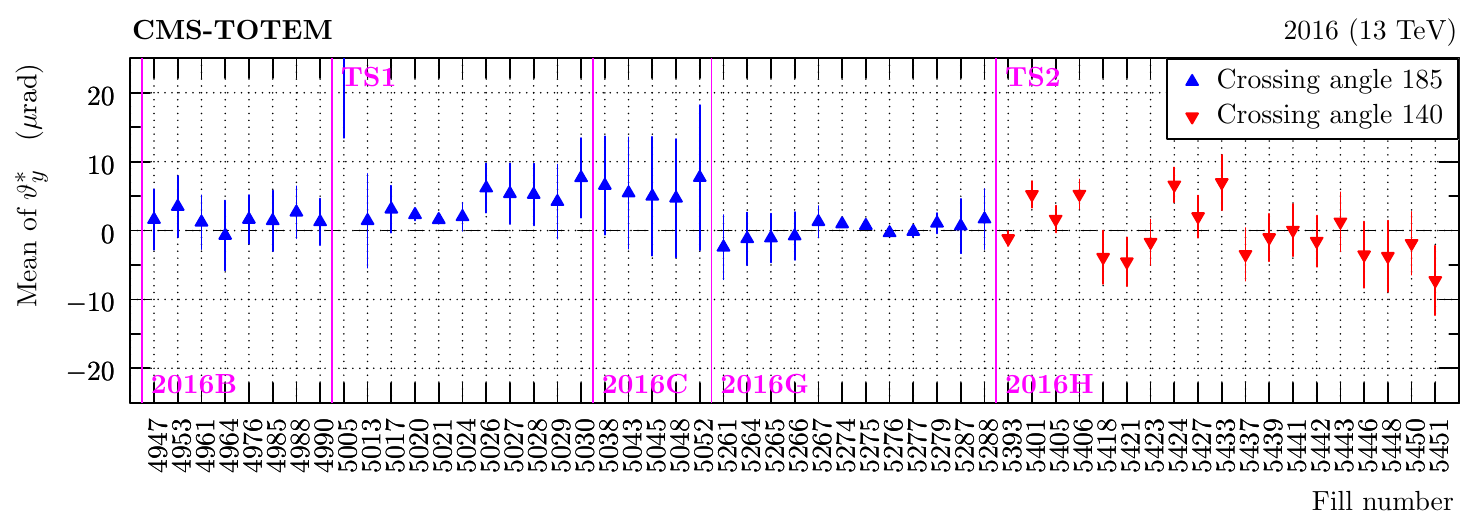}}
\caption{
Mean value of $\theta^*_y$ (in a safe region far from acceptance limitations) as a function of fill number (2016, sector 45). The markers in different colors represent data taken with different values of the crossing angle. The error bars represent the systematic uncertainty estimated as a difference of means evaluated at two different values of $\xi_\text{multi}$.
}
\label{fig:reco-thy-sum}
\end{figure}

Figure \ref{fig:reco-thy-ex} shows an example distribution of the vertical scattering angle, $\theta^*_y$, vs.~$\xi$ as reconstructed with the multi-RP method. The $\theta^*_y$ distribution is expected to be symmetric about zero. Except the low-$\xi$ region in the left plot (sector 45), which is affected by radiation damage (cf.~Section \ref{sec:efficiency}), we find this symmetry well maintained. A collection of $\theta^*_y$ mean values extracted from several fills is presented in Fig. \ref{fig:reco-thy-sum}. The mean is stable over time and close to zero (within $\pm 10\murad$). A single value of the crossing angle was used in the pre-TS2 period in 2016, and a different one in post-TS2 one.

The reconstructed proton objects provided for physics analyses combine:
\begin{itemize}
\item proton kinematics at the IP: deduced from tracking RP measurements (as discussed above) and
\item proton timing information: determined from timing RPs.
\end{itemize}
The timing information can be used to match PPS protons with a vertex in the central detector and thus for background suppression, cf.~Section \ref{sec:timing}.

\begin{figure}
\centerline{\includegraphics{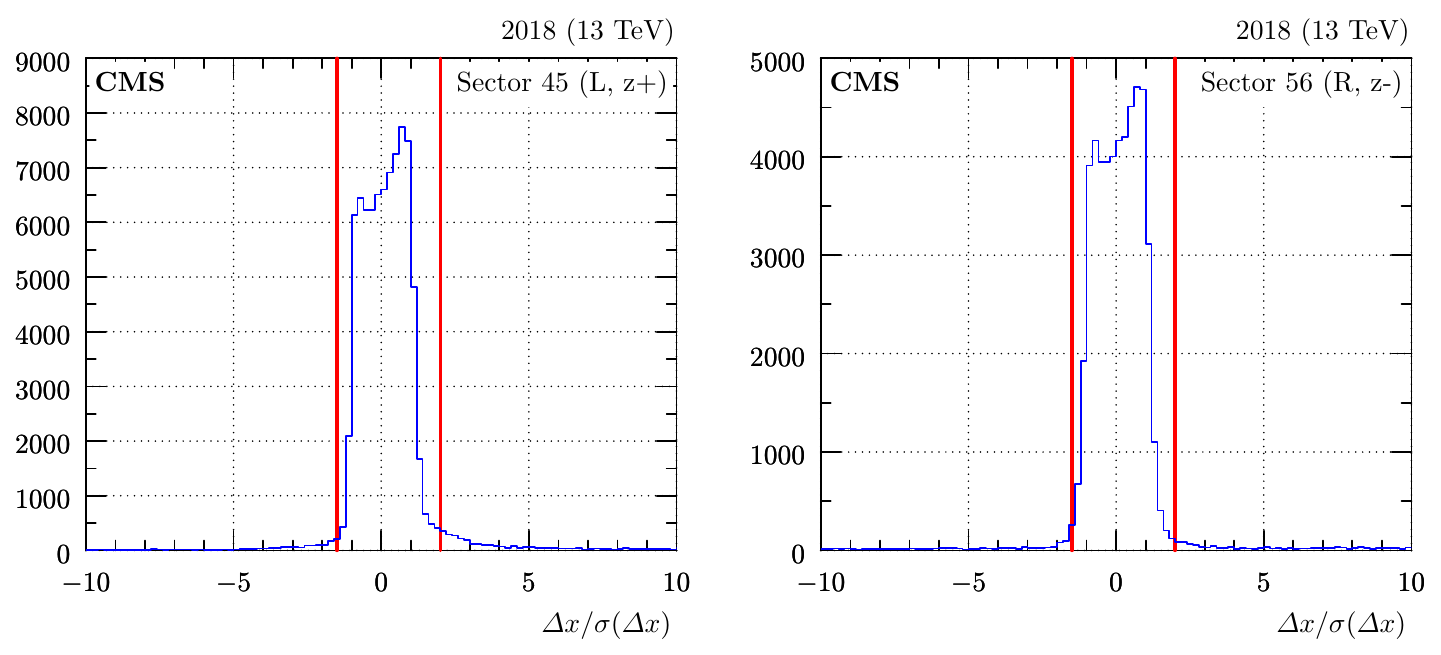}}
\caption{
Association of local tracks from tracking and timing RPs (fill 7039, 2018). $\Delta x$ refers to horizontal distance between the tracks from tracking and timing RPs, $\sigma(\Delta x)$ stands for the corresponding uncertainty. The vertical red lines delimit the tolerance window.
}
\label{fig:reco-timing}
\end{figure}

Tracks from the tracking and timing RPs are matched using $\Delta x$, the difference between the $x$ coordinate measured in the timing RP and that interpolated from the tracking RPs, cf.~Fig. \ref{fig:reco-timing}. The shape of the histograms effectively reveals the ``shape'' of the timing pad, somewhat smeared by the limited resolution of the tracking in the RPs. The tracking and timing tracks are matched if the ratio $\Delta x / \sigma(\Delta x)$ is between $-1.5$ and $+2.0$. This ratio range was determined empirically to provide good efficiency and purity.

\begin{figure}
\centerline{\includegraphics{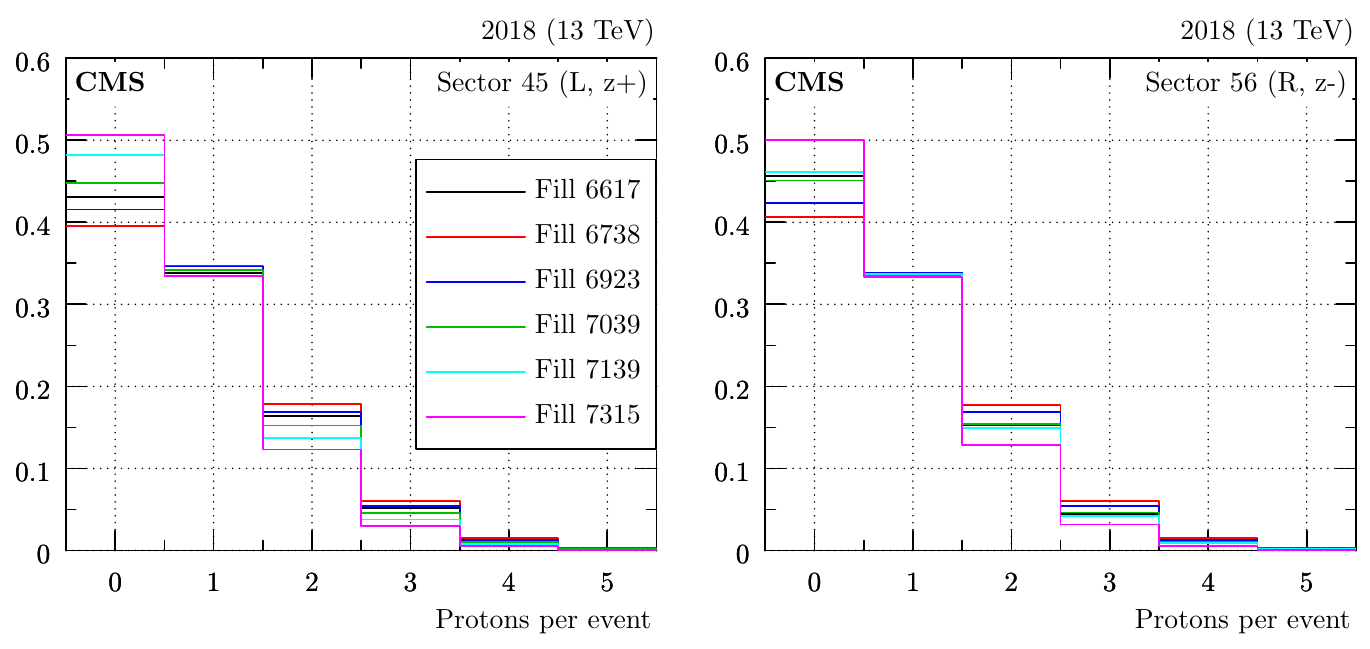}}
\caption{
Multiplicity of reconstructed protons per arm and per event (2018 data). The histograms are normalised to unit area. Different colors correspond to different fills as indicated in the legend. {Left}: sector 45, {right}: sector 56.
}
\label{fig:reco-prot-multiplicity}
\end{figure}

Figure \ref{fig:reco-prot-multiplicity} shows the multiplicity distributions of protons reconstructed per arm and per event. As expected, the probability decreases with increasing multiplicity. There are almost no events with five or more reconstructed protons.

\begin{figure}
\centerline{\includegraphics{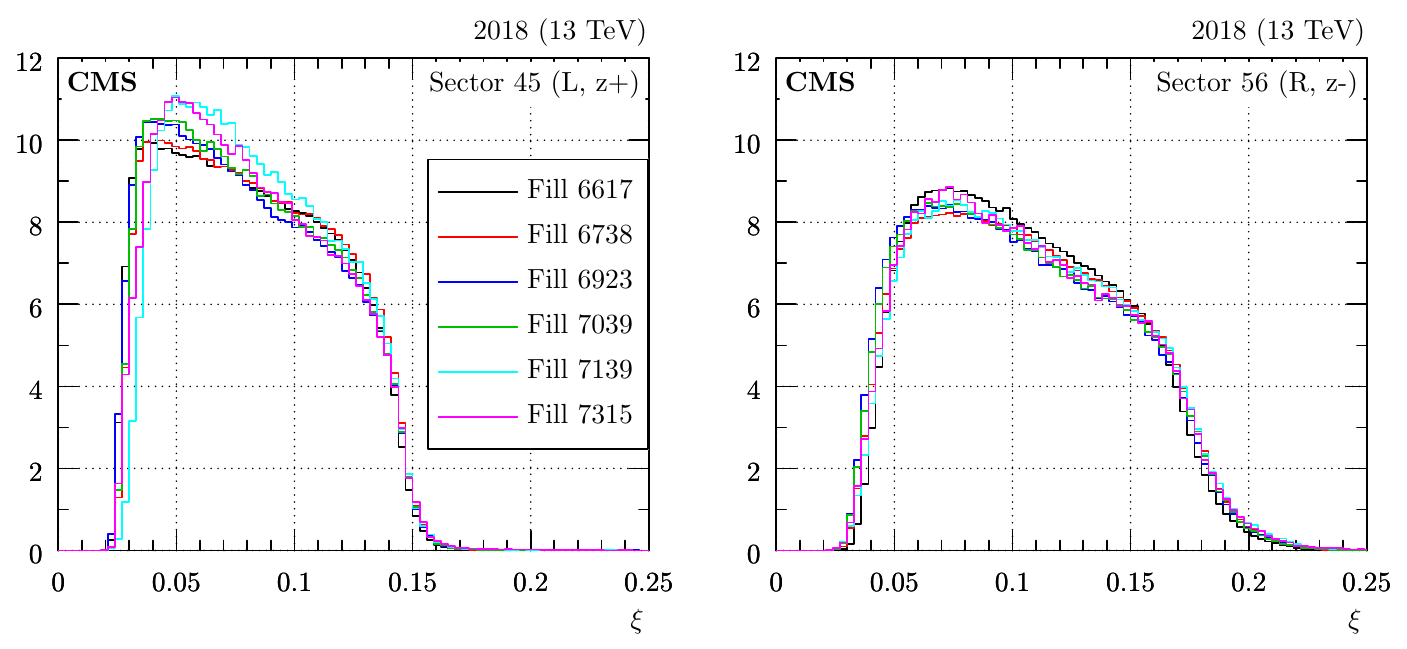}}
\caption{
$\xi$ distributions as extracted from reconstructed protons with no corrections (acceptance, efficiency, etc.), 2018 data. The histograms are normalised to unit area. Different colors correspond to different fills as indicated in the legend. {Left}: sector 45, {right}: sector 56.
}
\label{fig:reco-raw-xi-dist}
\end{figure}

Figure \ref{fig:reco-raw-xi-dist} shows the raw $\xi$ distributions as extracted from data with no selection based on reconstructed-proton observables. Since most of the protons detected in the RPs are due to pileup, they are not related to the triggering event in the central CMS, and the corresponding data set has essentially no bias due to the trigger. No corrections (acceptance, efficiency, unfolding or so) were applied to these distributions. The shape of the distributions is largely influenced by the acceptance, cf.~red curves in Fig. \ref{fig:acceptance-summary}. The differences between the left and right plots mostly follow from the difference in the optics between the sectors 45 and 56.

\section{Aperture constraints}
\label{sec:aperture}

Forward protons traveling from the IP to RPs may be intercepted by various LHC aperture limitations (collimators, beam screens, etc.), which result in detection inefficiency. These effects may be studied either by analyzing the aperture constraints of all LHC elements between the IP and the RPs or empirically by searching for discontinuities in the reconstructed distributions of the proton kinematic variables. This section presents a simple study with the latter approach, performed on zero-bias data (no trigger requirement) with limited statistics.

The study is based on the distributions of the reconstructed scattering angles vs.~$\xi$, cf.~Fig. \ref{fig:reco-thx-ex} and \ref{fig:reco-thy-ex}. In both projections the data are limited in the low- and high-$\xi$ region. The limitations at low $\xi$ mostly come from the distance of the RP from the beam. This effect can be modelled by considering the distance and the shape of the sensors, as done in Section \ref{sec:simulation}. The limitations at high $\xi$ are especially sharp in the $x$ projection, indicating that the edge arises because of horizontal constraints --- a consequence of the large horizontal dispersion. The slope of the constraint in the $\theta^*_x$ vs.~$\xi$ plane is given by the interplay of the horizontal dispersion and the effective length optical functions at the limiting LHC element.

\begin{figure}
\centering
\includegraphics{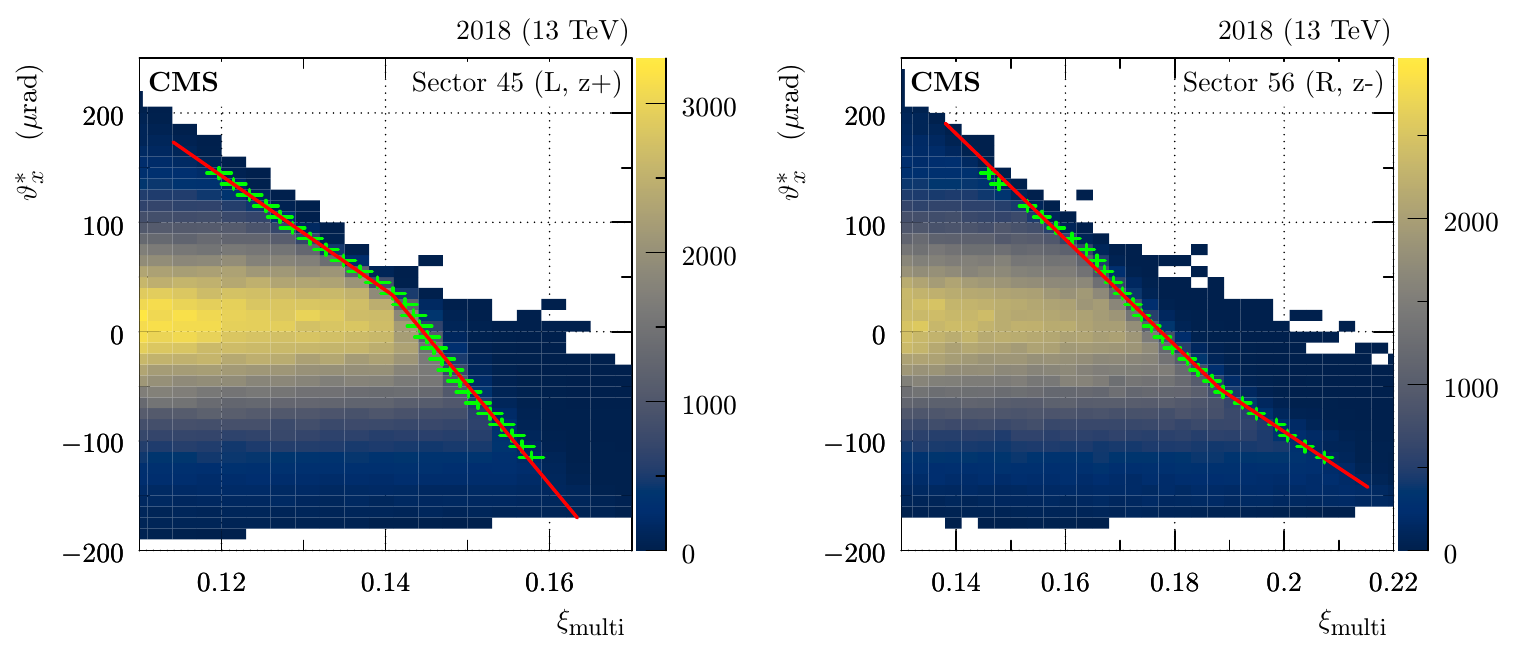}
\caption{
Distribution of $\theta^*_x$ vs.~$\xi$ reconstructed with the multi-RP method (fill 6617, 2018), zoomed at high $\xi$. The color code represents per-bin event counts. The green markers show the identified aperture cutoff, the red line the fit according to Eq.~(\ref{eq:aperture-param}). The green error bars vertically represent the bin size, horizontally a combination of statistical and systematic uncertainties. {Left}: sector 45, {right}: sector 56.
}
\label{fig:aper-method}
\end{figure}

Figure \ref{fig:aper-method} shows a typical high-$\xi$ pattern in the $\theta^*_x$ vs.~$\xi$ distribution that features a discontinuity (green markers), which is qualitatively similar for all fills in Run 2. The discontinuity is extracted by slicing the color-coded 2D histogram at constant $\theta^*_x$ and, for each slice, finding the $\xi$ position of the discontinuity (each green marker corresponds to one slice). In the left plot (sector 45), the results form a two-segment line indicating possibly the presence of two relevant aperture-limiting entities. The red line represents a two-segment line fit:
\begin{equation}
\theta^*_x = \theta_0 + a\,(\xi - \xi_0),\qquad a = a_0 \hbox{ for } \xi < \xi_0 \hbox{ or } a_1 \hbox{ for } \xi \ge \xi_0.
\label{eq:aperture-param}
\end{equation}
In the right plot (sector 56), this simple parametrisation is compatible with the green points within the estimated uncertainty.

Figure \ref{fig:aper-method} shows a significant asymmetry between sectors 45 and 56. This follows from the asymmetry of the optics; since in sector 45 the horizontal dispersion is larger, the aperture limitation is reached at smaller $\xi$ values.

\begin{figure}
\centering
\includegraphics{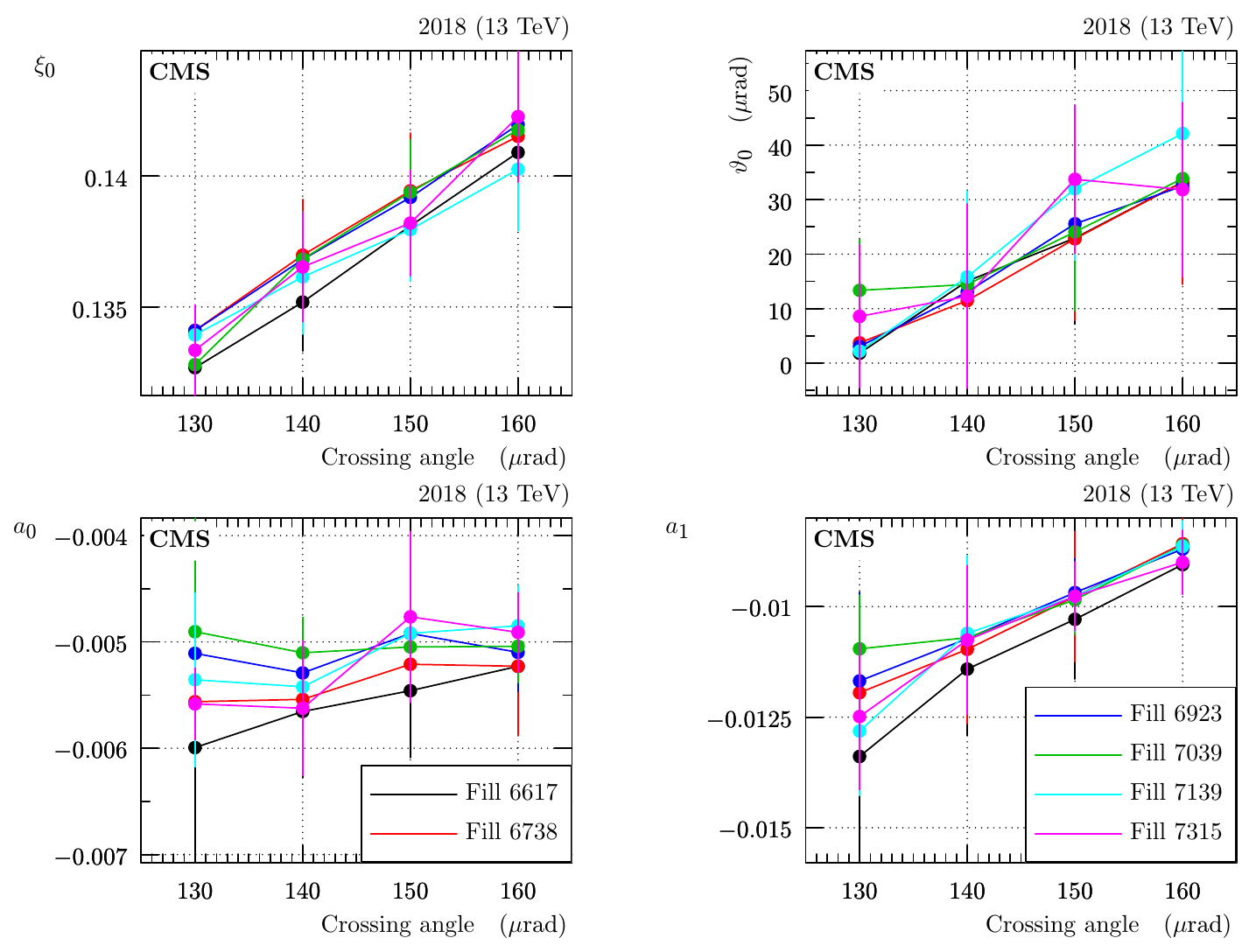}
\caption{
Summary of aperture-limitation parameters extracted from several LHC fills (different colors, 2018) and several crossing angle values (horizontal axis), for sector 45. The error bars represent a combination of statistical and systematic uncertainties.
}
\label{fig:aper-results}
\end{figure}

The fit according to Eq.~(\ref{eq:aperture-param}) has been performed independently on data from different fills, different crossing angle and $\beta^*$ values --- in order to assess a possible dependence on these parameters. An example of such a study is shown in Fig. \ref{fig:aper-results}. Within uncertainties, we observe almost no fill dependence (time stability) and a linear dependence on the crossing angle, which is expected from the optics dependence, cf.~Eq.~(\ref{eq:interpolation_function}). Equivalent conclusions have been reached for other data-taking periods in Run 2.

\section{Proton simulation}
\label{sec:simulation}

This section describes a fast simulation of forward protons in PPS. By design, it does not simulate details (interaction of protons with matter) but focuses on higher-level observables where the reproduction of features of the data is important. In particular, the simulation accounts for the following effects:
\begin{itemize}
\item beam smearing at the IP: vertex smearing and angular smearing (\ie beam divergence);
\item proton propagation from the IP to the center of each RP according to the LHC optics, cf.~Section \ref{sec:optics};
\item simulation of the LHC aperture limitations according to the model from Section \ref{sec:aperture};
\item proton propagation between sensors in each RP: linear propagation because of the lack of magnetic field in the RP region;
\item sensor efficiencies (optional): using efficiency maps extracted from data, cf.~Section \ref{sec:efficiency};
\item geometrical acceptance: check if the simulated protons pass through the sensitive area of each sensor;
\item digitisation: a software ``hit'' object is created at the nearest strip/pixel --- an effective pitch is used to reproduce the spatial RP resolution extracted from data;
\item for timing sensors, simulation of proton arrival time (with timing resolution extracted from data, cf.~Section \ref{sec:timing}).
\end{itemize}
The hit objects created in the simulation are then processed with the standard PPS reconstruction software.

The simulation can take into account realistic distributions of parameters of importance: $\beta^*$, crossing angle, optics, RP positions, apertures, resolution and efficiencies. The values of the crossing angle and $\beta^*$ are randomly sampled from the 2D histograms extracted from the data, cf.~Fig. \ref{fig:intro-xangle-beta-st}. The variations in RP positions reflect the movements performed during the LHC operation: \eg~vertical RP movements in the technical stops of 2018 to distribute the radiation damage. For consistency between simulation and data, the simulation conditions are randomly switched with the frequency extracted from data (following integrated luminosities).

The simulation can be used with any source of simulated forward protons. By default, the simulation uses a particle gun, which generates protons with a uniform $\xi$ distribution and Gaussian $\theta^*_x$ and $\theta^*_y$ distributions with zero mean and RMS of $60\murad$. These settings simulate minimum bias protons.

The beam divergence, $\sigma_\text{bd}$, used in the simulation was extracted from data using three complementary methods. First, the beam divergence can be estimated from the beam emittance, $\epsilon$, measured by the LHC: $\sigma_\text{bd} = \sqrt{\epsilon/\beta^*}$ . The second estimate is obtained from the beam spot size, $\sigma_\text{bs}$ measured by the CMS central detector: $\sigma_\text{bd} = \sigma_\text{bs} \sqrt{2} / \beta^*$. The factor of $\sqrt 2$ stems from converting the beam spot size (product of two beams) to the single-beam width, cf.~Eq.~(\ref{eq:beam_size}). The third method is the most direct, but can only be applied to data from the special ``alignment'' fills where a sample of elastically scattered protons can be selected. In the final state of elastic scattering there are two protons, ideally with exactly opposite directions. Since the direction fluctuations are predominantly caused by the beam divergence, the size of the latter is determined from the RMS of scattering angle differences between the two elastic protons. All the methods agree on a beam divergence of about $30\murad$.

\begin{figure}
\centerline{\includegraphics{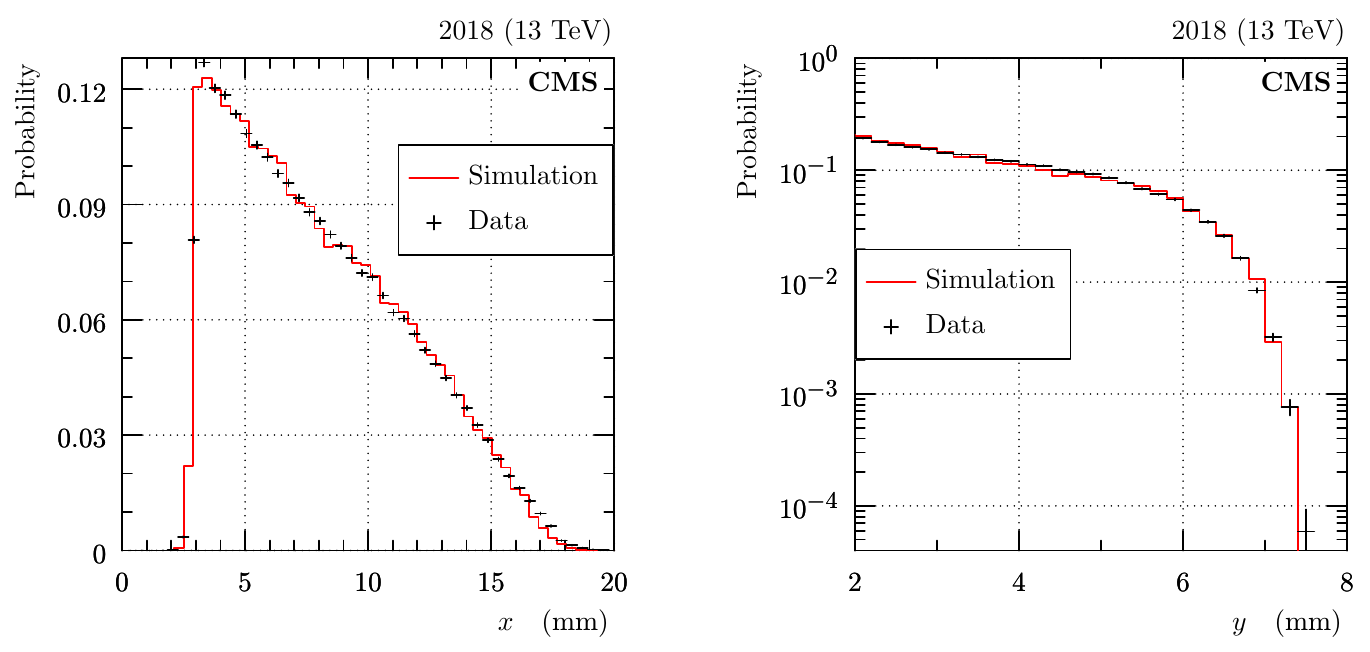}}
\caption{
Comparison of hit distributions from simulation (red) and LHC data (fill 6738, no explicit event/track selection, black), for the 2018 pre-TS1 configuration and the near RP in sector 56. The black error bars represent statistical uncertainties. {Left}: distribution of horizontal track positions, {right}: distribution of vertical track positions ($y$ range limited to the area around the upper sensor edge).
}
\label{fig:simu-validation-x}
\end{figure}

Multiple validations were performed to check whether the simulation reproduces observations; an example is shown in Fig.~\ref{fig:simu-validation-x}. In the left plot, the simulation describes well the cutoff at low $x$ (because of the sensor edge) and the smooth cutoff at large $x$ (because of the LHC aperture limitations). In the right plot, the simulation describes well the cutoff at large $y$ (because of the sensor edge).

\begin{figure}
\centerline{\includegraphics{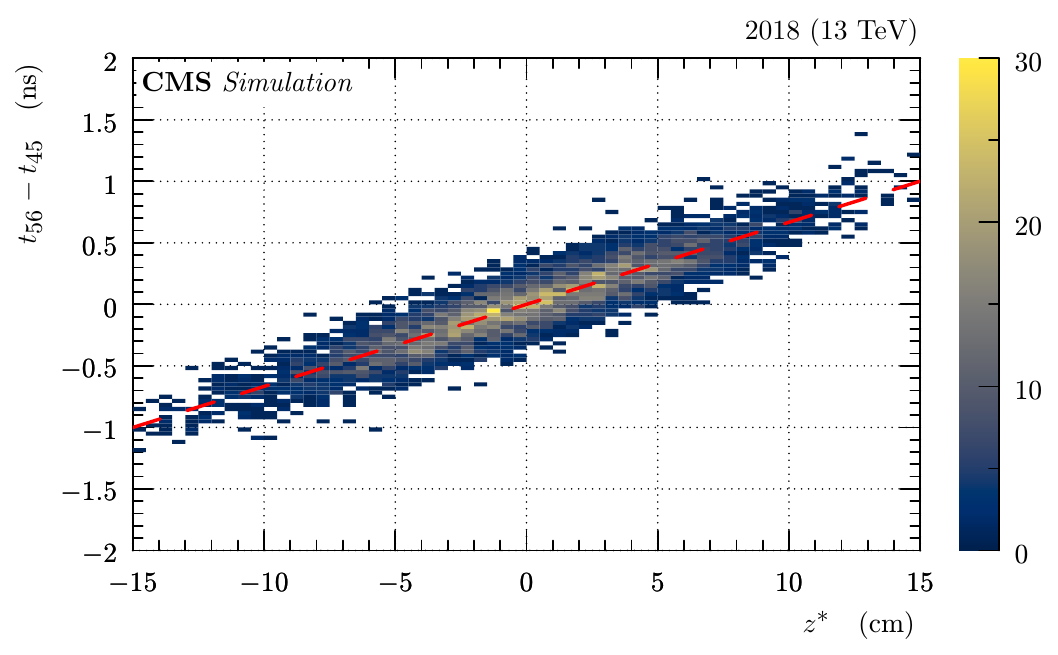}}
\caption{
Simulated correlation between vertex position along the beam, $z^*$, and the proton timing difference observed in LHC sectors 56 and 45 (2018 pre-TS1 configuration). The color code represents per-bin event counts. Reconstruction resolution of $z^*$ is not included in this plot. The red dashed line indicates the ideal correlation.
}
\label{fig:simu-timing}
\end{figure}

An example of the timing simulation is shown in Fig. \ref{fig:simu-timing}. Here, a realistic timing resolution is used for the reconstructed protons (vertical axis), but perfect vertex $z$ (horizontal) reconstruction is assumed. 

\begin{figure}
\centering
\includegraphics[width=0.49\textwidth]{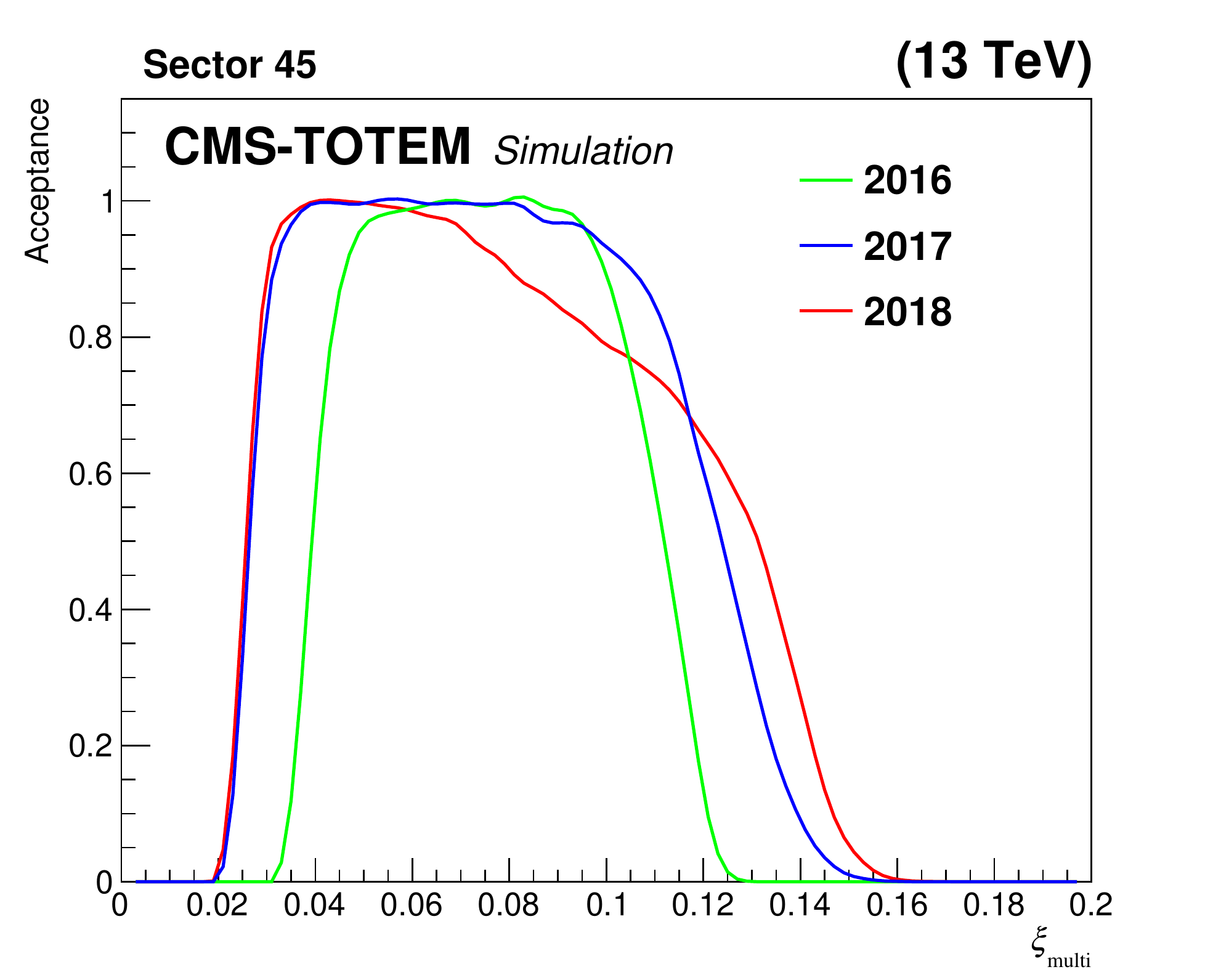}
\includegraphics[width=0.49\textwidth]{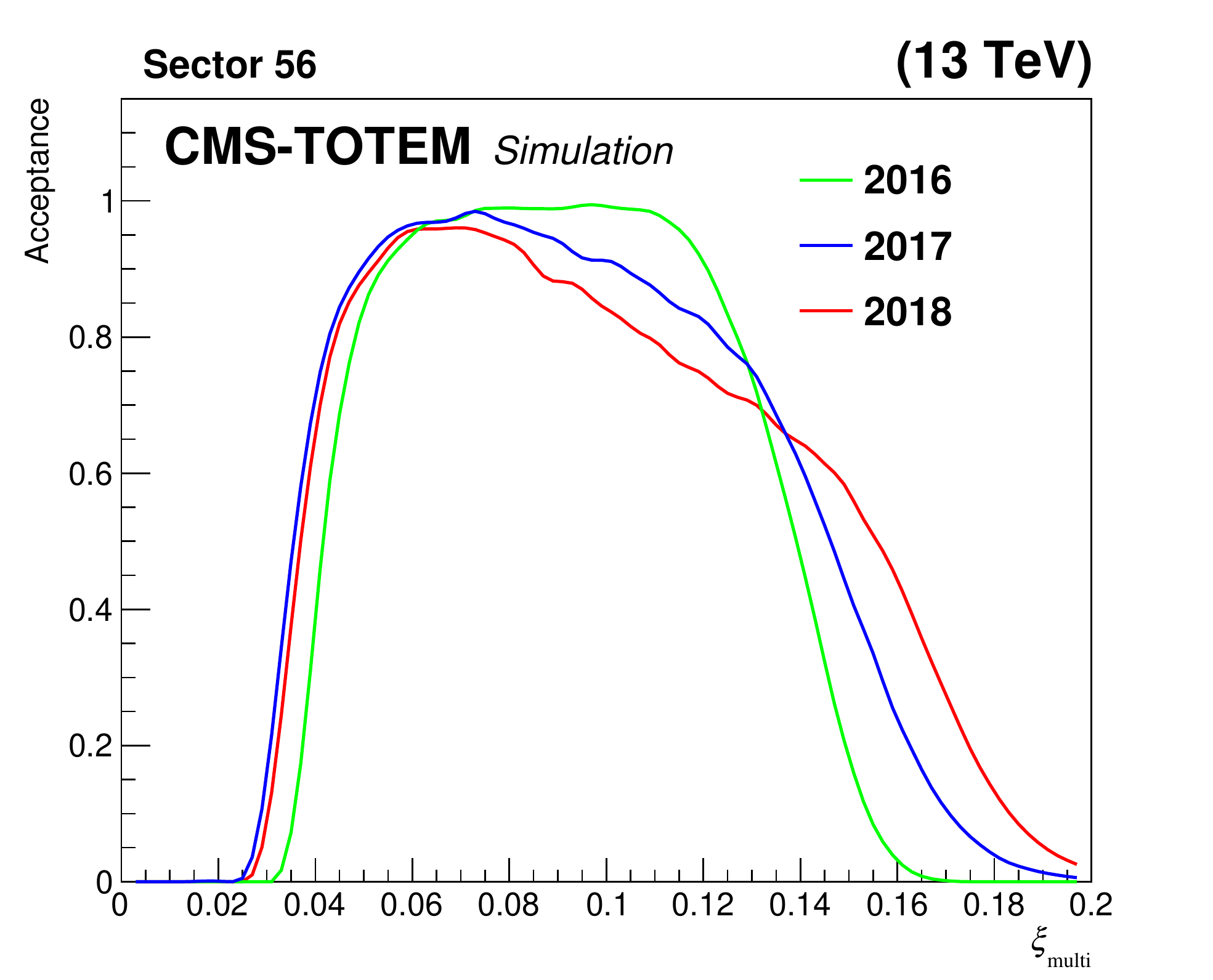}
\caption{Fraction of reconstructed multi-RP protons, as a function of $\xi_\text{multi}$, for a proton sample produced with the PPS direct simulation. Since perfect detector efficiency was assumed in this simulation, the results reflect mostly the geometrical acceptance of the apparatus.}
\label{fig:acceptance-summary}
\end{figure}

Figure~\ref{fig:acceptance-summary} shows the effect of the LHC aperture limitations (discussed in Section \ref{sec:aperture}) on PPS acceptance, which is estimated with the proton simulation. The differences between the left and right plots stem primarily from the differences in the optics in the LHC sectors 45 and 56. The differences between the colors (representing different years) are related to the sensor types used in different years. In 2016, very wide Si strip sensors were used, thus limiting potential loss of protons because of the vertical displacement from the beam. Consequently, the green curve presents a plateau close to full acceptance at the central $\xi$ range. In 2018, vertically narrower Si pixel sensors were used, thus unable to detect protons with sizable vertical displacement from the beam. The proton loss rate increases with $\xi$ due to the optics: in particular due to the nonzero value of $D_y$ (cf.~Section \ref{sec:calibration_of_the_vertical_dispersion}) and $L_y$ increasing (in absolute value) with $\xi$ (cf.~Fig. \ref{fig:Lx_Ly_vs_xi}). In 2017, a hybrid configuration with strip (pixel) sensors in the near (far) RP was used (cf.~Table \ref{tab:rp-config}) and consequently the acceptance shape is in between the two extremes. The acceptance in sector 56 (right plot) is more sensitive to the reduced size of the pixel sensors because of the larger (absolute) value of $L_y$ in this sector.

\section{Uncertainties}
\label{sec:uncertainties}

Since the simulation described in Section \ref{sec:simulation} reproduces the data well (cf.~\eg~Fig. \ref{fig:simu-validation-x}), it can be used to validate the performance of the proton reconstruction presented in Section \ref{sec:reconstruction}. The performance will be characterized in terms of the three quantities below.
\begin{itemize}
\item ``{Bias}'' = mean of reconstruction - truth. This may occur because of effects neglected by the reconstruction formula; a notable example is the single-RP reconstruction of $\xi$, Eq.~(\ref{eq:reco-single-rp}), which is unable to correct for the effect of the horizontal scattering angle $\theta^*_x$. The ``bias'' may be nonzero even with a perfect knowledge of the conditions (alignment, optics, etc.).

\item ``{Resolution}'' = RMS of reconstruction - truth. This may occur because of random event-to-event fluctuations, \eg~from finite sensor resolution or imperfect separation of kinematics variables in the reconstruction. A notable example of the imperfect separation could be the single-RP reconstruction of $\xi$; since this reconstruction is biased by a term proportional to $\theta^*_x$, the event-to-event fluctuations in the scattering angle effectively lead to a degraded $\xi$ resolution. The ``resolution'' may be nonzero even with a perfect knowledge of the conditions.

\item ``{Systematics}'' = effect of biased conditions. The ``systematics'' may be nonzero even if ``bias'' and/or ``resolution'' vanish.
\end{itemize}

The considered sources of conditions bias include:
\begin{itemize}
\item alignment: following the uncertainties from Table \ref{tab:alig-unc}, variations of the horizontal and vertical alignment were studied separately. Furthermore, symmetric (same sign in near and far RP) and anti-symmetric (opposite sign in the two RPs) shifts have been studied.
\item Optics: uncertainties of the horizontal effective length, $L_x$, (cf.~Fig. \ref{fig:Lx_u}, left), its derivative $\rd  L_x/\rd  s$ (cf.~Fig. \ref{fig:Lx_u}, right) and the horizontal dispersion.
\end{itemize}

The results presented here were obtained with the fast simulation described in Section \ref{sec:simulation} and its default settings, which reproduce well the zero bias data. Specifically, the $\theta^*_x$ distribution is given by a convolving of two Gaussian functions, one representing the physics scattering (with an RMS of $60\murad$) and one representing the beam divergence (with an RMS of $30\murad$).
 
The MC-based results from the fast simulation are compared with semianalytic calculations. These provide a validation (good agreement is found) and a detailed insight in the mechanisms producing certain trends in results, as discussed later.

Below, we show results for the period 2018 pre-TS1 and for the detector arm in sector 56. These are typical since the results for other periods and the other arm are qualitatively similar. We systematically show separately the results for single- and multi-RP reconstruction since rather different characteristics are expected. For brevity we focus on the results of $\xi$ reconstruction. Some results for the reconstructed four-momentum transfer squared, $t$, are shown at the end of this section.

\begin{figure}
\centerline{\includegraphics{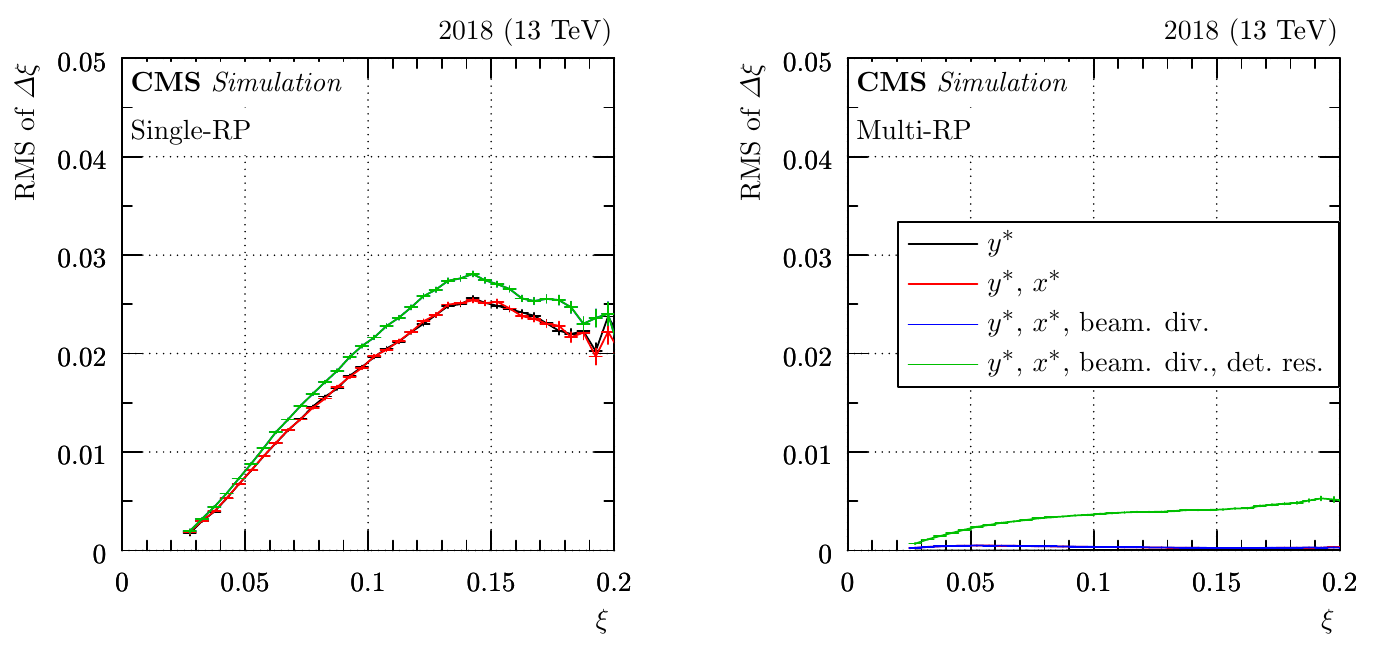}}
\caption{
Example of resolution studies for $\xi$ (2018 pre-TS1, sector 56). On the vertical axes, $\Delta\xi$ refers to the difference between the reconstructed and simulated $
\xi$. On the horizontal axes, $\xi$ denotes the simulated value. The different colors refer to different smearing effects considered. Black: only vertical vertex smearing, red: in addition also horizontal vertex smearing, blue: in addition also beam divergence, green (the most complete scenario): in addition also detector spatial resolution. The error bars represent statistical uncertainties. Note that some curves are superimposed. {Left}: single-RP, {right:} multi-RP reconstruction.
}
\label{fig:unc-resol}
\end{figure}

\begin{figure}
\centerline{\includegraphics{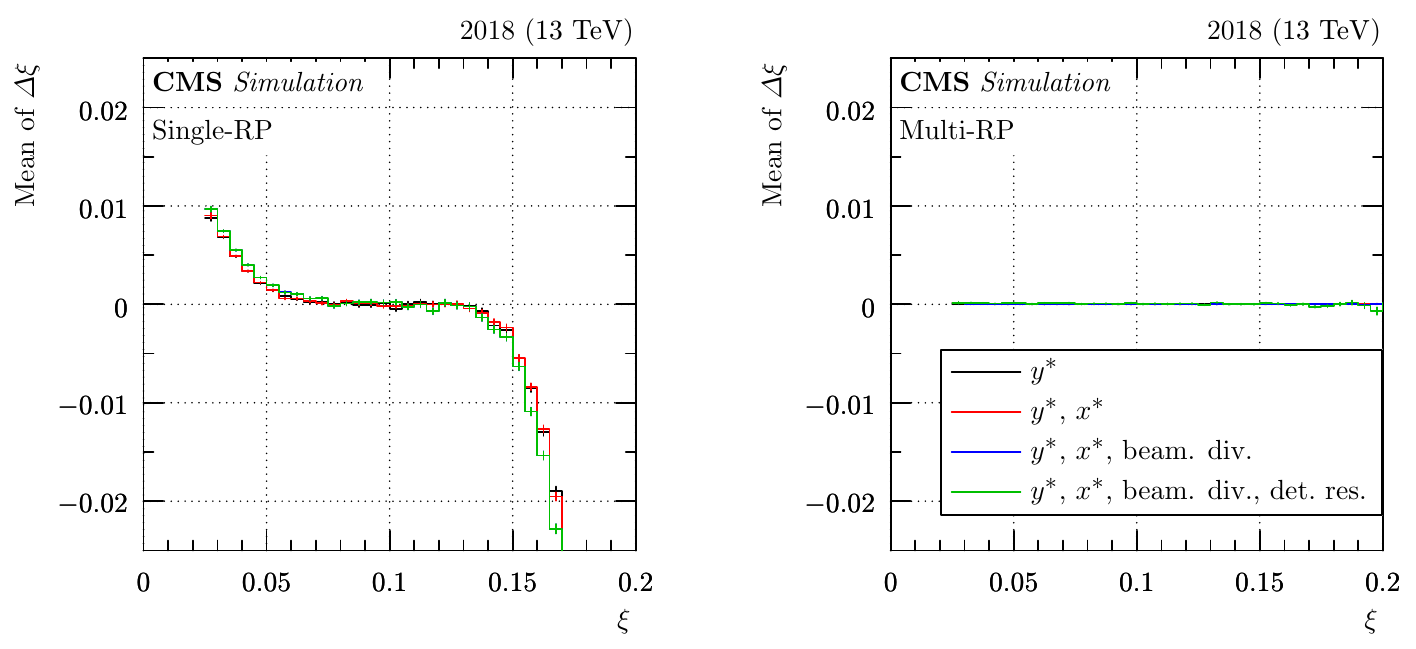}}
\caption{
Example of bias studies for $\xi$ (2018 pre-TS1, sector 56). The different colors refer to different smearing effects considered (see caption of Fig. \ref{fig:unc-resol}). The error bars represent statistical uncertainties. {Left}: single-RP, {right:} multi-RP reconstruction.
}
\label{fig:unc-bias}
\end{figure}

Figure \ref{fig:unc-resol} shows an example of the resolution studies. For single-RP reconstruction (left plot), the resolution is dominated by the neglected angular term ($L_x(\xi)\,\theta^*_x$) in the proton propagation. The RMS grows with $\xi$ because the horizontal effective length, $L_x(\xi)$, grows (in absolute value) with $\xi$ (cf.~Fig. \ref{fig:Lx_Ly_vs_xi}). At very high $\xi$, the width of the $\theta^*_x$ distribution within detector acceptance is reduced by the LHC collimators (cf.~Section \ref{sec:aperture}). Therefore, fluctuations in reconstruction are reduced, which however leads to a bias (quantified in Fig. \ref{fig:unc-bias}). For the multi-RP reconstruction (right plot), the only sizable contribution to the resolution comes from the detector spatial resolution. This explicitly justifies the statement that neglecting the horizontal vertex, $x^*$, in the reconstruction has a negligible effect, cf.~Section \ref{sec:reconstruction}.

Figure \ref{fig:unc-bias} shows an illustration of the bias studies. The single-RP reconstruction (left plot) is significantly biased close to the acceptance edges (very low and very high $\xi$). At these edges the accepted $\theta^*_x$ range becomes strongly asymmetric and since the $\theta^*_x$ term is neglected in single-RP reconstruction, the bias appears. The bias is negligible for multi-RP reconstruction (right plot).

\begin{figure}
\centerline{\includegraphics{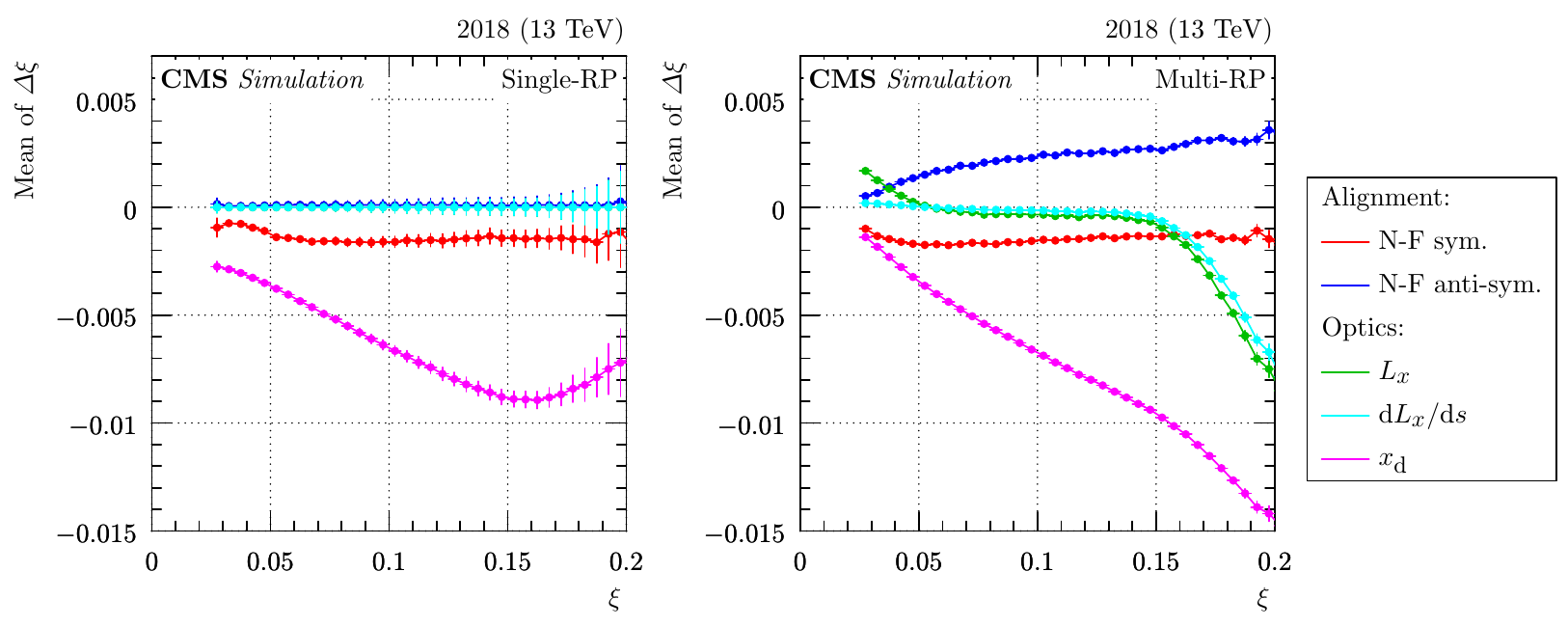}}
\caption{
Example of systematic studies for $\xi$ (2018 pre-TS1, sector 56). Each curve corresponds to a perturbation at $1\,\sigma$ level. The red and blue curves represent alignment variations: in the former both the near and far RP are shifted in the same direction, in the latter opposite directions are considered. The remaining scenarios cover perturbations of the optical functions. The error bars represent statistical uncertainties. {Left}: single-RP, {right:} multi-RP reconstruction.
}
\label{fig:unc-systematics}
\end{figure}

Figure \ref{fig:unc-systematics} shows an example of the biased-conditions studies. The individual curves show the systematic error in the reconstruction of $\xi$ caused by various conditions biases applied at the $1\,\sigma$ level (cf.~the list above). For both single-RP (left plot) and multi-RP (right plot), the leading contribution (magenta) stems from the uncertainty in the horizontal dispersion. The change of behavior at large $\xi$ is due to the LHC aperture limitations, which modify/restrict the distribution of protons within the RP acceptance. The single-RP reconstruction (left plot) has very low sensitivity to certain scenarios, \eg~the blue and cyan one. The multi-RP reconstruction (right plot) is more sensitive to systematic errors, especially at very high $\xi$.

\begin{figure}
\centerline{\includegraphics{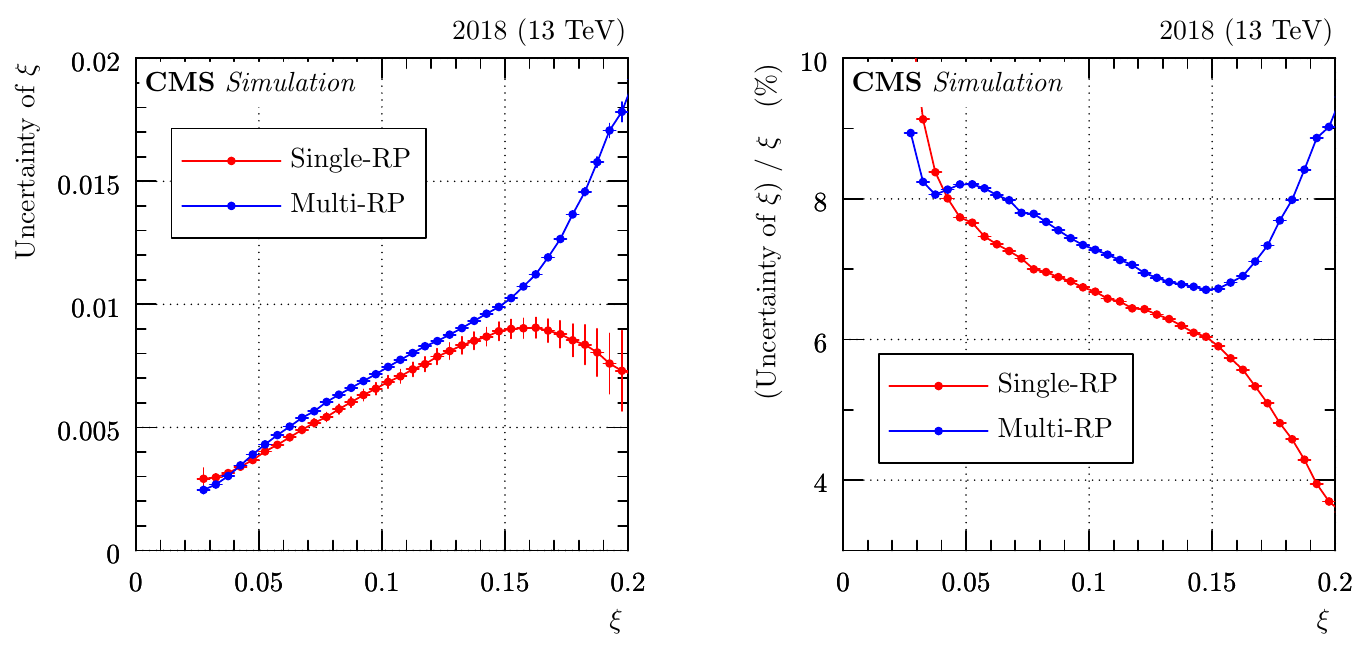}}
\caption{
Example of combined systematic uncertainties of proton $\xi$ (2018 pre-TS1, sector 56). The results for the single-RP and multi-RP reconstructions are shown in red and blue, respectively. The error bars represent statistical uncertainties. {Left}: absolute, {right:} relative uncertainty.
}
\label{fig:unc-systematics-abs-rel}
\end{figure}

Since the contributions shown in Fig. \ref{fig:unc-systematics} are statistically independent, they can be combined in quadrature to obtain the total uncertainty, as shown in Fig. \ref{fig:unc-systematics-abs-rel}. Up to $\xi\approx 0.15$, the uncertainties of the single-RP (red) and the multi-RP (blue) reconstruction are very similar.

\begin{figure}
\centerline{\includegraphics{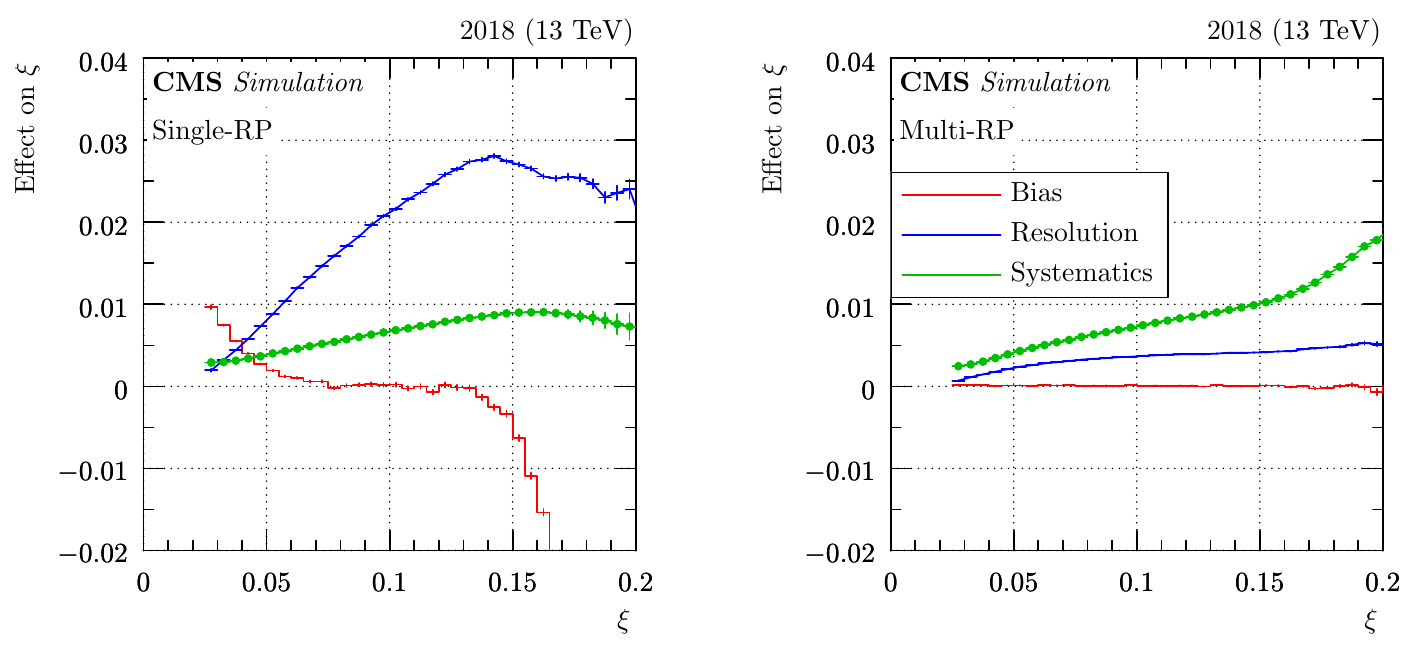}}
\caption{
Comparison of bias, resolution and systematics characteristics (2018 pre-TS1, sector 56). For the bias and resolution curves, all considered smearing effects are included. The systematics curves represent the combination of all contributions. The error bars represent statistical uncertainties.
}
\label{fig:unc-summary}
\end{figure}

A summary of all the studies presented in this section is provided in Fig. \ref{fig:unc-summary}. The comparison of the single-RP (left plot) to the multi-RP reconstruction (right plot) shows that the former has significantly larger bias, significantly worse resolution, and almost comparable systematics; it is better only in the high-$\xi$ region. This plot justifies the general preference for the multi-RP reconstruction.

\begin{figure}
\centerline{
	\includegraphics{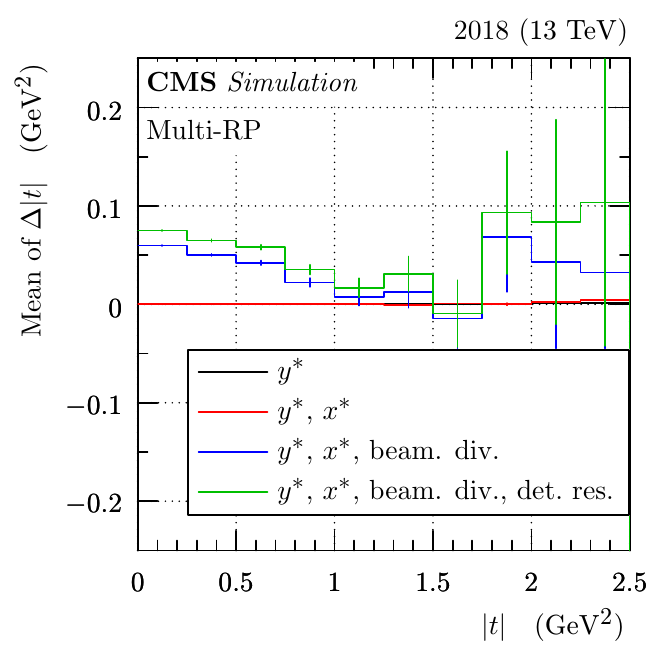}
	\hfil
	\includegraphics{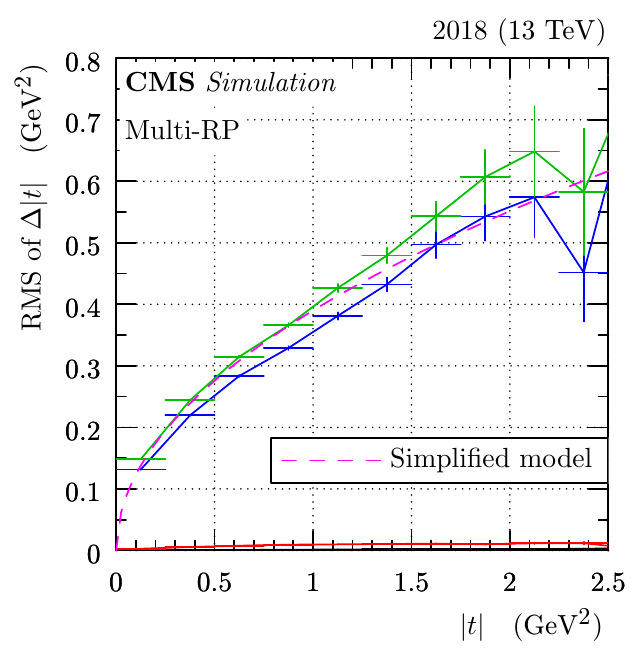}
}
\caption{
Example of resolution bias (left) and resolution (right) studies for four-momentum transfer squared, $t$, with multi-RP reconstruction (2018 pre-TS1, sector 56). On the vertical axes, $\Delta \abs{t}$ refers to the difference between the reconstructed and simulated value of $\abs{t}$. On the horizontal axes, $\abs{t}$ denotes the simulated value. The different colors refer to different smearing effects considered. Black: only vertical vertex smearing, red: in addition also horizontal vertex smearing, blue: in addition also beam divergence, green (the most complete scenario): in addition also detector spatial resolution. In the right plot, the dashed magenta curve represents the simplified analytic model from Eq.~(\ref{eq:sigma delta t}). The error bars represent statistical uncertainties.
}
\label{fig:unc-t-bias-resolution}
\end{figure}

Besides $\xi$, PPS can also estimate the four-momentum transfer squared, $t$, of protons reaching the RP detectors. Formally, this quantity is defined as $(P' - P)^2$, where the four momenta $P$ and $P'$ are those before and after the collision, respectively. It can be related to other kinematic variables:
\begin{equation}
\begin{split}
t        &= t_0(\xi) - 4 p_\text{nom}^2 (1 - \xi) \sin^2\left(\frac{\sqrt{{\theta_x^*}^2 + \theta_y^*}^2}{2}\right),\\
t_0(\xi) &= 2 \left( m^2 + p_\text{nom}^2 (1 - \xi) - \sqrt{ (m^2 + p_\text{nom}^2) (m^2 + p_\text{nom}^2 (1 - \xi)^2)} \right).
\end{split}
\label{eq:t}
\end{equation}
Since $t$ depends strongly on the scattering angles, it only makes sense to estimate it with the multi-RP reconstruction (with the single-RP approach $\theta^*_x$ is not available at all and for $\theta^*_y$ only a crude estimate is made). Typical examples of $t$ reconstruction bias and resolution are shown in Fig. \ref{fig:unc-t-bias-resolution}. The smearing effect with the largest impact is the beam divergence (cf.~the difference between the red and blue curves), followed by the spatial resolution of the sensors (cf.~the difference between the blue and green curves).

As shown in Fig. \ref{fig:unc-t-bias-resolution}, left, there is a nonzero bias in $t$ reconstruction, mostly due to the beam divergence. Formally, the beam divergence causes a smearing in scattering angles: $\theta^*_{x,y} \to \theta^*_{x,y} + \Delta\theta^*_{x,y}$, where the standard deviation of $\Delta\theta^*_{x, y}$ is given by the beam divergence, $\sigma_\text{bd}$. Inserting this into Eq.~(\ref{eq:t}) one can obtain the beam-divergence effect on $\abs{t}$ -- the difference in $\abs{t}$ with and without beam divergence in the approximation of small scattering angles:
\begin{equation} 
\Delta \abs{t} \approx p_\text{nom}^2 (1 - \xi) \left[ 2\theta^*_x\Delta\theta^*_x + 2\theta^*_y\Delta\theta^*_y + {\Delta\theta^*_x}^2 + {\Delta\theta^*_y}^2 \right].
\label{eq:delta t}
\end{equation} 
Since $\Delta\theta^*_{x,y}$ are expected to fluctuate symmetrically about zero, the first two terms in the square brackets yield a strongly suppressed contribution to the mean value of $\Delta \abs{t}$. Conversely, the last two terms are always positive and therefore give rise to the reconstruction bias: $\Delta \abs{t} \approx 2p_\text{nom}^2 (1-\xi)\sigma_\text{bd}^2$. For $\xi = 0$, this simple model gives mean $\Delta \abs{t} \approx 0.08\GeV^2$, thus well comparable with results in the figure. The nonflat shape reported in the figure is due to the limited acceptance of the RP detectors and the near-far association constraints (cf.~Section \ref{sec:reconstruction}) applied in the proton reconstruction.

Figure \ref{fig:unc-t-bias-resolution}, right, shows the $\abs{t}$ resolution, which deteriorates with increasing $\abs{t}$. This is expected from Eq.~(\ref{eq:delta t}), particularly from the first two terms in the square brackets where the beam divergence fluctuations are scaled with the scattering angles. Neglecting the other terms in the square brackets, one can derive the functional dependence of the $\abs{t}$ resolution
\begin{equation} 
\hbox{RMS of } \Delta \abs{t} \approx 2p_\text{nom} \sqrt{1-\xi}\ \sqrt{\abs{t}}\ \sigma_\text{bd},
\label{eq:sigma delta t}
\end{equation}
which is consistent with the plot.

\section{Validation with dimuon sample}
\label{sec:dilepton}

\begin{figure}[ht]
\centering
\includegraphics[width=8 cm]{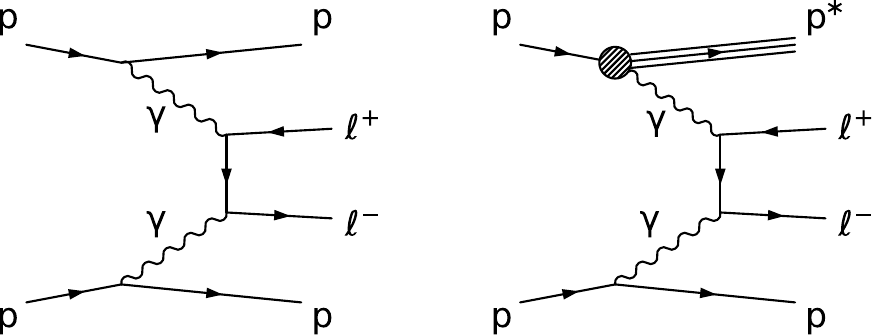}
\caption{Diagrams for $\PGg\PGg\to\PGmp\PGmm$ production with intact protons. Left: fully exclusive production, with both protons
remaining intact. Right: Single proton dissociation, with one of the two protons remaining intact. In the right plot, the three lines labelled $p^*$ 
indicate that the proton does not remain intact, but dissociates into an undetected low-mass system.}
  \label{fig:feynmanmuons}
\end{figure}

As a final check of the proton reconstruction, the calibrations and reconstruction algorithms described in the
previous sections are applied to a control sample of $\PGg\PGg\to\PGmp\PGmm$ events
with at least one intact proton (Fig.~\ref{fig:feynmanmuons}), using the 2017 and 2018 data.

As described in Refs.~\cite{Cms:2018het,Aad:2020glb}, the value of $\xi$ in signal events can 
be inferred from the muon pair via the expression:

\begin{equation}
  \xi(\PGmp\PGmm) = \frac{1}{\sqrt{s}} \left[\pt(\PGmp) \re^{\pm \eta(\PGmp)} + \pt(\PGmm) \re^{\pm \eta(\PGmm)}\right],
\end{equation}

\noindent with the $\pm \eta$ solutions corresponding to the case where the protons are moving in the ${\pm}z$ direction, respectively.

The offline event selection in the central detectors is identical to that of Ref.~\cite{Cms:2018het}. Two oppositely charged muons are required with $\pt > 50$\GeV that pass standard tight
identification criteria. In order to exclude the region dominated by resonant $Z \to \PGmp\PGmm$ production,
an invariant mass requirement of $m(\PGmp\PGmm) > 110$\GeV is also imposed. Finally, in order to enhance the (semi) exclusive production processes,
selections are applied to the track multiplicity at the dimuon vertex, and to the acoplanarity ($a = 1 - \abs{\Delta \phi(\PGmp\PGmm)}/\pi$) of the
muons. The track multiplicity selection is applied by fitting the two muons to a common vertex, and requiring that no additional charged tracks are
present within 0.5\mm of the vertex position. Back-to-back muons, characteristic of the signal process, are selected by requiring $a < 0.009$.

The protons reconstructed with the single-RP and multi-RP algorithms in these events are then examined, to look for correlations with the muons. In each event, the
two solutions, corresponding to the two arms of the spectrometer, are considered separately. In the 2018 data it is possible to reconstruct more than one proton per arm;
for this study, in order to limit the combinatorial backgrounds, we require no more than
one proton is reconstructed in the arm of interest. Backgrounds are expected to arise from real dimuon production (from Drell--Yan or
$\PGg\PGg\to\PGmp\PGmm$ events with double proton dissociation), in combination with unrelated protons from other collisions in the same bunch
crossing (``pileup'').

In Ref.~\cite{Cms:2018het}, this procedure was applied to the 2016 data, in both the $\PGmp\PGmm$ and $\Pep\Pem$ final states. Although the smaller integrated luminosity did not allow 
detailed studies, a combined $>5\,\sigma$ excess of correlated events was observed using the single-RP algorithm, compatible with the predicted signal. With the 2017 and 2018 data, 
approximately 10 times more single-RP $\PGmp\PGmm$ events are available, permitting more refined studies with this sample. 

\begin{figure}[!ht]
\centering
\includegraphics[width=7 cm]{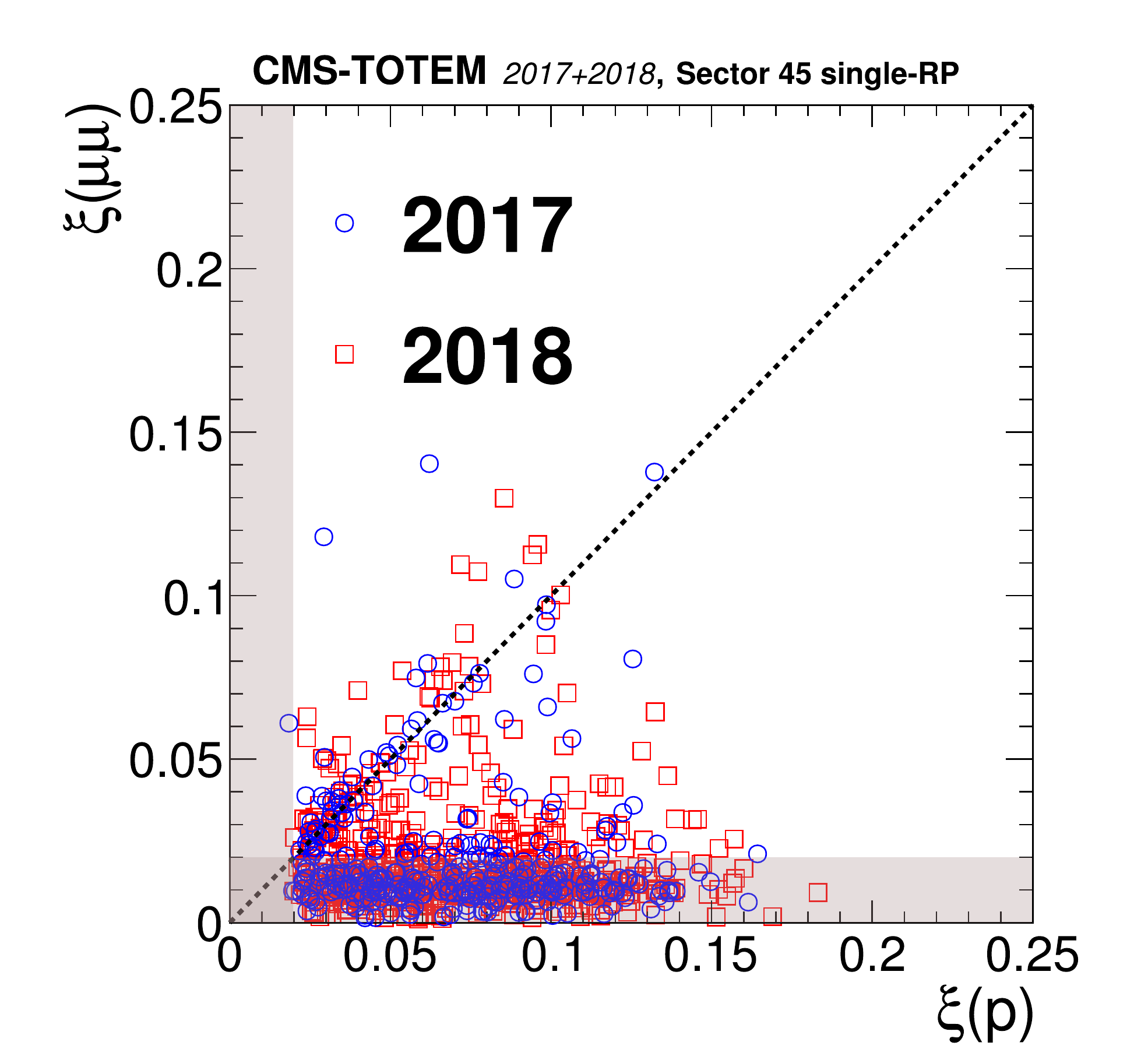}
\includegraphics[width=7 cm]{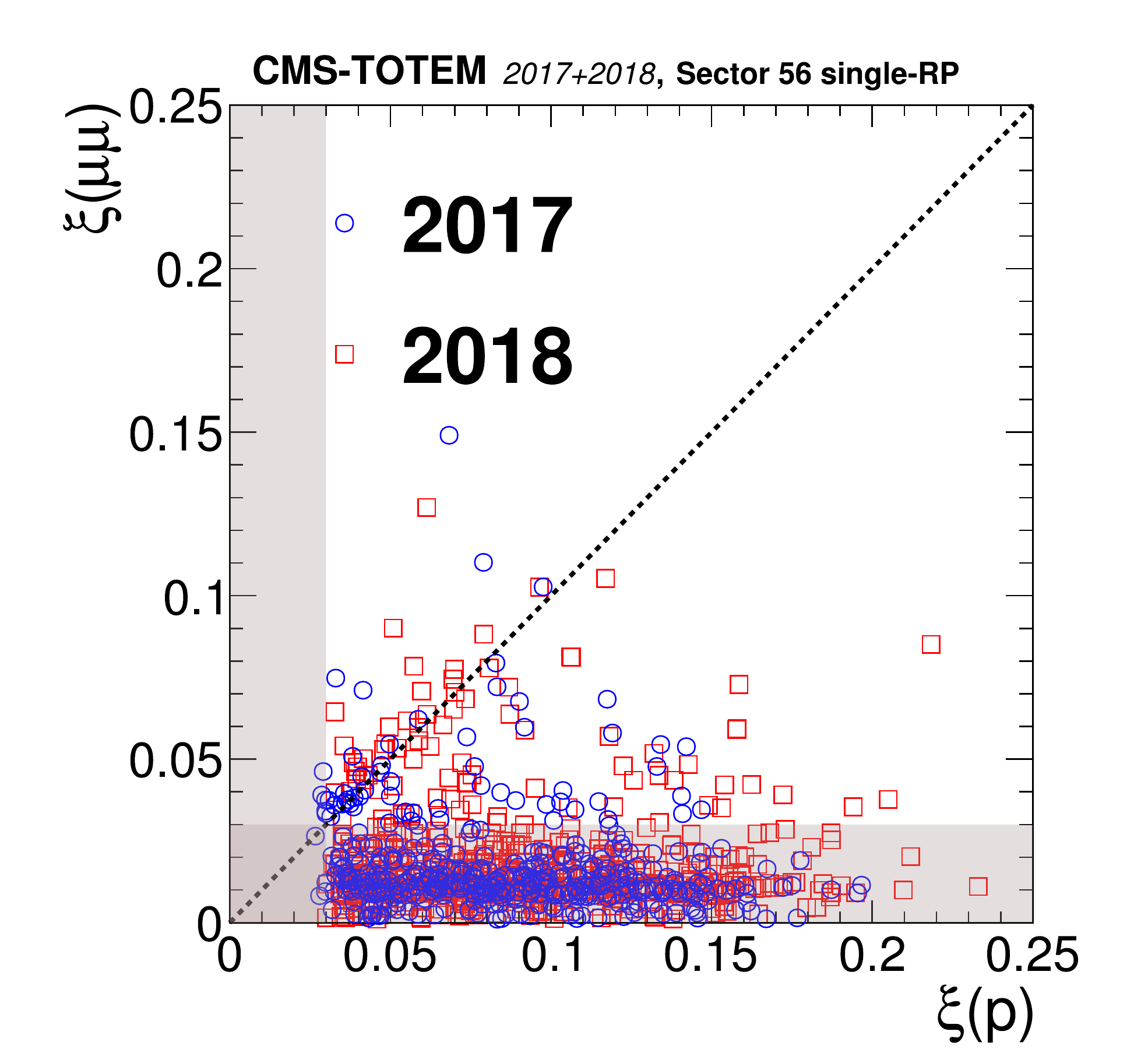}

\includegraphics[width=7 cm]{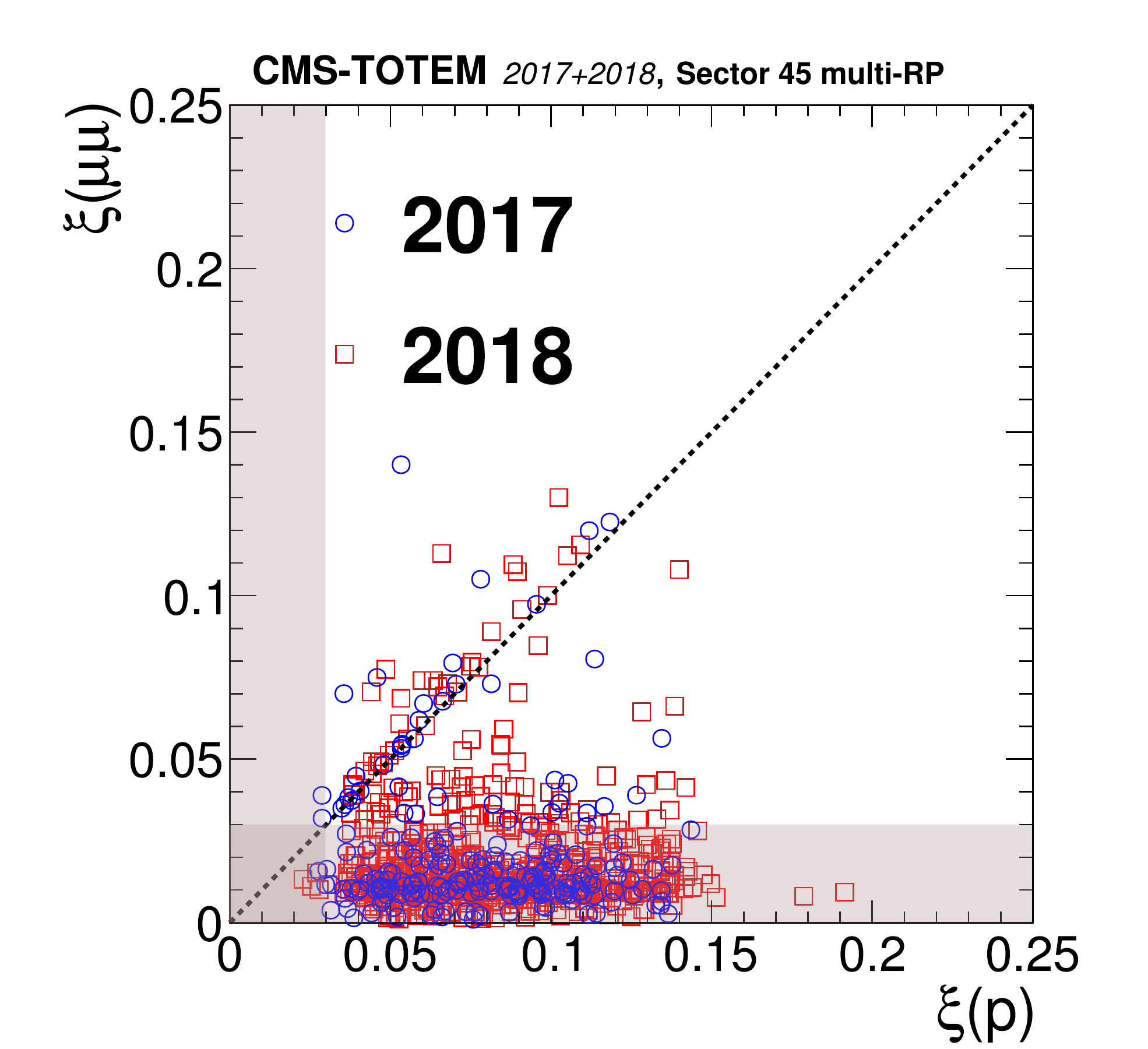}
\includegraphics[width=7 cm]{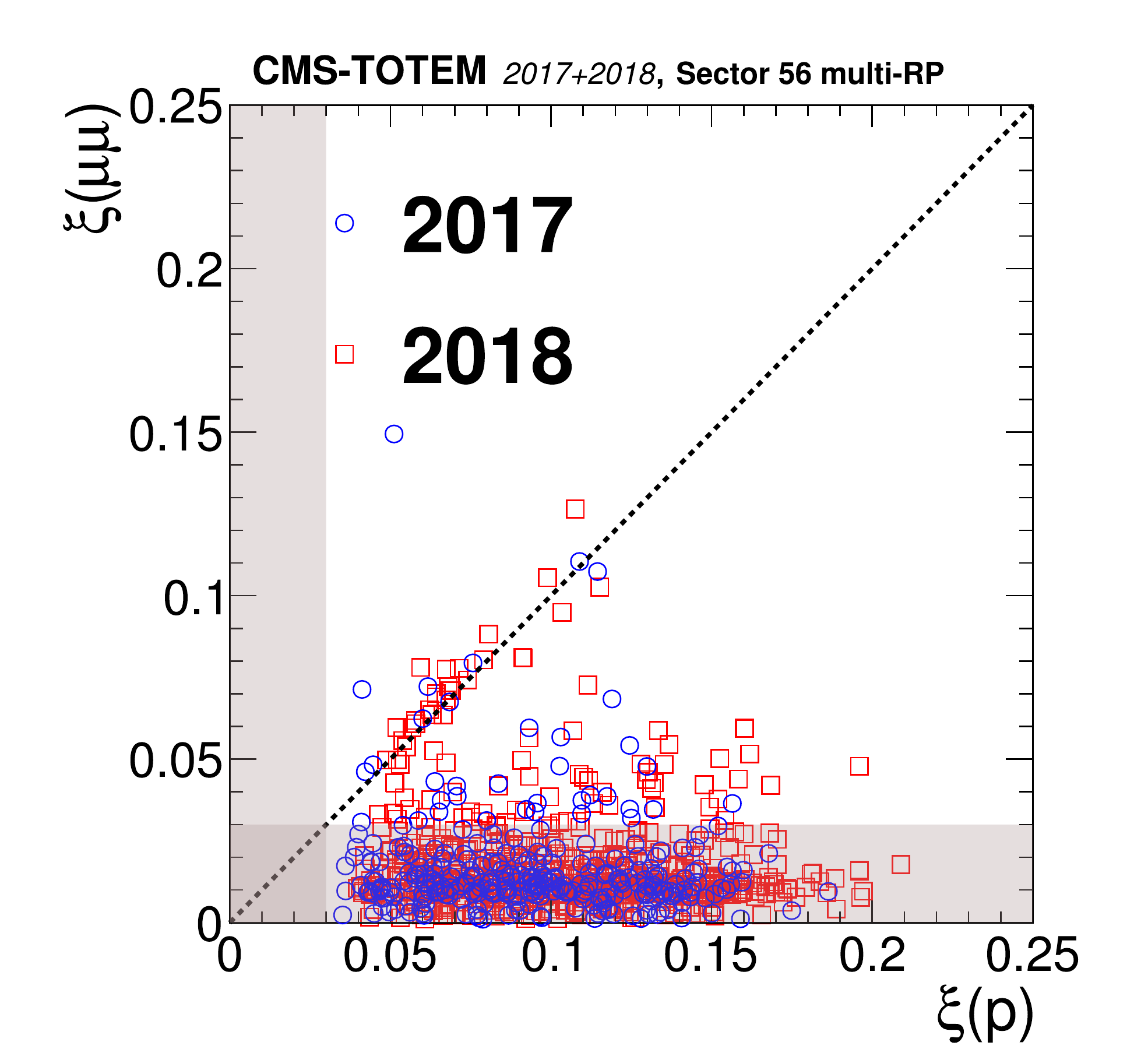}
\caption{Distribution of $\xi$(p) vs.  $\xi(\PGmp\PGmm)$ for the $z>0$ (LHC sector 45) and $z<0$ (LHC sector 56) directions in the CMS coordinate system.
The two styles of points represent the data collected during 2017 and 2018. The shaded bands represent the region incompatible with the PPS acceptance for signal events;
events in this region are expected to arise from random combinations of muon pairs with protons from pileup interactions. The upper plots show the results of the single-RP reconstruction algorithm, and the lower plots show the multi-RP results. The dotted line illustrates the case of a perfect
correlation, where signal events are expected.}
  \label{fig:muon2dcorr}
\end{figure}

Figure~\ref{fig:muon2dcorr} shows the resulting two-dimensional scatter plots from the 2017 and 2018 data, separately for the two arms and the two years. The
shaded bands indicate the approximate region that is kinematically inaccessible for signal events, since the protons would be outside the acceptance. These regions can be populated by background events where a dimuon event is combined with an unrelated proton from a pileup interaction. In
the remaining area of the plots, a clear clustering of events around the diagonal, where a fully correlated signal would be expected, is visible for both arms and years.
The samples extend to $\xi \sim 0.12$; no significant deviation from the diagonal is observed in this region. The difference between the two proton reconstruction algorithms can be seen from the plots. The multi-RP algorithm gives a narrower distribution around the diagonal and fewer off-diagonal background events, whereas the single-RP algorithm extends the coverage to lower $\xi$ values. 

In order to compare more quantitatively the data with simulation, a one-dimensional projection onto the variable $1 - \xi(p)/\xi(\PGmp\PGmm)$ is performed,
combining both arms and years, for events with $\xi(\PGmp\PGmm) > 0.04$. The expected shape of the residual background is obtained from a sideband region with the
acoplanarity ($0.009 < a < 0.1$) and extra track multiplicity ($5 < N < 10$). The expected signal shape is obtained from a simulated sample of
$\PGg\PGg \to \PGmp\PGmm$ events with both protons intact. A full simulation of the central CMS detectors is performed, and the direct simulation
described earlier is used for the protons. For the simulation, a mixture of LHC crossing angles and PPS configurations reflecting the integrated luminosity of each data taking 
condition is used. The background
shape is normalised to the data in the range $\abs{1 - \xi(p)/\xi(\PGmp\PGmm)} > 0.5$. The signal simulation is then normalised to the difference between the data and the
background in the range $\abs{1 - \xi(p)/\xi(\PGmp\PGmm)} < 0.5$.

\begin{figure}[ht]
\centering
\includegraphics[width=16 cm]{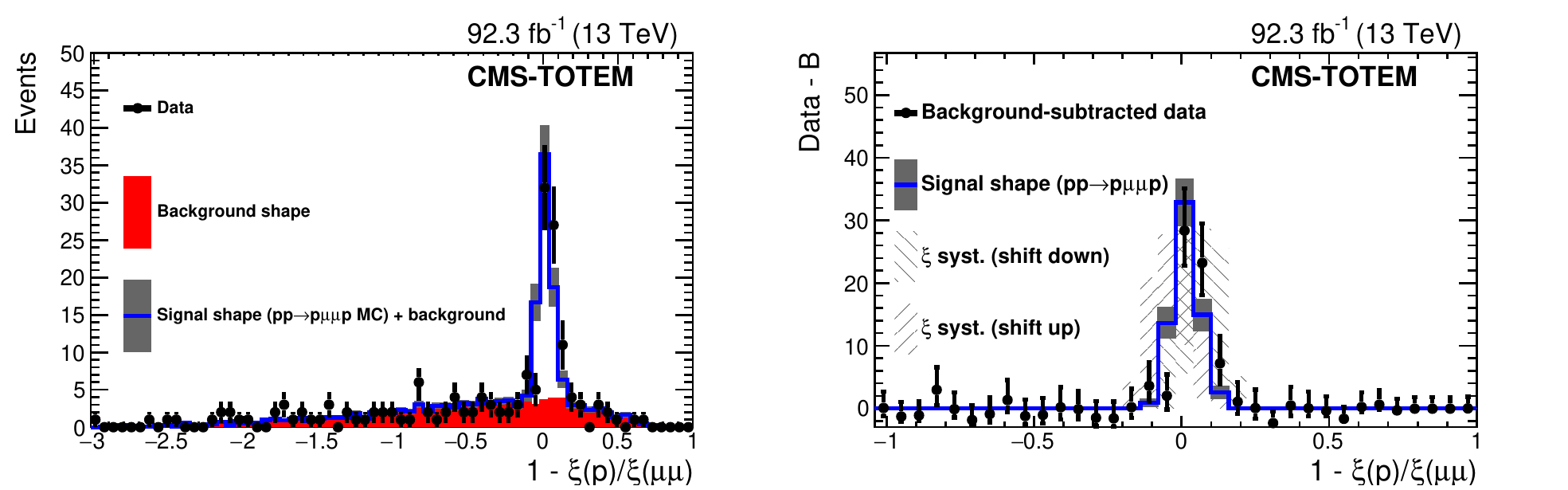}
\caption{One-dimensional projections of the correlation between  $\xi$(p) and $\xi(\PGmp\PGmm)$, for the full 2017+2018 data sample and both arms combined, using the multi-RP algorithm. A minimum
requirement of $\xi(\PGmp\PGmm) > 0.04$ is applied. The left plot shows the data compared with the background shape (solid histogram) estimated from
sideband regions, and the signal shape obtained from simulation (open histogram). The right plot shows the data and signal shape in a narrower region, after subtracting
the background component from the data. In the right plot the dark bands represent the statistical uncertainty due to the number of simulated events, whereas the two light 
shaded bands represent the effect of shifting the distribution up or down by the systematic uncertainty of the proton $\xi$ reconstruction. The vertical 
bars on the data points represent statistical uncertainties.}
  \label{fig:muon1dprojection}
\end{figure}

The resulting projections are shown in Fig.~\ref{fig:muon1dprojection}, with the data first compared with the sum of the signal and background components, and then 
to the signal shape after subtracting the background, in a narrower range. In the background-subtracted plot, the systematic uncertainties in $\xi$ are
indicated by light shaded bands on the simulation, corresponding to the cases where the reconstructed $\xi$ is shifted up or down by the systematic uncertainty. The width
of the signal peak in the data is well reproduced by the simulation ($\sim 4.8\%$, including a subleading contribution of $\sim 1.8\%$ from the muon resolution, estimated from simulation), indicating that the $\xi$ resolution is well described. The peak position is slightly shifted (by $\sim 4\%$), but well within the error bands, indicating that any residual 
effect is compatible with the known systematics.

In summary, the PPS multi-RP reconstruction was used to study $\PGg\PGg\to\PGmp\PGmm$ events with at least one final-state proton, in the kinematic
range $m(\PGmp\PGmm) > 110$\GeV and $\xi > 0.04$. A good correlation between  $\xi(\PGmp\PGmm)$ and the $\xi$ of the protons is observed in the data up to $\xi \sim 0.12$; the mean and width of the signal distribution are reproduced by the simulation, within the known systematic uncertainty. This indicates that the optics, alignment, and
related systematics of the proton $\xi$ reconstruction are well understood for the data collected during 2017 and 2018, in addition to the previously studied 2016 data~\cite{Cms:2018het}.

\section{PPS tracking efficiency}
\label{sec:efficiency}

The efficiency of the PPS tracking detector needs to be closely monitored, as radiation-induced effects can degrade the performance during the LHC operation. 

Multiple factors need to be taken into account: the efficiency of the detectors, the reconstruction algorithm efficiency, and the probability that the proton interacts with the material between the two tracking stations, and cannot be detected. 

PPS used multiple detector technologies during data-taking, and the definition of the efficiency changes accordingly. 

In 2016, with only the strip detectors used, allowing only one  proton track to be reconstructed in each station, the reconstruction algorithm efficiency is close to 100\%, since loose association constraints can be used (see Section \ref{sec:reconstruction}). The dominant role is played by detector effects, such as radiation damage and multi-tracking inefficiency. A more detailed description is given in Section \ref{sec:strips-eff-alternative}. 

In 2017 and 2018, the pixel detectors could resolve multiple tracks in the same station, and a different approach for the efficiency estimation is used. The reconstruction efficiency for multi-RP protons was split into two independent multiplicative factors: the efficiency of the ``near'' detector and the so-called multi-RP efficiency. The former takes into account only the detector-related effects for the near RP, whereas the latter accounts for detector-related efficiency in the far RP, the reconstruction algorithm efficiency, and the proton propagation. The first factor is derived as described in Sections \ref{sec:strips-eff-alternative} and \ref{sec:pixels-eff-alternative}, and the second is discussed in Section \ref{sec:multiRP-eff-alternative}.

Efficiency corrections are computed for each RP and data-taking period separately. 

\subsection{Silicon strip detector efficiency}
\label{sec:strips-eff-alternative}

Two main sources of inefficiency affect the PPS strip detectors: radiation damage and the presence of multiple tracks in the same event. These effects were studied separately and are described with two efficiency factors. 

If more than one particle produces a signal in the strip detectors, track candidates that do not correspond to a real particle, so-called ghost tracks, will be found. Because of this, strip detectors can only be used in events where one track is present \cite{strips-multitrack-note}. 

\begin{figure}
\centering
\includegraphics{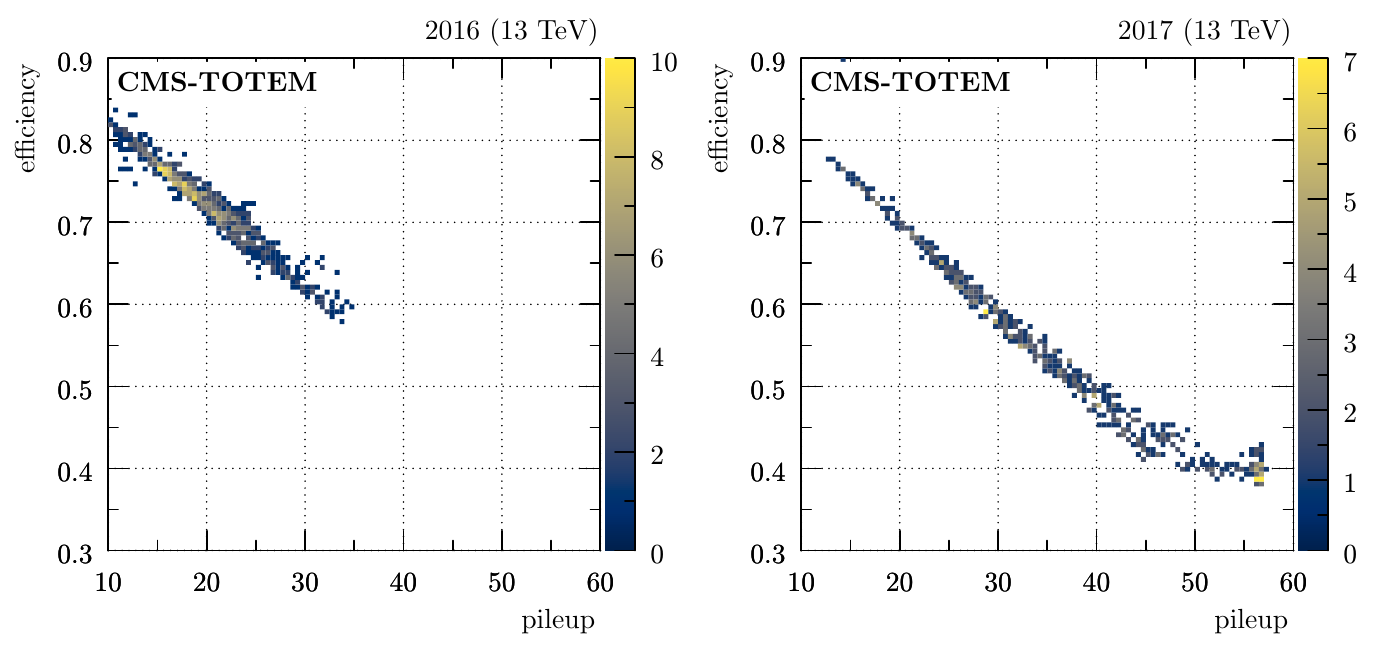}
\caption{Strip multiple track efficiency component versus pileup in the sector 45 near RP. The color code represents per-bin counts. {Left:} data-taking period between the first and the second 2016 Technical Stops. {Right:} data-taking period between the second and the third 2017 Technical Stops.}
\label{fig:strips-multitrack-efficiency}
\end{figure}

In minimum-bias samples, events with one or more protons in the strip detectors are selected. This is done by requiring either at least one track pattern in both strip orientations, or a number of detector hits greater than the maximum allowed by the pattern recognition algorithm, which is tuned to accept a single proton track with some tolerance for detector noise. The selected events are used to compute the efficiency factor, which is the ratio between the number of reconstructed tracks and the number of selected events. This efficiency factor is inversely related to the pileup, and ranges between 40 and 80\%. Consistent results are observed in both 2016 and 2017, and across different sectors, and illustrated in Fig.~\ref{fig:strips-multitrack-efficiency}.

The second factor takes into account time-dependent effects produced by radiation, and it has been studied with a tag-and-probe method \cite{strips-radiation-note}. In order to probe the efficiency of the strip detectors in one station, minimum-bias events with one reconstructed track in the other RP (tag) of the same arm, passing loose quality criteria, are selected. Events with more than one recognized track pattern in the strip detector being probed are excluded, together with events with multiple tracks in the tag RP, in case of pixel detectors. A matching window of $\abs{\Delta\xi} < 0.01 $ is defined, where $\Delta\xi$ represents the difference between the single-RP $\xi$ measurement associated with the track in the tag RP, and the measurement in the RP being probed, if a track is detected. 

The efficiency correction factor is defined as the ratio between the number of events in which a strips track satisfies this matching criterion, and the total number of events selected.
Statistical uncertainties are negligible, and two sources of systematic uncertainty were evaluated. A 1\% uncertainty is associated to the choice of the minimum-bias sample used for the estimation; an uncertainty of the same size is associated to the variation of the quality criteria applied to the tagging track. A larger (10\%) conservative systematic uncertainty is applied to 2016 efficiency factors because a different method is used. Efficiencies are derived by comparing $\xi$ distributions in data with respect to the ones observed in the alignment fill, when the detectors had not suffered any radiation damage yet. The uncertainty is estimated by comparing with results obtained with the tag-and-probe method. 

\begin{figure}
\centering
\includegraphics[width=.45\textwidth]{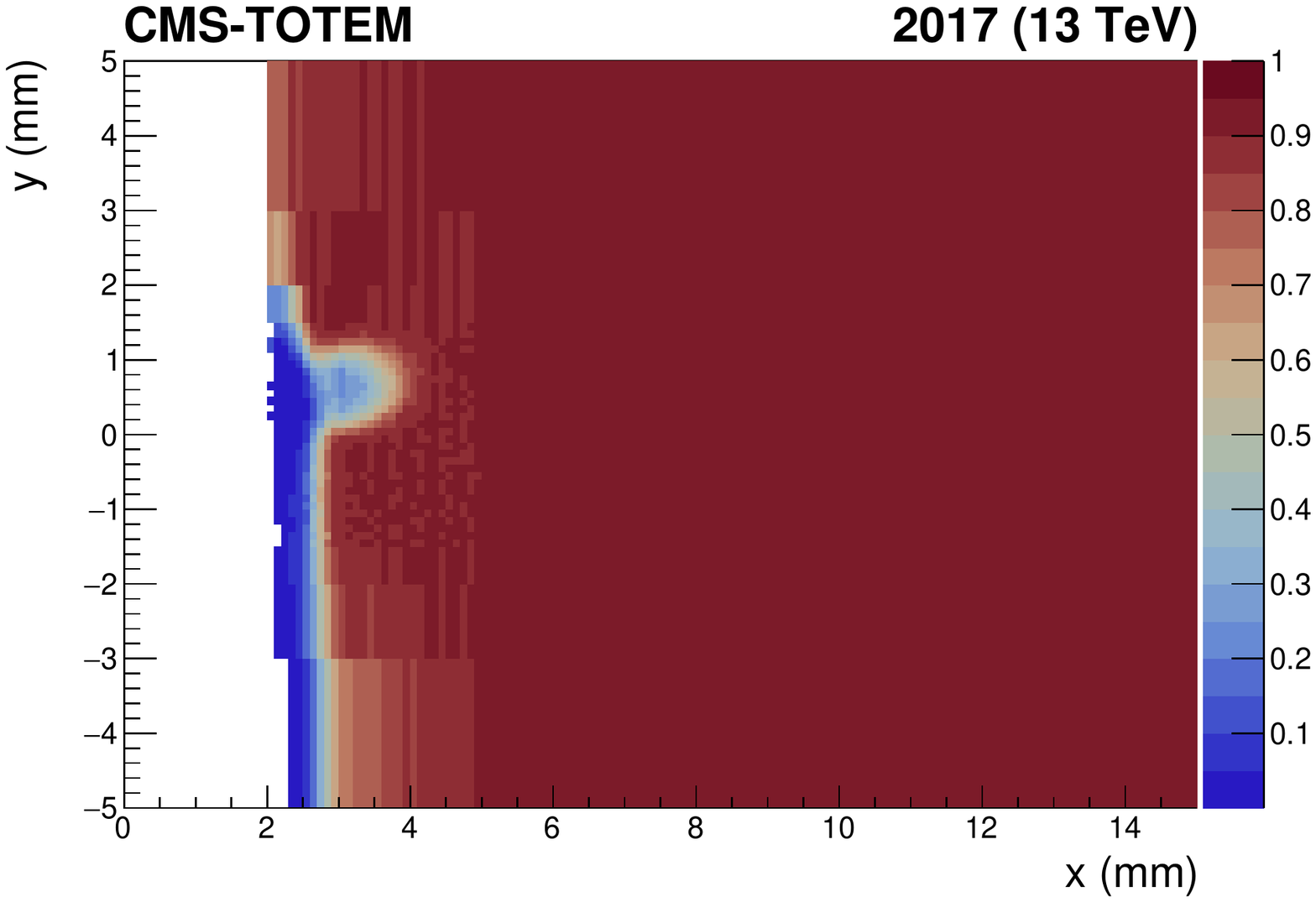}
\includegraphics[width=.45\textwidth]{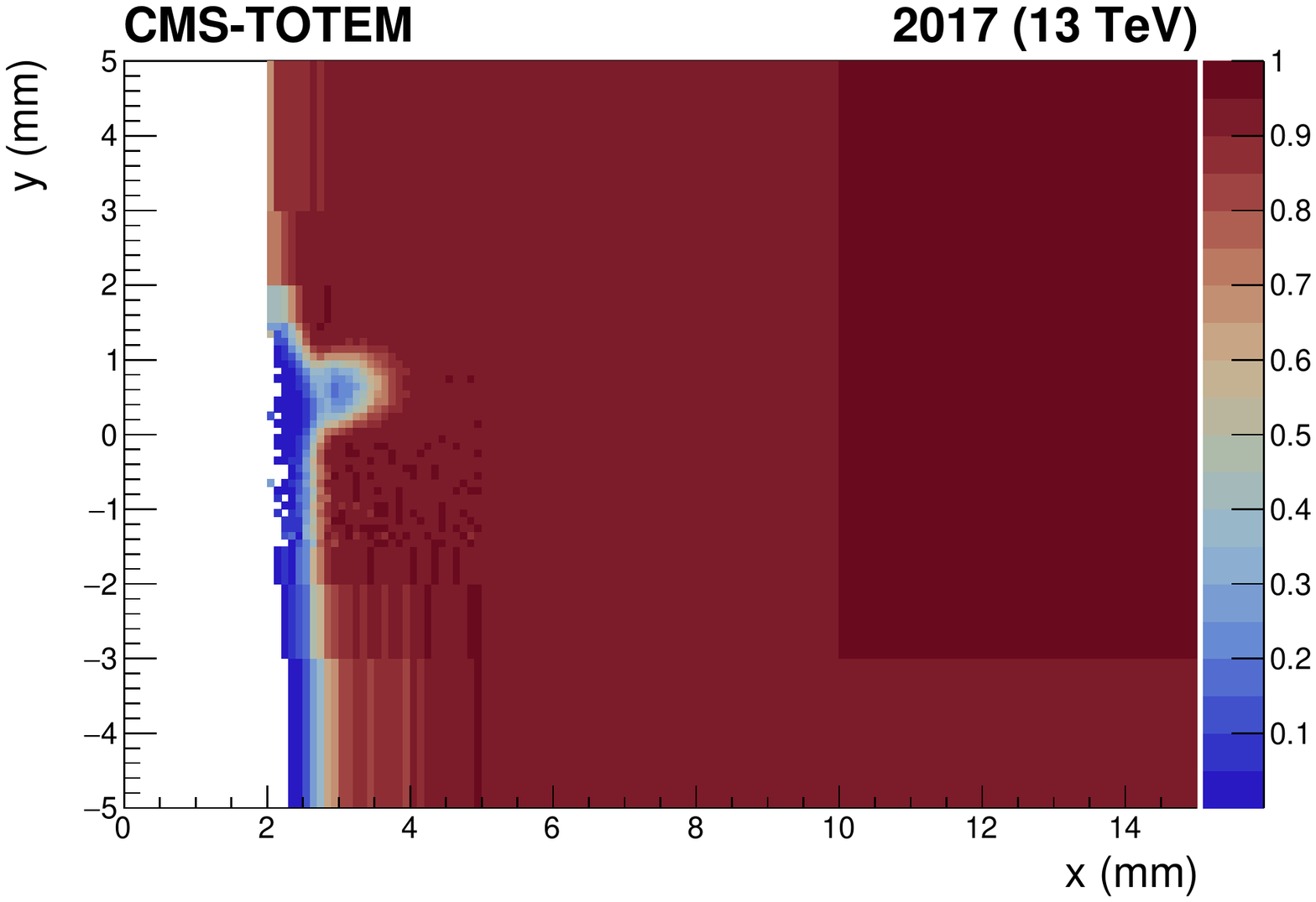}
\includegraphics[width=.45\textwidth]{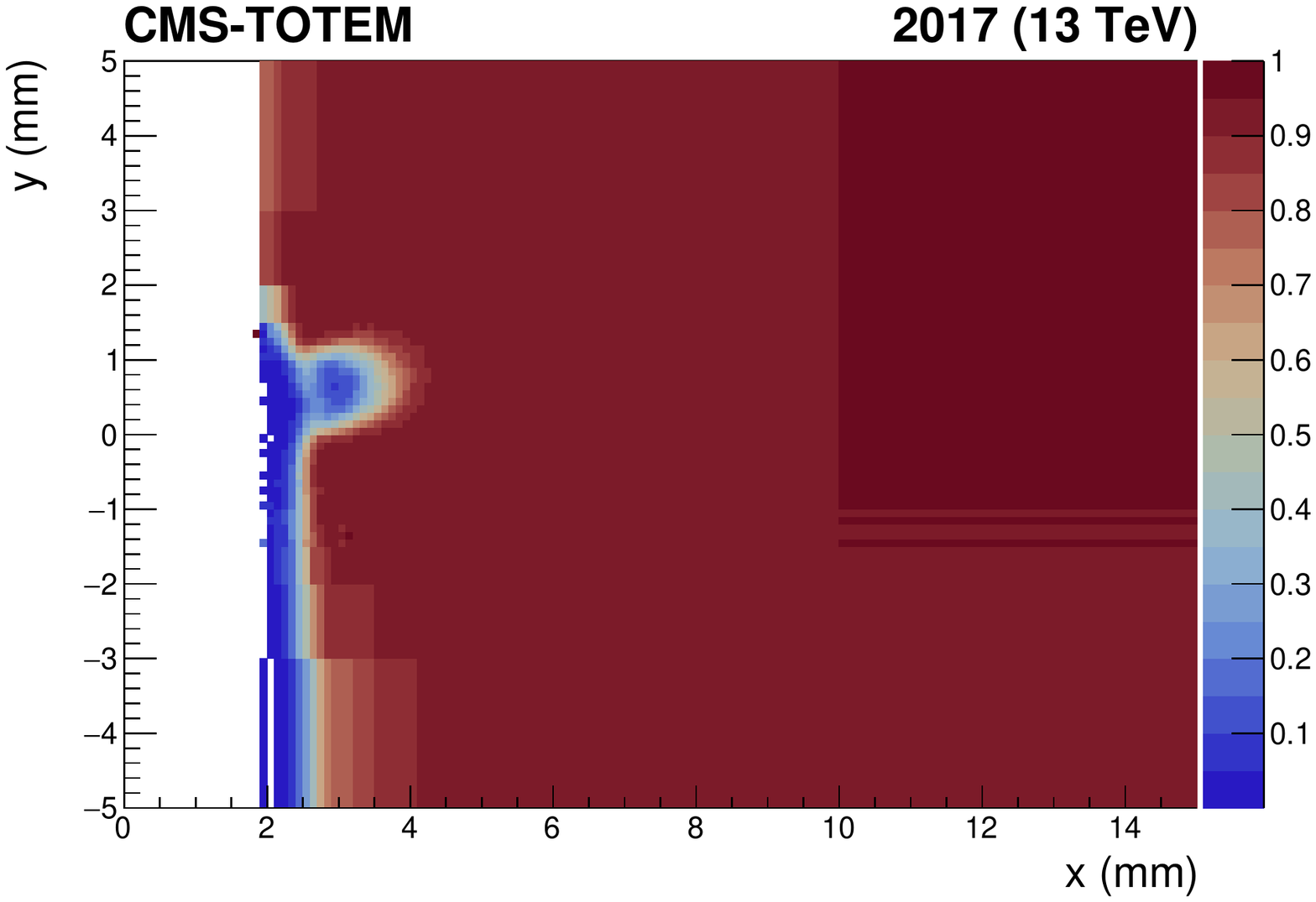}
\includegraphics[width=.45\textwidth]{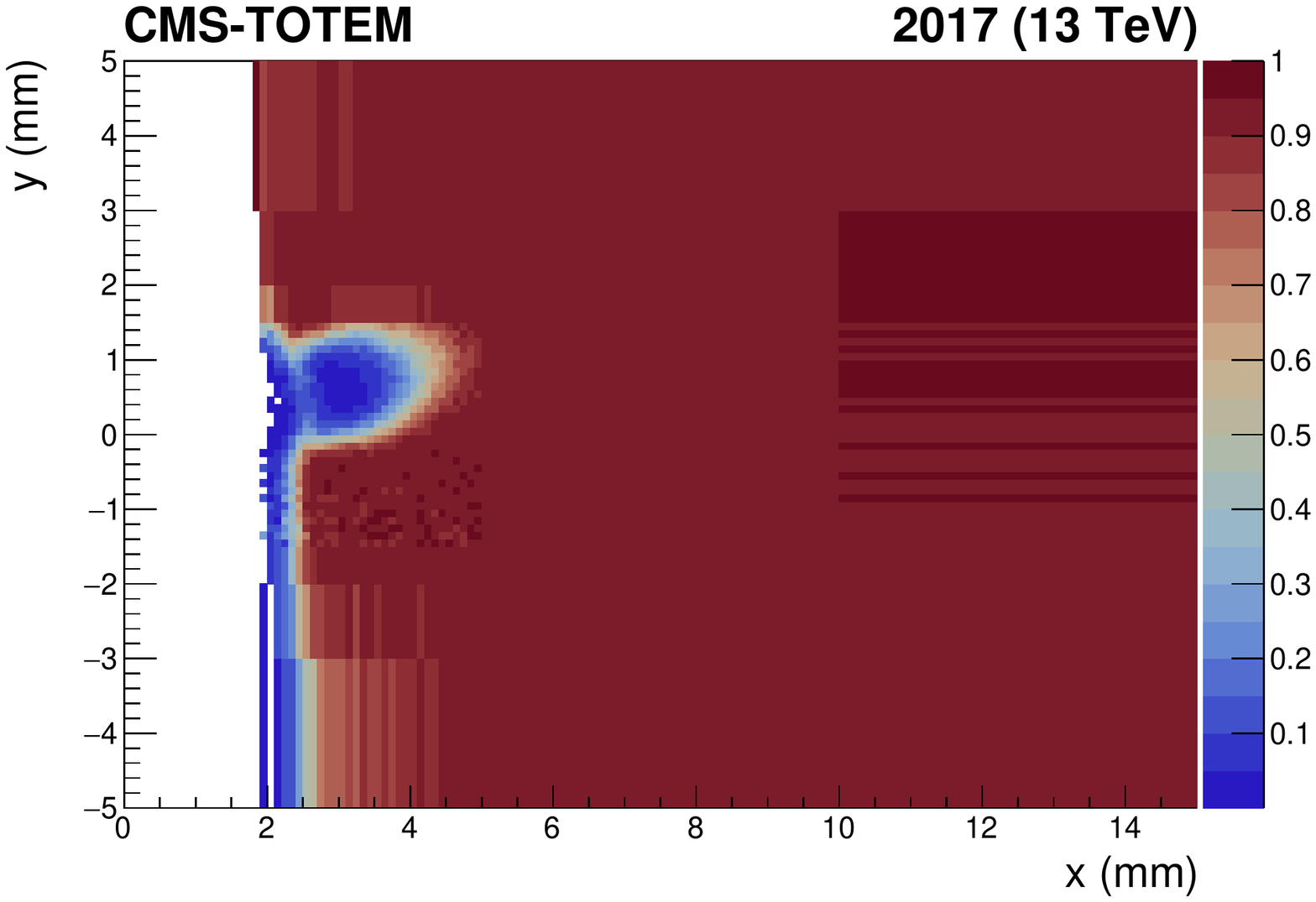}
\caption{Strip detector tracking efficiency component (in color code) related to radiation effects, computed with data collected in 2017, over different consecutive data-taking periods. The figure shows the results for the sector 56 near station in 2017, as a function of the $x$-$y$ coordinates of the track measured in the tagging far station, for different periods. Each period is defined as an interval in integrated luminosity computed since the detector installation. Upper left: $L_\text{int} = 0$--9\fbinv. Upper right: $L_\text{int} = 9$--10.7. Lower left: $L_\text{int} = 10.7$--18.5\fbinv. Lower right: $L_\text{int} = 18.5$--22.2 \fbinv. }
\label{fig:strips-radiation-efficiency}
\end{figure}

Figure~\ref{fig:strips-radiation-efficiency} shows the results as a function of the $x$-$y$ coordinates of the track measured in the tagging RP, for the region covered by the detector acceptance and below the collimator aperture limits. The area damaged by radiation is clearly visible and its size and inefficiency grows with the integrated luminosity. However, efficiency measurements show average values higher than 95\% in the rest of the detector area. Similar results are observed in the 2016 data, although the lower collected integrated luminosity reduced the radiation effects. Data-taking periods in which strips detectors were not inserted or operational are excluded from the presented results. They mainly affect the last period of 2017 (lower right plot of Fig. \ref{fig:strips-radiation-efficiency}), where they account for $\approx$10\% efficiency loss.

In 2016, the near-far RPs correlation between inefficiency factors due to multiple tracks in strip detectors has been measured between 50\% and 80\%. The complete tracking inefficiency can therefore be computed as the product of the following factors: the multiple-track inefficiencies (taking into account their correlation), the radiation damage inefficiency for both the near and far stations, and the proton interaction probability. The latter has been measured by the TOTEM experiment to be approximately 2\% (see Section 7.5.1 in Ref. \cite{Niewiadomski:1131825}).

\subsection{Pixel detector efficiency}
\label{sec:pixels-eff-alternative}

The main contribution to the pixel detector inefficiency comes from radiation effects. The method used to derive the efficiency is described in detail in Ref. \cite{pixel-radiation-note} and is based on the measurement of the efficiency of each detector plane during data taking. A minimum bias sample collected at the beginning of the detector operation is used to model the track distribution; the track efficiency is quantified as the probability of having at least three efficient detector planes out of six.

\begin{figure}
\centering
\includegraphics[width=0.45\textwidth]{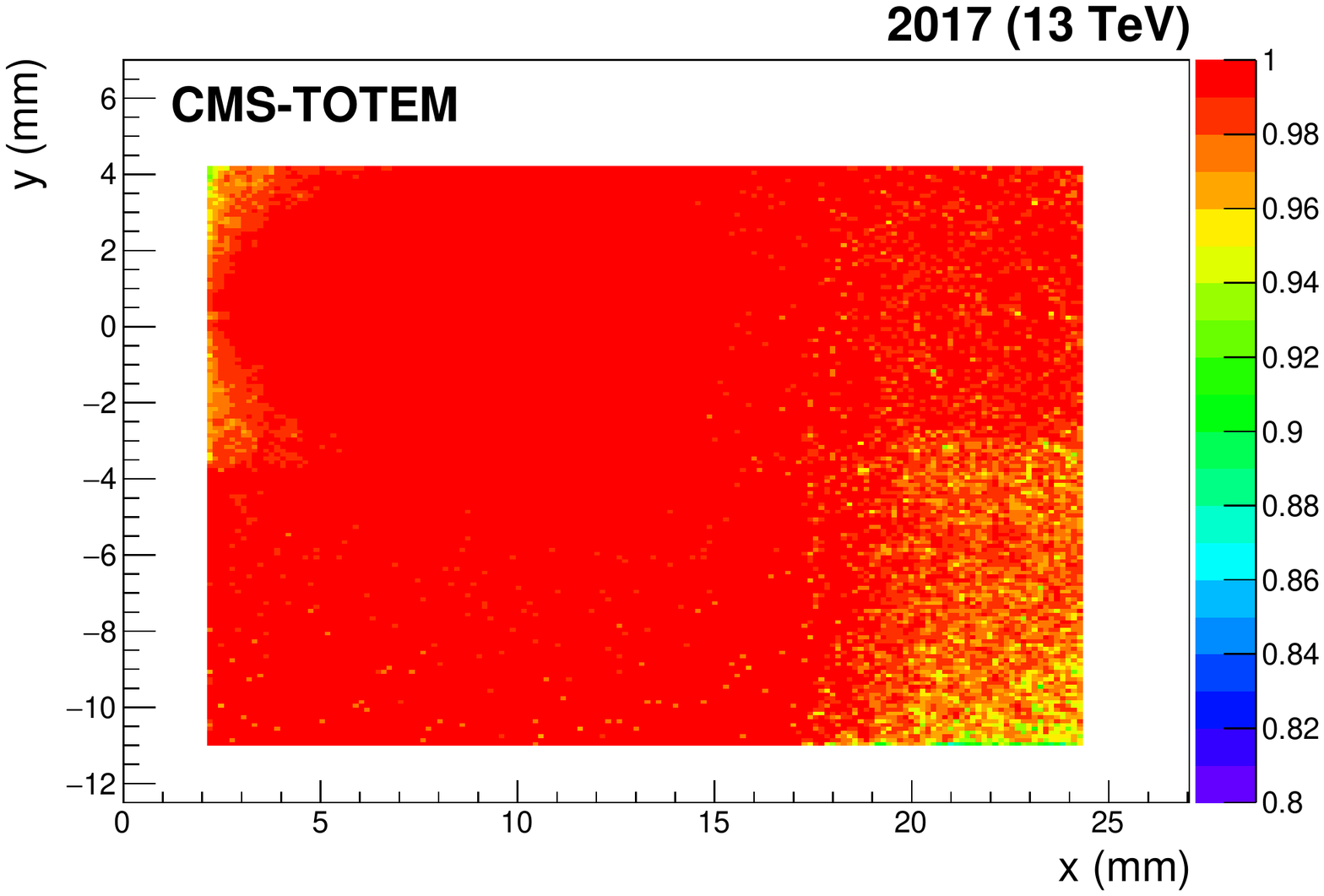}
\includegraphics[width=0.45\textwidth]{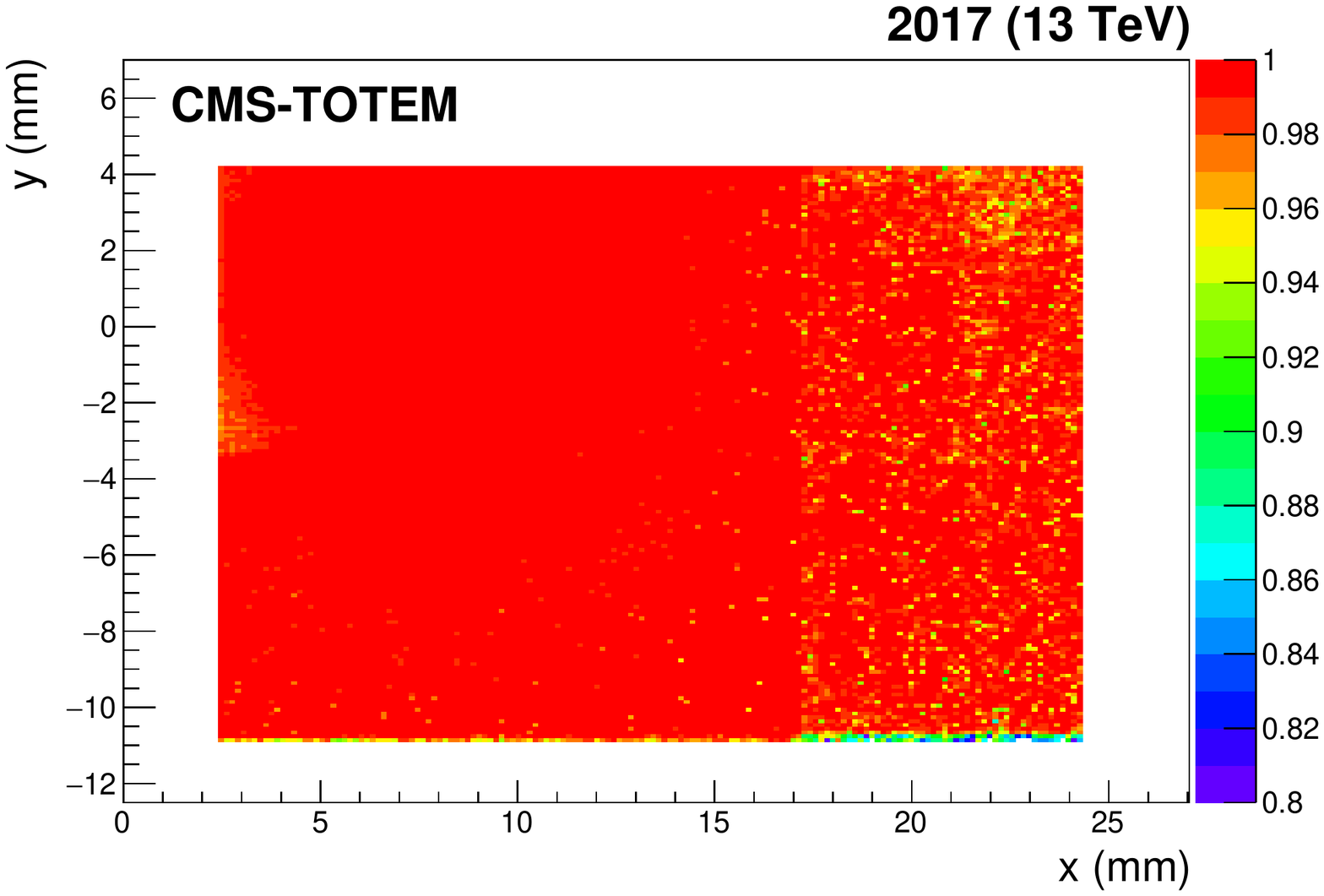}
\caption{Pixel detector efficiency map, computed on the first data collected in 2017 for sector 45 (left) and 56 (right), and shown as a function of the $x$-$y$ coordinates. The color code represents the efficiency value. The slightly lower efficiency on the bottom-right corner of the sector 45 far station is due to suboptimal detector configuration.}
\label{fig:pixels-radiation-efficiency}
\end{figure}

The results are represented as a function of the $x$-$y$ coordinates on a scoring plane perpendicular to the beam, as in Fig.~\ref{fig:pixels-radiation-efficiency}. Statistical uncertainties are estimated to be negligible using Monte Carlo simulations, and a  1\% systematic uncertainty linked to the minimum bias sample choice is assigned. 

\begin{figure}
\centering
\includegraphics[width=0.24\textwidth]{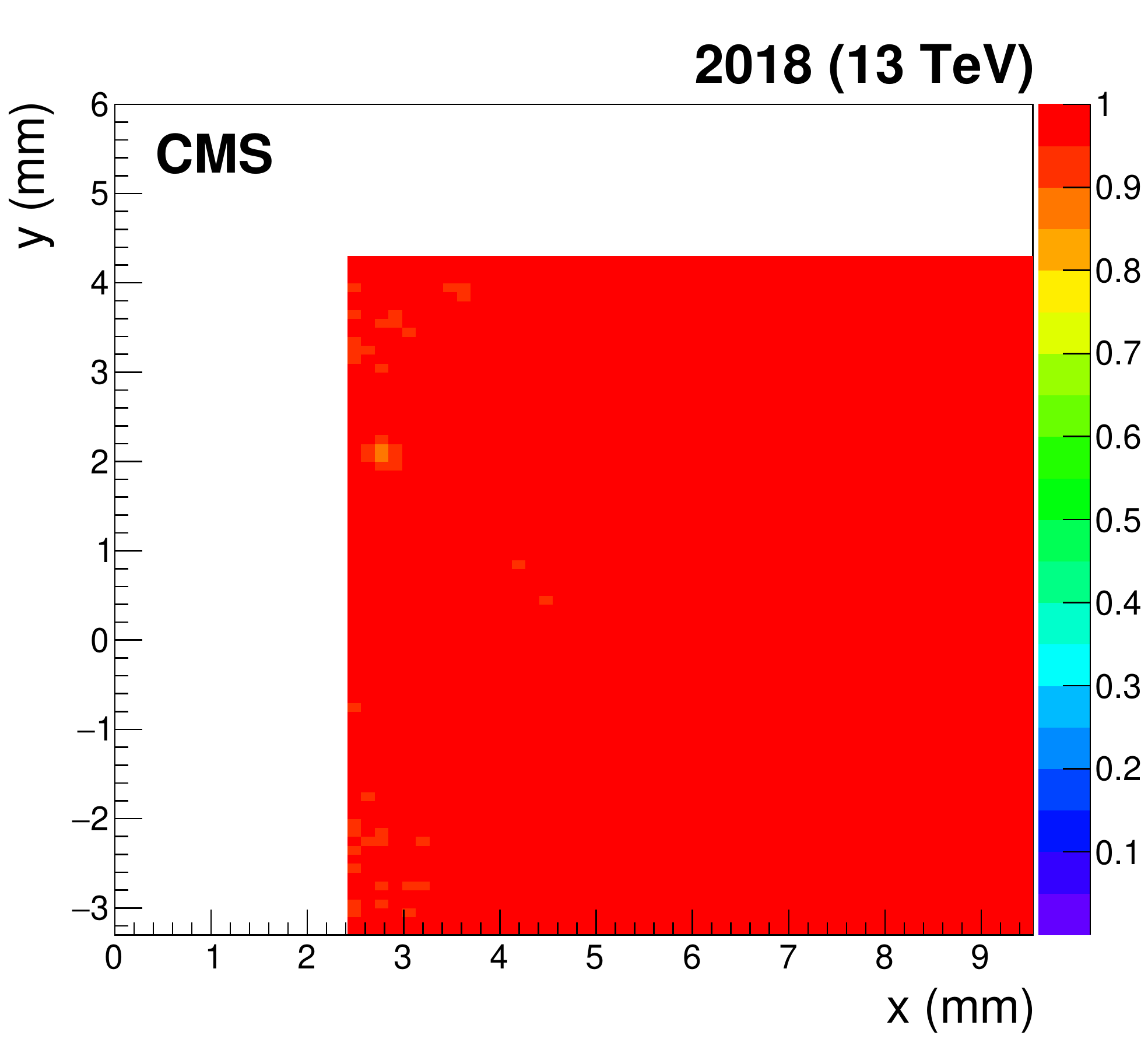}
\includegraphics[width=0.24\textwidth]{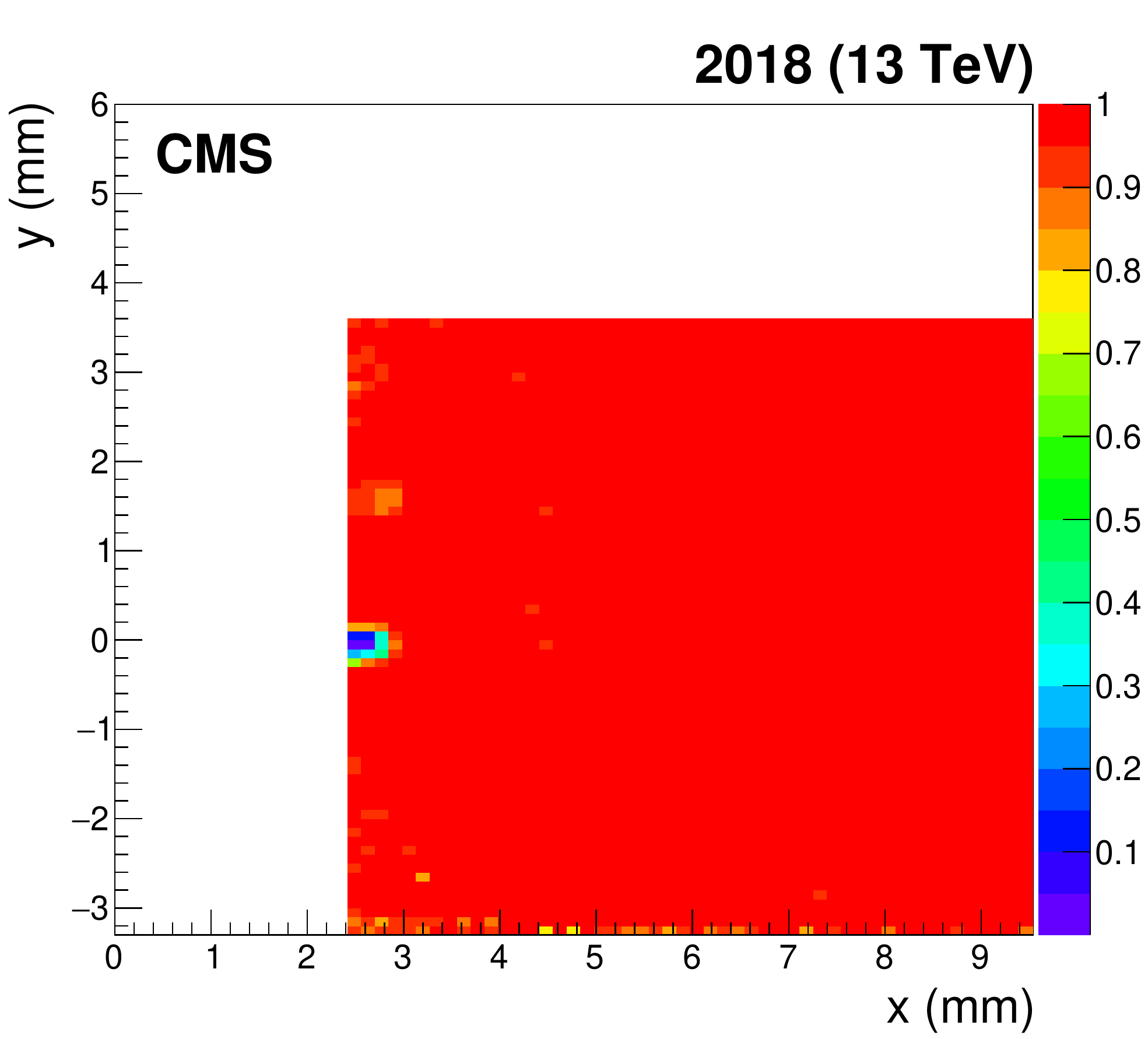}
\includegraphics[width=0.24\textwidth]{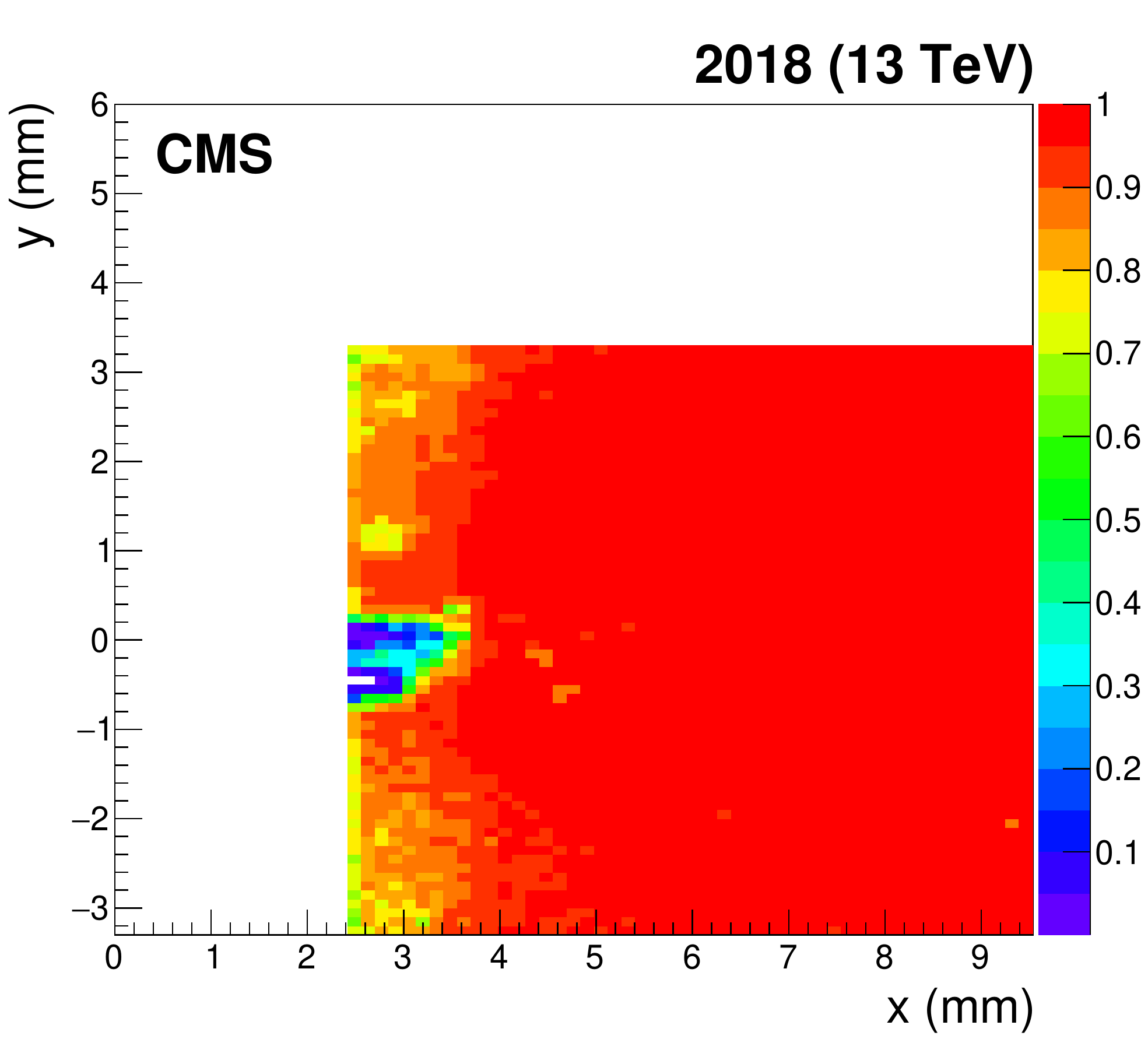}
\includegraphics[width=0.24\textwidth]{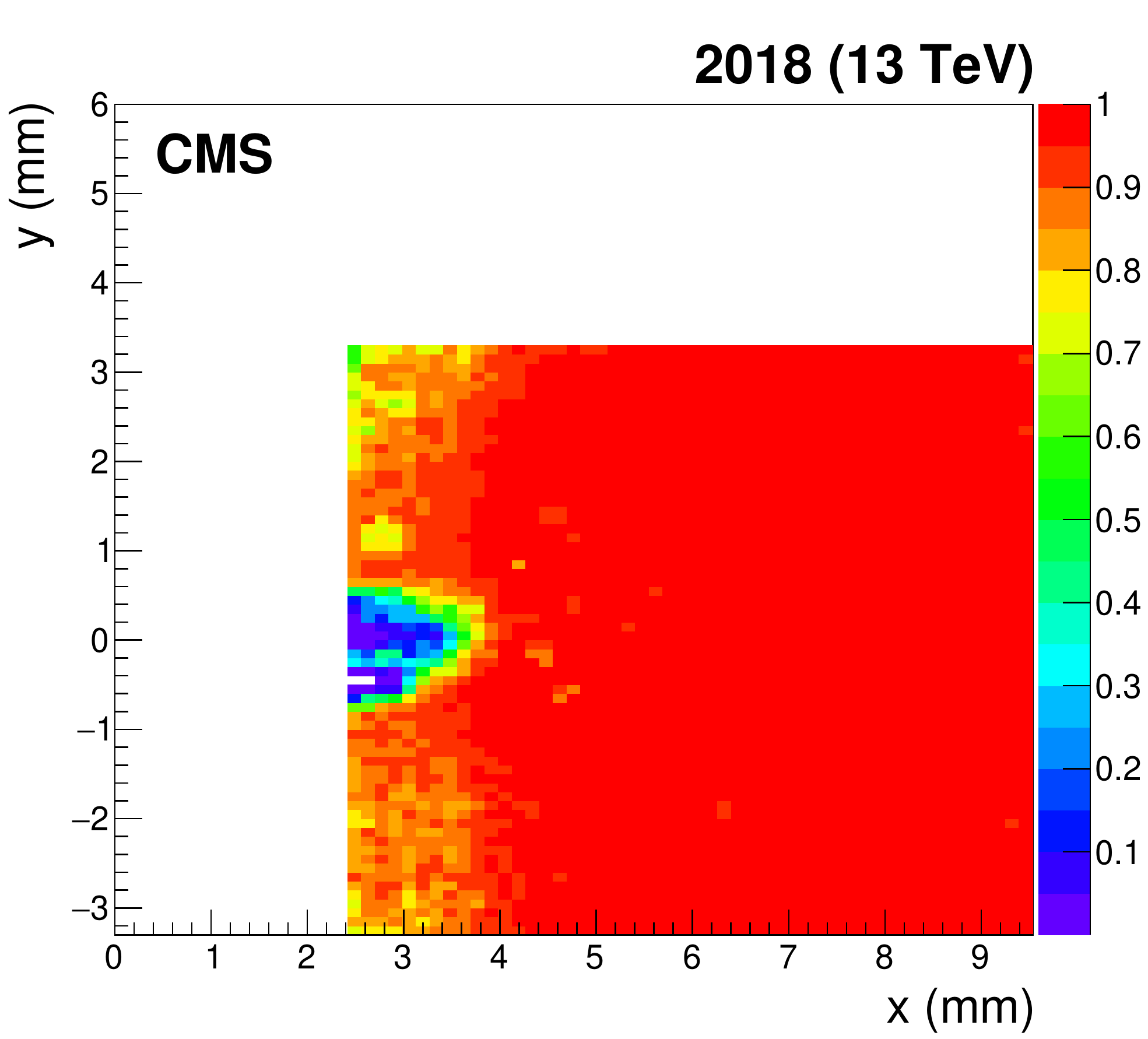}
\caption{Evolution of the pixel detector package efficiency (in color code) in the detector region closest to the beam for the sector 45 far station, computed with data collected in 2018. During each TS, detectors in both sectors were vertically shifted by 0.5 mm downwards.
From left to right: efficiency computed after the detector collected $L_\text{int} =  $,  21.0,  50.3,  57.8\fbinv, respectively. Each efficiency map is produced using a small data sample of $\sim0.5 $\fbinv.}
\label{fig:pixels-radiation-efficiency-vs-lumi}
\end{figure}

In 2017 and 2018 the efficiency $x$-$y$ maps exhibit a small damaged region where the sensors are most irradiated, as shown in Fig.~\ref{fig:pixels-radiation-efficiency-vs-lumi}. This inefficiency is actually due to the radiation damage of the electronics and not of the sensor itself. This region expands progressively with integrated luminosity. Outside the damaged region, the efficiency reaches a plateau higher than 98\%. During each technical stop the RPs were shifted vertically by 0.5\mm, so as to spread the radiation damage over a wider region, and thus mitigate its effects. 

Another effect can cause inefficiency in pixel detectors. If one of the protons coming from the interaction point interacts upstream of the near RPs, it can generate a shower of secondary particles. If the number of tracks exceeds the reconstruction capabilities of the pixels in the near RPs, a shower may cause the detectors to become inefficient.

This inefficiency factor has been quantified by studying the number of hits measured in pixel detectors. The track reconstruction algorithm is tuned to reconstruct a maximum of ten track candidates, in order to save computation time and storage. When this threshold is exceeded, no track is reconstructed. Shower events are thus identified as events with no tracks where the number of detected hits is significantly higher than that expected from detector noise. 

The fraction of events identified as showers scales linearly with pileup, and is highly correlated, as expected, between the near and far detectors in the same sector. A conservative inefficiency upper limit of 1.5\,(1.7)\% for sector 45\,(56) was measured with a 0.1\% systematic uncertainty, which accounts for the dependence on pileup. 

\subsection{Multi-RP efficiency factor}
\label{sec:multiRP-eff-alternative}

The multi-RP efficiency factor is evaluated in the same way in 2017 and 2018, and includes the efficiency of the detectors installed in the far RPs, the efficiency of the multi-RP reconstruction algorithm, and the probability that a proton propagates from the near RPs to the far without interacting. 
These multiple components are evaluated together using a tag-and-probe method. For each data-taking period, minimum-bias samples are selected for this purpose. Each single-RP proton reconstructed with the near RPs is used as tag, provided that its track angle measured with that tracking station is lower than 20 mrad. This selection excludes very inclined background tracks that do not originate from the interaction point.

The efficiency is evaluated as the ratio between the number of times in which a multi-RP proton is reconstructed using the single-RP tag proton, and the number of tag protons. The systematic uncertainties related to the sample choice for the efficiency estimation are  $\approx$1\%. Asymmetric statistical uncertainties are evaluated with the Clopper--Pearson frequentist approach \cite{ClopperPearson}. 

\begin{figure}
\centering
\includegraphics[width=0.49\textwidth]{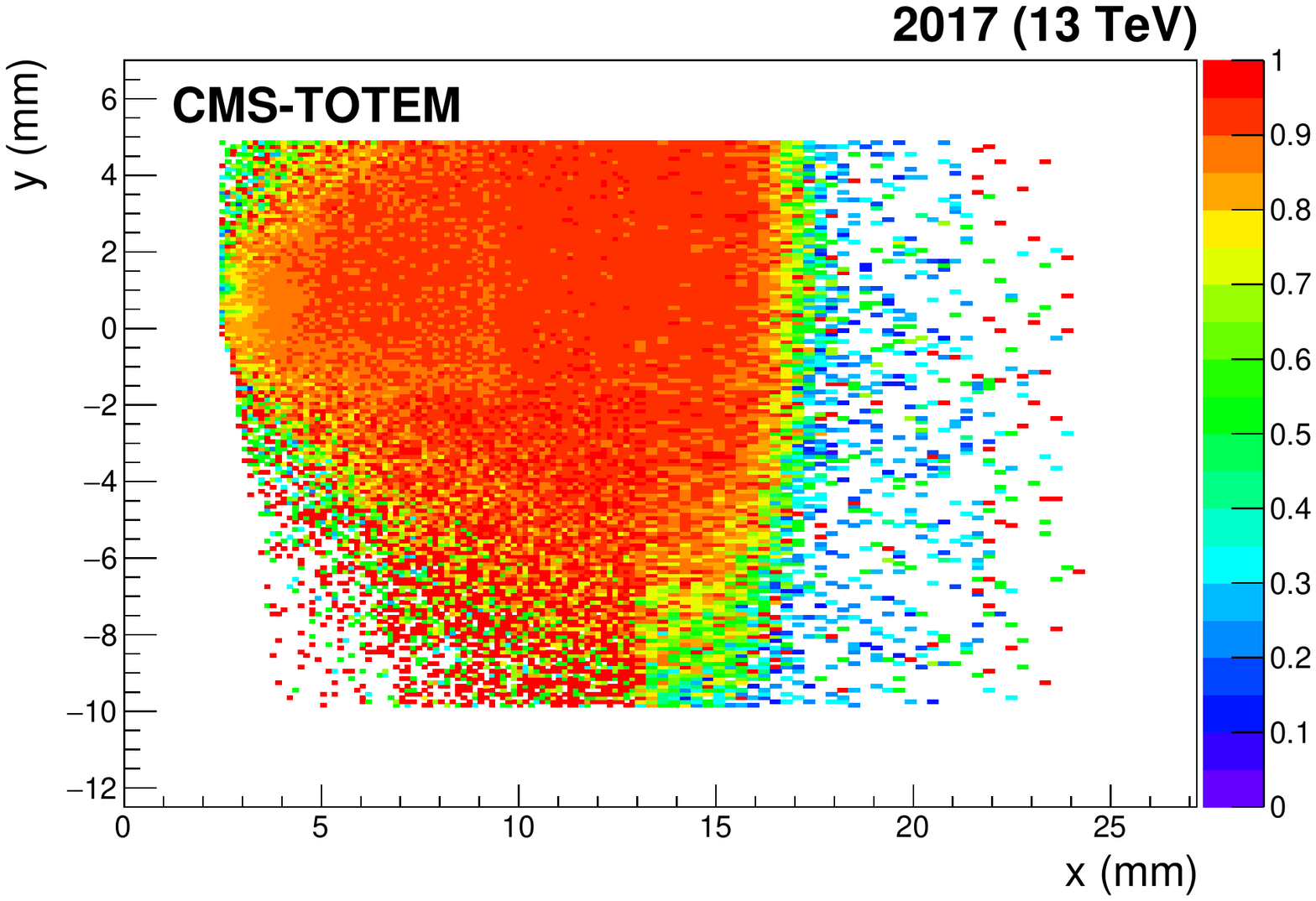}
\includegraphics[width=0.49\textwidth]{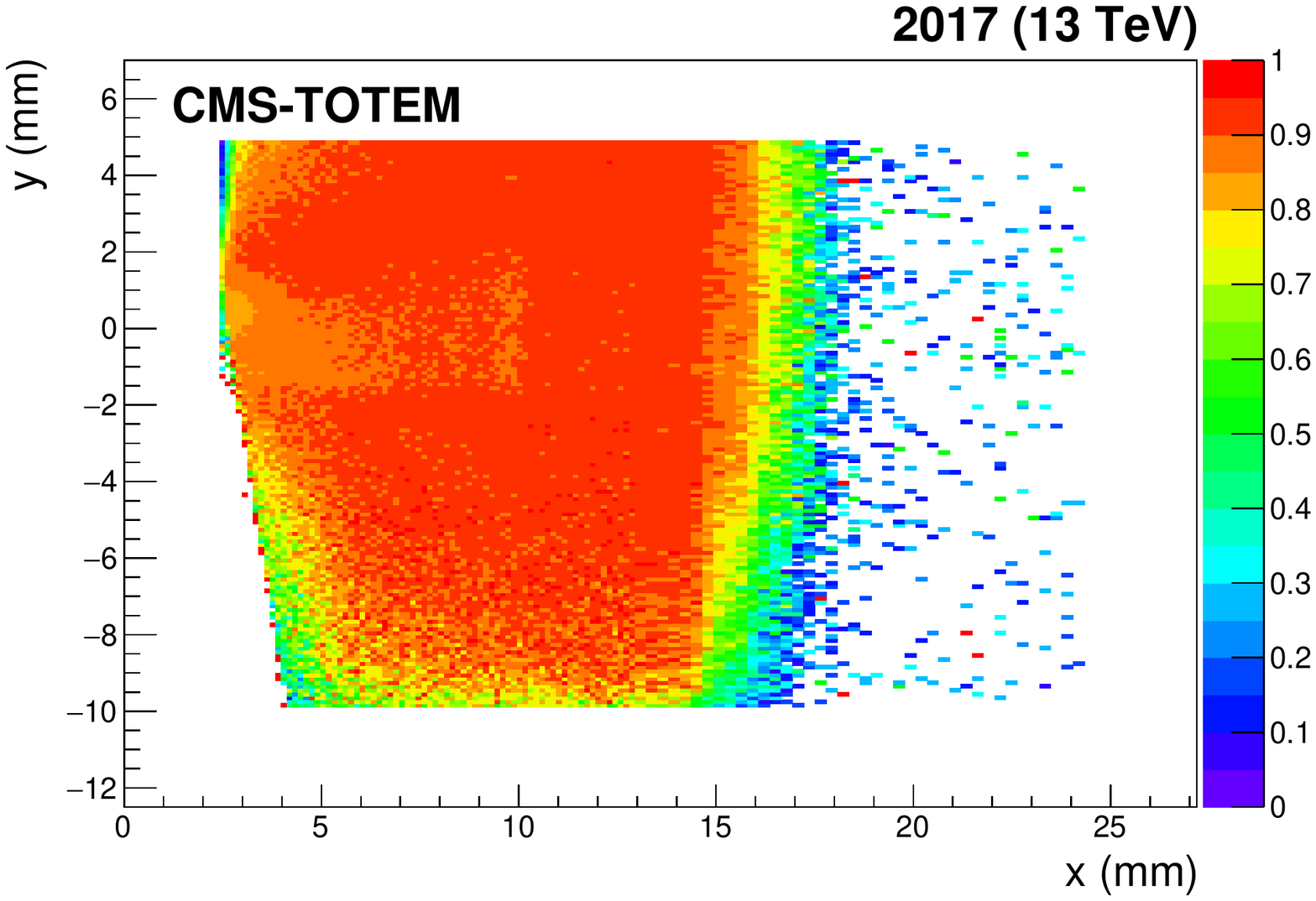}
\includegraphics[width=0.49\textwidth]{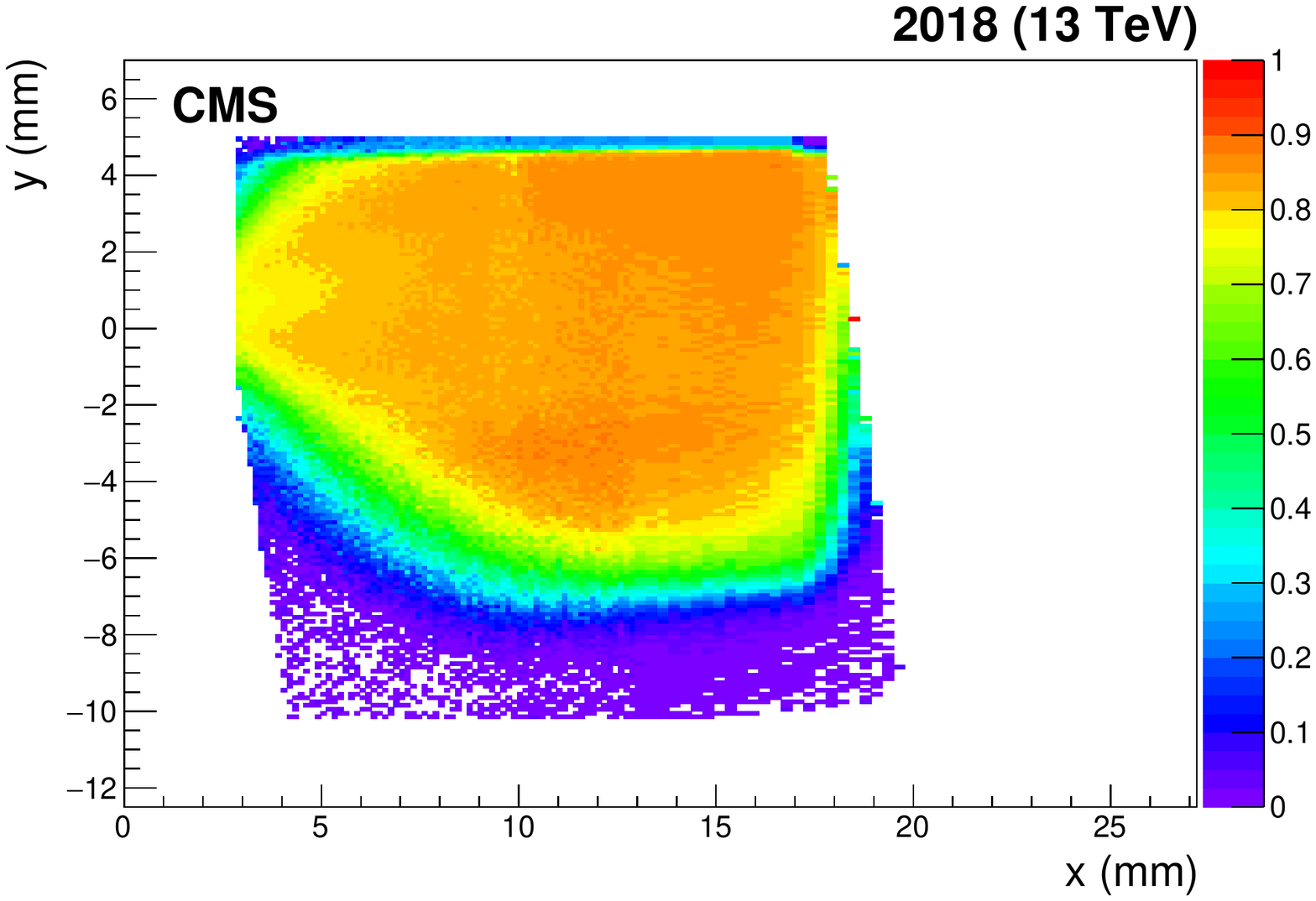}
\includegraphics[width=0.49\textwidth]{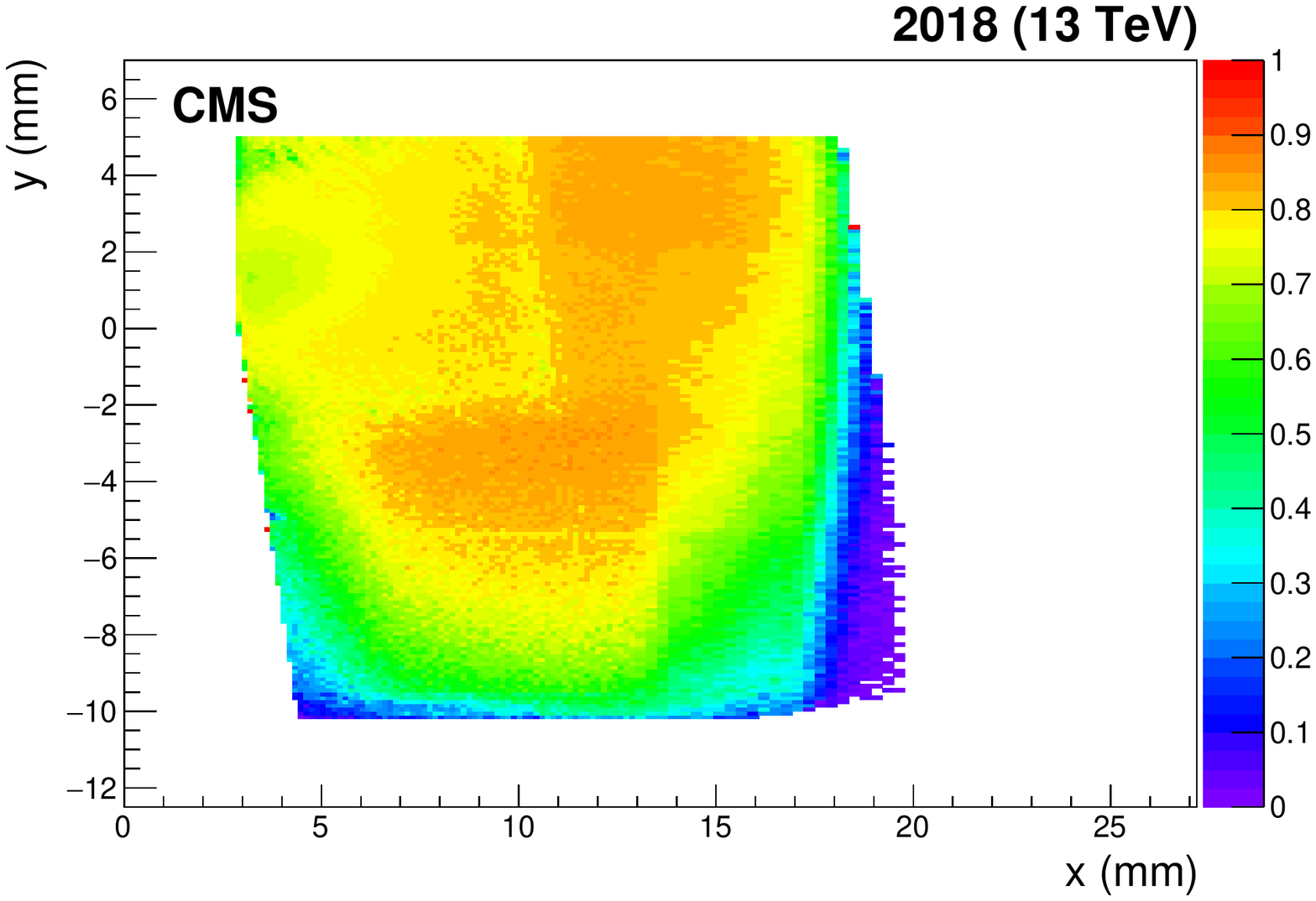}
\caption{Multi-RP efficiency factor as a function of the $x$-$y$ coordinates that includes the efficiency of the detectors installed in the far RPs, the efficiency of the multi-RP reconstruction algorithm, and the probability that a proton propagates from the near RPs to the far ones without interacting. The color code represents the efficiency value. Top: multi-RP efficiency in sector 45 (left) and 56 (right) at the beginning of the 2017 data-taking. Bottom: multi-RP efficiency in sector 45 (left) and 56 (right) at the beginning of the 2018 data-taking.}
\label{fig:multiRP-efficiency}
\end{figure}

The efficiency is plotted in Fig.~\ref{fig:multiRP-efficiency} as a function of the $x$-$y$ coordinates of the near RP scoring plane. The overlap between the acceptances of the RPs in the same sector, combined with the collimator aperture limits, defines the shape of the efficiency map. 

This efficiency has generally a plateau value higher than 90\% in 2017, and slightly lower in 2018. These high values reflect the good performance of the detectors and of the reconstruction algorithm.  Lower performance can be observed in the most irradiated region because of radiation damage and multiple tracks. The latter takes place when more than one track in the far RPs satisfies the association requirements with the near RPs track. Under these circumstances the multi-RP reconstruction cannot choose among the far RP tracks, and fails, causing inefficiency. 

Consistent results are observed in 2017 and 2018 when restricting the analysis to events with a single track in the near RP. A small loss in the 2018 multi-RP reconstruction algorithm performance is observed when including events with multiple tracks in the near station, because of the higher multiple-match probability, as mentioned above.

\begin{figure}
\centering
\includegraphics[width=0.49\textwidth]{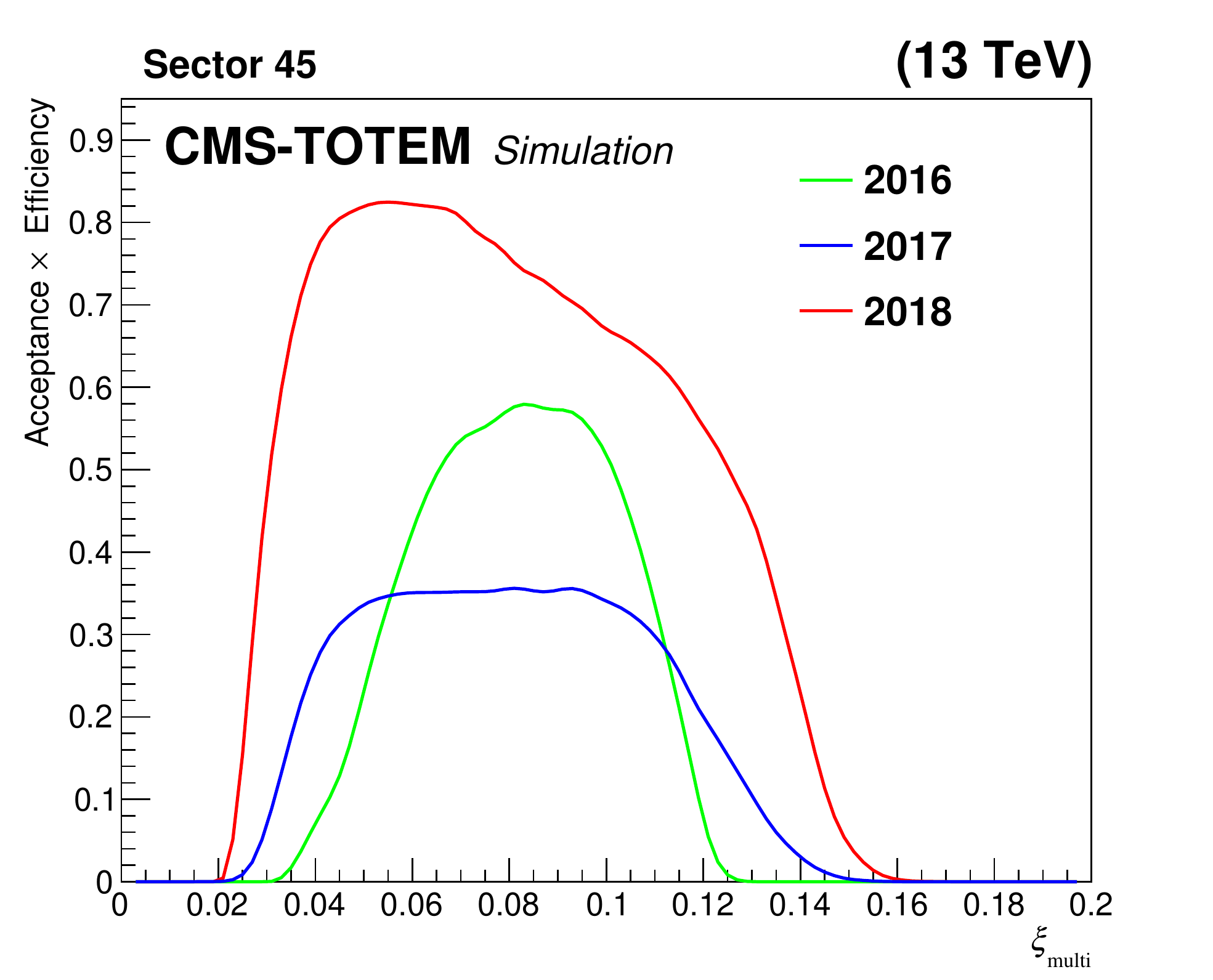}
\includegraphics[width=0.49\textwidth]{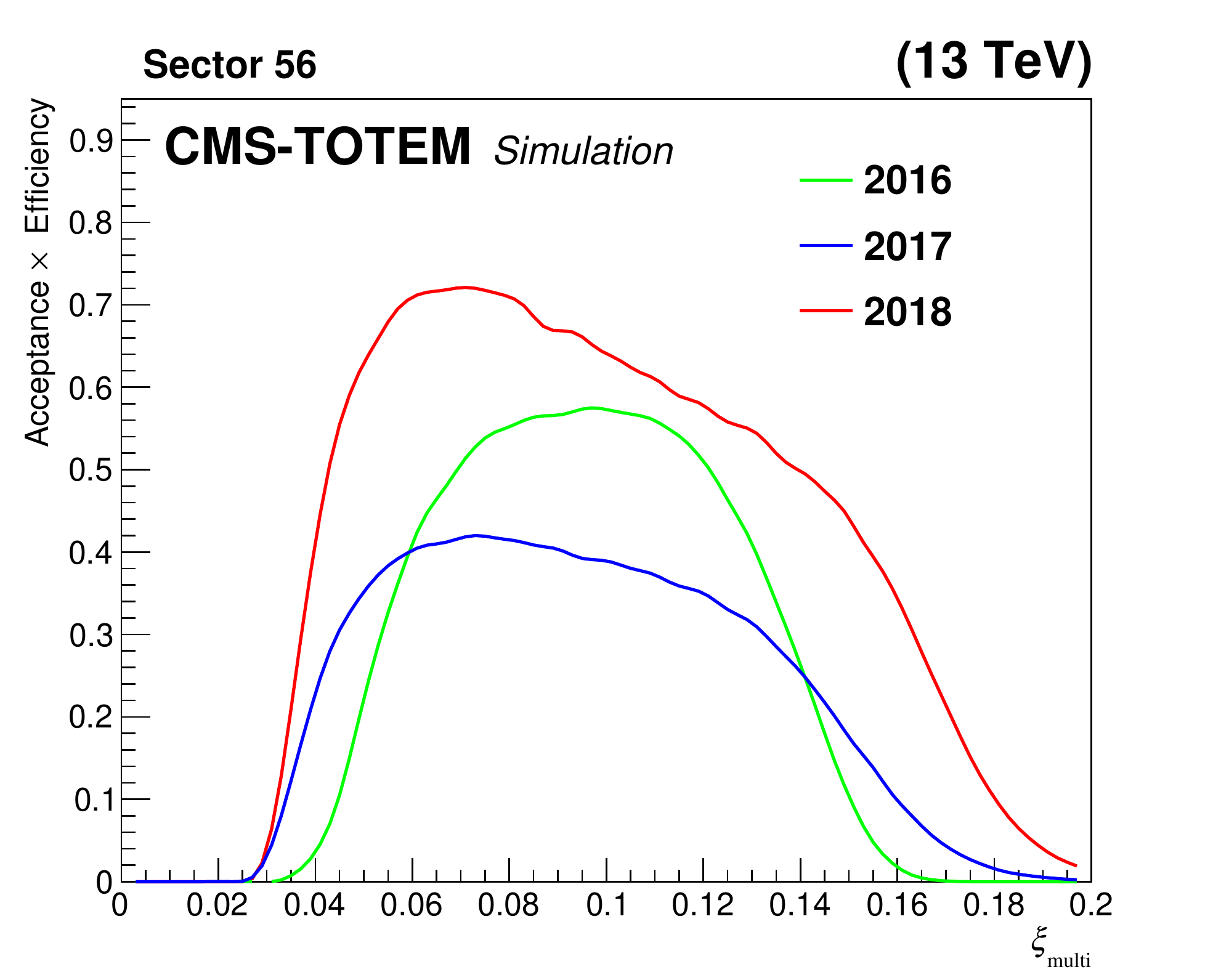}
\caption{Fraction of reconstructed multi-RP protons, as a function of $\xi_\text{multi}$, for a proton sample produced with the PPS direct simulation. Acceptance and efficiency effects are taken into account.  The left and right plots show results for sector 45 and 56, respectively. The efficiency systematic uncertainties, computed by combining in quadrature the systematic uncertainties estimated for each efficiency factor, are 10, 2.7, and 2.1\% for 2016, 2017, and 2018, respectively.}
\label{fig:efficiency-summary}
\end{figure}

Figure \ref{fig:efficiency-summary} shows the fraction of reconstructed multi-RP protons predicted by the fast simulation, which includes both efficiency and acceptance effects. The difference in the shape of the plots for the three years is mainly due to the different acceptances (cf.~Fig.~\ref{fig:acceptance-summary}). The higher value of the fraction in 2018 reflects the presence of the pixel detectors (as opposed to the strip ones) in both RP stations. 

The difference between the 2016 and 2017 performance is caused by multiple factors: the average pileup in 2017 was significantly higher than in 2016, producing a higher strip multi-tracking inefficiency (Fig.~\ref{fig:strips-multitrack-efficiency}). The integrated luminosity accumulated in 2016 was about one fourth that in 2017, resulting in less severe radiation damage. Furthermore, the sector 45 near RP was not available for a significant portion of 2017 ($\approx$24\% of the whole data-taking), thus effectively lowering the overall efficiency, since downtime is included as an inefficiency component. 

\section{Timing}
\label{sec:timing}

In order to study the performance of the proton vertex matching provided by the PPS timing detectors, a special data set collected with low instantaneous luminosity is used, where
the mean number of inelastic interactions per bunch crossing was $\mu \sim 1$. In this data set, a sample of events is studied with exactly one reconstructed vertex built from a maximum of 10 tracks in the 
central CMS tracker and exactly one multi-RP proton in each arm of the PPS detectors. This provides a control sample enriched in central diffraction (or double-Pomeron exchange)~\cite{Chew:1974vu} events.

In signal events, the z position of the vertex as determined with the central CMS tracker and the time-of-flight difference between the two protons ($\Delta \mathrm{t_{PPS}}$) are linearly 
correlated with a slope of $c/2$ (where c is the speed of light). In practice, even in low-pileup data, there is a nonzero probability of combining unrelated
pileup protons with the central vertex. Since the pileup protons are uncorrelated with the central vertex, this background may be modeled using event-mixing
techniques, where either one or both protons are chosen from different events than those of the central vertex.

The correlation is quantified using a one-dimensional projection of $z_{\mathrm{PPS, timing}} - z_\text{vertex}$,
where $z_\text{PPS, timing} = \Delta{t_\text{PPS}} \, \frac{c}{2}$, and $z_\text{vertex}$ is the position measured by the central tracker.
To estimate the resolution for the signal events, a fit is performed to the sum of signal plus background, using two Gaussian shapes. For the signal component, the mean and width of the Gaussian are 
left as free parameters. The resolution of $z_\text{vertex}$ in the central tracker is estimated to be 50--150\mum for the selection applied here~\cite{Chatrchyan:2014fea}, and thus can be neglected. 

\begin{figure}[ht]
\centering
\includegraphics[width=16 cm]{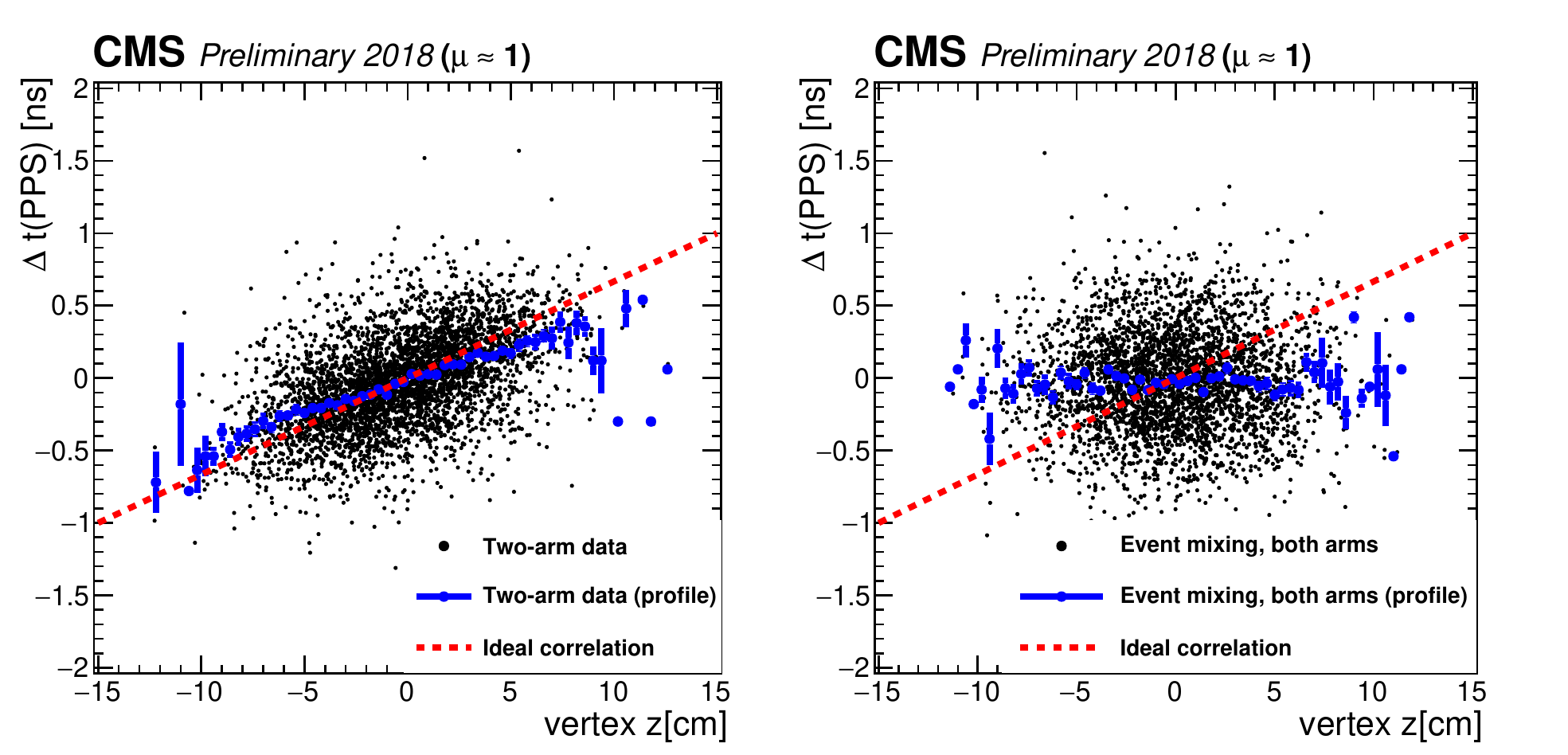}
\caption{Correlation between the z vertex position measured in the central CMS tracker, and the time difference of the protons measured in the PPS detectors. Left: low-pileup data with protons on both arms. 
Right: mixed background sample, with both protons chosen from a different event than that of the central vertex. The red dashed line indicates the ideal slope of c/2, which would be expected with 
zero background. The blue points with error bars show the profile of the data, with the mean and RMS of the time difference in each bin of the position z.}
  \label{fig:timingcorrelation}
\end{figure}

To test the sensitivity to the background shape, three different approaches are tried. First, the background mean and width are treated as free parameters in the fit.
Second, the mean and width are constrained from a fit to an event mixing sample, where both protons are chosen from different events than that of the central vertex.
Third, the mean and width are constrained from a fit to an event-mixing sample, where one proton is chosen from the same event as the central vertex, and the second proton
is chosen from a different event.

\begin{figure}[ht]
\centering
\includegraphics[width=7 cm]{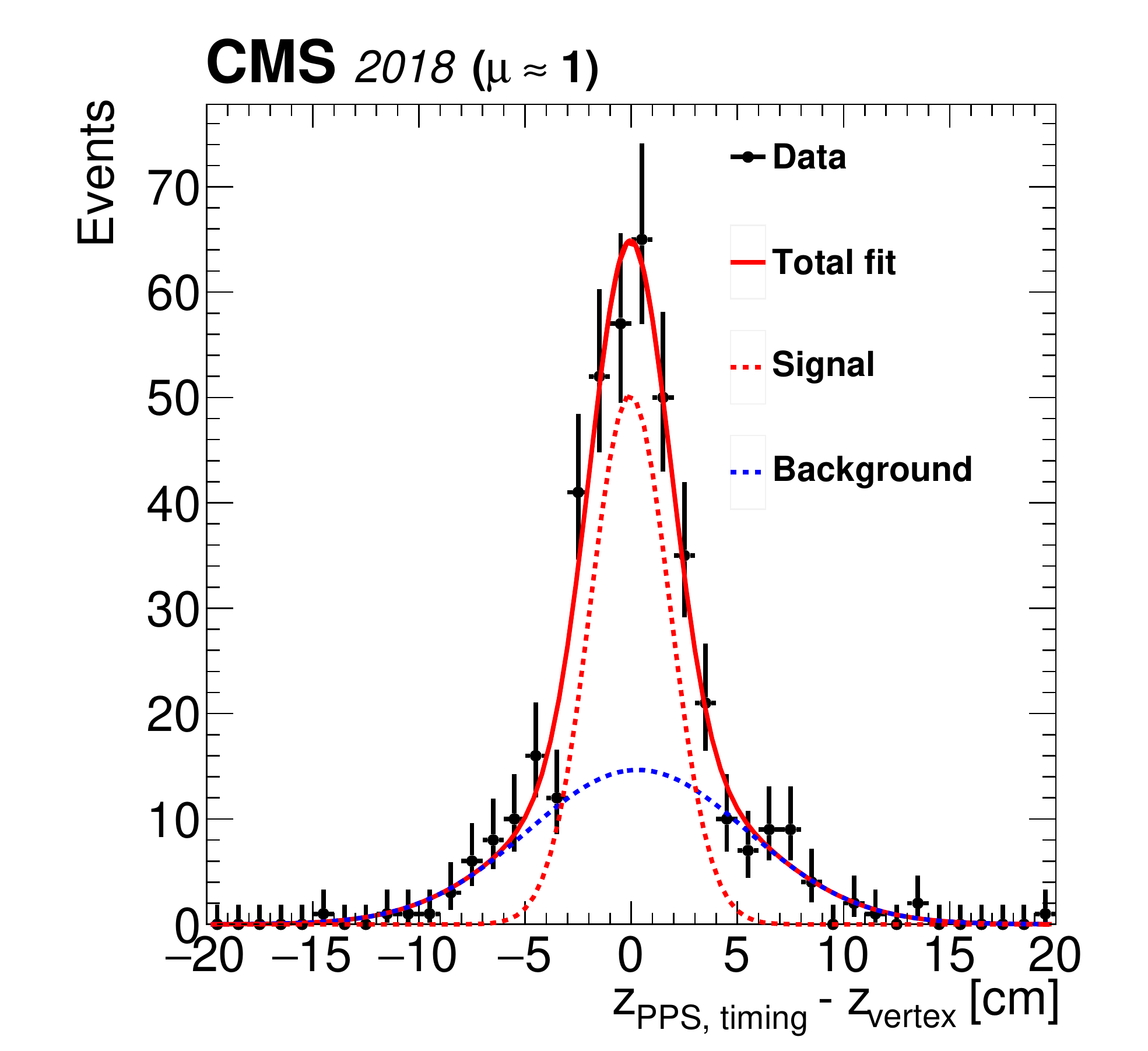}
\includegraphics[width=7 cm]{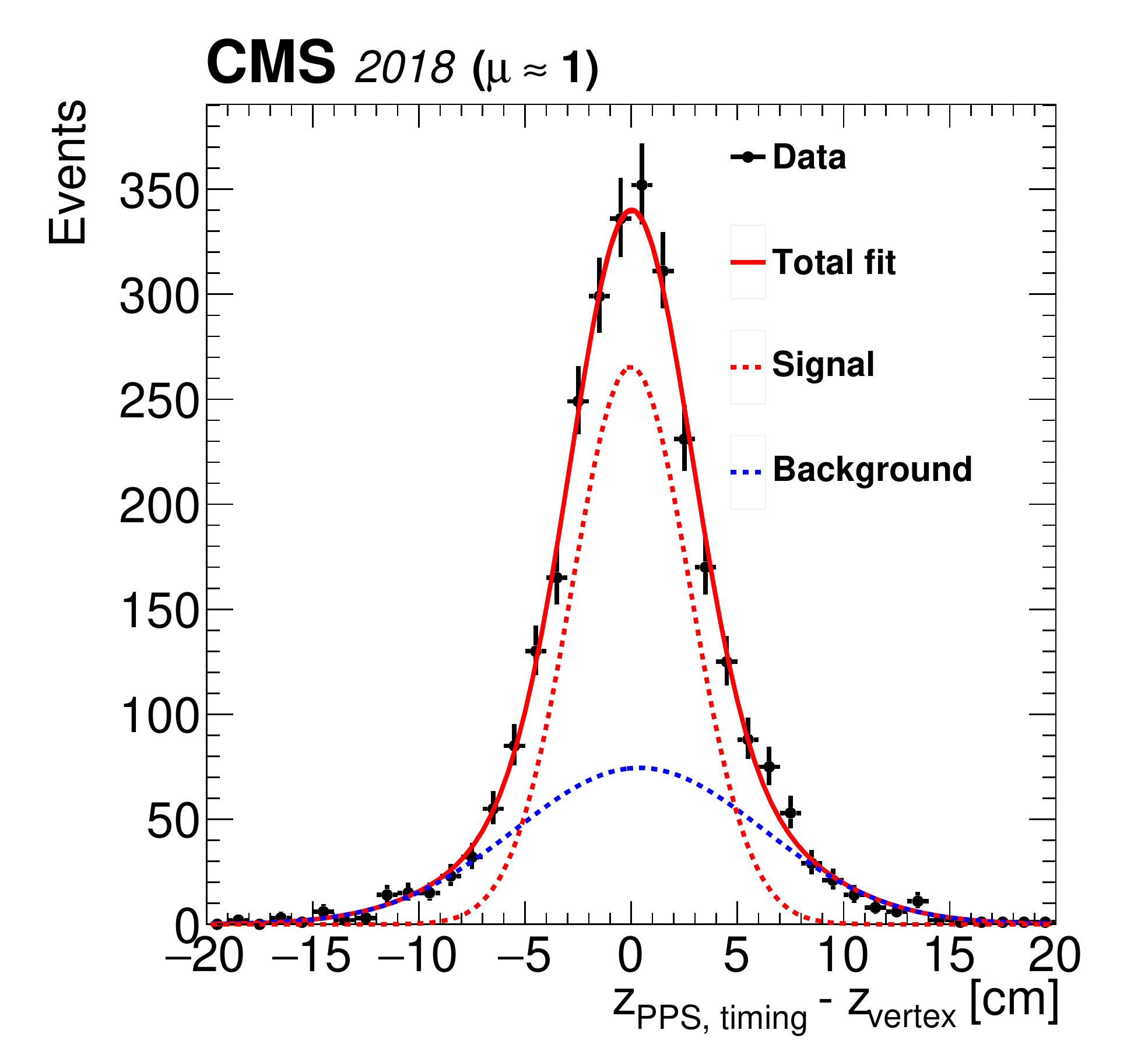}
\caption{Vertex resolution obtained from the difference of proton arrival times, using data collected during low-pileup runs. Left: resolution for two-arm multi-RP events, using the subset of events with a predicted resolution $<$100\,ps for both arms. Right: resolution for two-arm events using all events with exactly one multi-RP proton in each arm. The fitted signal (red dashed line) and  background (blue dashed line) are shown separately, with the means and widths of both components treated as free parameters. The bars on the data points indicate the statistical uncertainties.}
  \label{fig:timing1dres}
\end{figure}

The correlation between the vertex position and the proton time difference is shown in Fig.~\ref{fig:timingcorrelation}.
The sample is further subdivided into a ``high-resolution'' selection, with $<$100\,ps timing resolution predicted on both arms (corresponding to the case with timing measurements on
all 4 planes of each arm), and a ``high-efficiency'' selection, with no requirement on the predicted timing resolution of each arm. The spatial resolutions obtained from 
the fits (Fig.~\ref{fig:timing1dres}) for the two categories are $1.87 \pm 0.21~(1.87 - 1.93)$\,cm and $2.77 \pm 0.17~(2.45 - 2.86)$\,cm, where the first value and uncertainty 
correspond to the central value and statistical uncertainty obtained using a Gaussian background shape with free parameters, and the numbers in brackets represent the range of central 
values obtained under the three different background approaches. The complete list of values obtained is shown in Table~\ref{tab:timingresolutions}.

\begin{table}                                                                                                                                                
\begin{center}                                                                                                                                               
\begin{tabular}{llc}                                                                                                                             
Selection & Background & Resolution $\pm$ stat. [cm] \\\hline
High resolution & Free & $1.87 \pm 0.21$ \\
High resolution & Both arms mixed & $1.93 \pm 0.18$ \\
High resolution & One arm mixed & $1.92 \pm 0.18$\\[\cmsTabSkip]
High efficiency & Free & $2.77 \pm 0.17$ \\
High efficiency & Both arms mixed & $2.86 \pm 0.10$ \\
High efficiency & One arm mixed & $2.45 \pm 0.13$ \\
\end{tabular}                                                                                                                                                
\end{center}                                                                                                                                                 
  \caption{Vertex position resolutions obtained from the proton times measured in the PPS timing detectors, using different selection criteria and background shape assumptions. 
The sample of events in the high resolution and high efficiency categories is always the same, therefore the statistical uncertainties are highly correlated.}
  \label{tab:timingresolutions}                                                                                                                          
\end{table}

The resolutions obtained are consistent with the quadrature sum of single-arm timing resolutions, estimated independently~\cite{CMS-DP-2019-034}. This 
indicates that the overall vertex matching resolution is dominated by the single-arm detector and electronics performance, without large contributions correlated between 
the two arms. In the high resolution category, a time resolution below 100 ps per arm is confirmed, with the full PPS timing system in LHC collisions.
The results further indicate that the single-arm resolutions may be used to predict the overall resolution in high-pileup data, where the two-arm technique described 
here cannot be used.

\section{Summary}
\label{sec:summary}

The procedures developed to reconstruct the proton tracks from the signals detected in the CMS-TOTEM Precision Proton Spectrometer have been described.
The performance of the reconstruction is studied with data from the LHC Run 2 of proton-proton collisions at 13\TeV energy, corresponding to an integrated luminosity of 107.7\fbinv.

A multi-step alignment of the detectors is performed. Alignment with respect to the LHC collimators is followed by relative alignment of the sensor planes within a Roman Pot (RP) and among all RPs. Then, global alignment is performed with respect to the LHC beam with elastic events collected in low luminosity runs, and is extrapolated to the RP's positions in the high luminosity fills. 
Finally, the timing detectors are aligned with respect to the tracking detectors.
The alignment uncertainties are $150\,\mu$m and $100\,\mu$m in the horizontal and vertical projections, respectively. The precision of the relative alignment between near and far RPs is better than $10\,\mu$m.

A precise modelling of the LHC optics is a necessary precondition for the reconstruction. 
The track horizontal and vertical positions at the RPs can be obtained from the proton kinematics at the interaction point via the optical functions. The horizontal dispersion, $D_x$, is calibrated using the $L_y$ = 0 constraint from the data and a sample of (semi)exclusive dimuon events. 
The horizontal dispersion carries information on the dependence of the optics model on the horizontal crossing angle.
The parameters of the optics model (half crossing angle, quadrupole positions and magnet strengths) are determined from a fit to the beam position obtained from the beam position monitors and RPs, and the measured horizontal dispersion.
The vertical dispersion is estimated from the vertical vertex position and the vertical scattering angle. 
The effective length and its derivative with respect to the position along the beam are calibrated at $\xi=0$ using elastic events, where $\xi$ is the relative momentum loss of a forward proton. 

An approximate determination of $\xi$ and of the vertical scattering angle can be performed with the information of a single RP.
A more accurate and complete determination of the proton kinematics is obtained by combining the information from both tracking RPs in each arm. 
The two reconstruction methods are referred to as single- and multi-RP. 
The single-RP reconstruction has significantly lower resolution especially because of the neglected term proportional to the horizontal scattering angle.
A large bias at small and large $\xi$ is hence observed given the asymmetric acceptance in the horizontal scattering angle. 
The multi-RP reconstruction has a much better resolution, negligible bias and comparable systematic uncertainties at small and intermediate $\xi$. 
At large $\xi$, the effect of the systematic uncertainty in the optics calibration is larger for the multi-RP reconstruction. 

A fast simulation of the proton propagation along the beam line and of the PPS detectors has been developed. It includes realistic beam parameters and beam smearing effects; the calibrated optics model; the LHC aperture limitations; the simulation of the detector planes and sensor geometry, acceptance and spatial resolution; and a realistic simulation of the proton arrival time.

A sample of (semi)exclusive dimuon events has been analyzed in order to validate the proton reconstruction. 
A good correlation between the measured proton $\xi$ values and those inferred from the dimuon system is observed. 
The data are well described by the simulation. 
As expected, the multi-RP reconstruction shows a better resolution than the single-RP method.

The proton reconstruction efficiency has been measured for the different data taking periods. It depends on different multiplicative factors describing the sensor efficiency, the reconstruction algorithm efficiency, and the effect of interactions along the proton path. 
The silicon pixel detector efficiency is caused by the radiation damage. 
The silicon strip detector efficiency loss is caused by radiation damage and in addition by multiple tracks in the same event. 
The effect of radiation damage is studied as a function of the integrated luminosity and is significant in the region closest to the detector edge facing the beam. 
The efficiency of the multi-RP reconstruction is smaller than that for the single-RP reconstruction, because of the sensor efficiency of the extra RP, and the effect of multiple, ambiguous proton combinations between tracks from the near and far detectors.

The correlation between the difference in arrival time of protons in the two detector arms and the $z$ vertex position has been studied using low pileup data.
The width of the $z$ position residuals is consistent with the single-arm timing resolutions. 
For part of the data taking period they are better than 100\unit{ps}, corresponding to $\approx$2\cm.

The performance of the Precision Proton Spectrometer in Run 2 has proven the feasibility of continuously operating a near-beam proton spectrometer at a high luminosity hadron collider. PPS has had no impact on the operation of LHC in terms of background, heating, or impedance. The success of PPS has been made possible by two independent collaborations, CMS and TOTEM, joining forces to pursue a common physics interest.

\begin{acknowledgments}
We congratulate our colleagues in the CERN accelerator departments for the excellent performance of the LHC and thank the technical and administrative staffs at CERN and at other CMS and TOTEM institutes for their contributions to the success of the common CMS-TOTEM effort. In addition, we gratefully acknowledge the computing centers and personnel of the Worldwide LHC Computing Grid and other centers for delivering so effectively the computing infrastructure essential to our analyses. Finally, we acknowledge the enduring support for the construction and operation of the LHC, the CMS and TOTEM detectors, and the supporting computing infrastructure provided by the following funding agencies: BMBWF and FWF (Austria); FNRS and FWO (Belgium); CNPq, CAPES, FAPERJ, FAPERGS, and FAPESP (Brazil); MES and BNSF (Bulgaria); CERN; CAS, MoST, and NSFC (China); MINCIENCIAS (Colombia); MSES and CSF (Croatia); RIF (Cyprus); SENESCYT (Ecuador); MoER, ERC PUT and ERDF (Estonia); Academy of Finland, Magnus Ehrnrooth Foundation, MEC, HIP, and Waldemar von Frenckell Foundation (Finland); CEA and CNRS/IN2P3 (France); BMBF, DFG, and HGF (Germany); GSRI (Greece); the Circles of Knowledge Club and NKFIH (Hungary); DAE and DST (India); IPM (Iran); SFI (Ireland); INFN (Italy); MSIP and NRF (Republic of Korea); MES (Latvia); LAS (Lithuania); MOE and UM (Malaysia); BUAP, CINVESTAV, CONACYT, LNS, SEP, and UASLP-FAI (Mexico); MOS (Montenegro); MBIE (New Zealand); PAEC (Pakistan); MES and NSC (Poland); FCT (Portugal); MESTD (Serbia); MCIN/AEI and PCTI (Spain); MOSTR (Sri Lanka); Swiss Funding Agencies (Switzerland); MST (Taipei); MHESI and NSTDA (Thailand); TUBITAK and TENMAK (Turkey); NASU (Ukraine); STFC (United Kingdom); DOE and NSF (USA).
        
\hyphenation{Rachada-pisek} Individuals have received support from the Marie-Curie program and the European Research Council and Horizon 2020 Grant, contract Nos.\ 675440, 724704, 752730, 758316, 765710, 824093, 884104, and COST Action CA16108 (European Union); the Leventis Foundation; the Alfred P.\ Sloan Foundation; the Alexander von Humboldt Foundation; the Belgian Federal Science Policy Office; the Fonds pour la Formation \`a la Recherche dans l'Industrie et dans l'Agriculture (FRIA-Belgium); the Agentschap voor Innovatie door Wetenschap en Technologie (IWT-Belgium); the F.R.S.-FNRS and FWO (Belgium) under the ``Excellence of Science -- EOS" -- be.h project n.\ 30820817; the Beijing Municipal Science \& Technology Commission, No. Z191100007219010; the Ministry of Education, Youth and Sports (MEYS) of the Czech Republic; Svenska Kulturfonden (Finland); the Hellenic Foundation for Research and Innovation (HFRI), Project Number 2288 (Greece); the Deutsche Forschungsgemeinschaft (DFG), under Germany's Excellence Strategy -- EXC 2121 ``Quantum Universe" -- 390833306, and under project number 400140256 - GRK2497; the Hungarian Academy of Sciences, the New National Excellence Program - \'UNKP, the NKFIH research grants K 124845, K 124850, K 128713, K 128786, K 129058, K 131991, K 133046, K 138136, K 143460, K 143477, 2020-2.2.1-ED-2021-00181, and TKP2021-NKTA-64 (Hungary); the Council of Science and Industrial Research, India; the Latvian Council of Science; the Ministry of Education and Science, project no. 2022/WK/14, and the National Science Center, contracts Opus 2021/41/B/ST2/01369 and 2021/43/B/ST2/01552 (Poland); the Funda\c{c}\~ao para a Ci\^encia e a Tecnologia, grant CEECIND/01334/2018 (Portugal); MCIN/AEI/10.13039/501100011033, ERDF ``a way of making Europe", and the Programa Estatal de Fomento de la Investigaci{\'o}n Cient{\'i}fica y T{\'e}cnica de Excelencia Mar\'{\i}a de Maeztu, grant MDM-2017-0765 and Programa Severo Ochoa del Principado de Asturias (Spain); the Chulalongkorn Academic into Its 2nd Century Project Advancement Project, and the National Science, Research and Innovation Fund via the Program Management Unit for Human Resources \& Institutional Development, Research and Innovation, grant B05F650021 (Thailand); the Kavli Foundation; the Nvidia Corporation; the SuperMicro Corporation; the Welch Foundation, contract C-1845; and the Weston Havens Foundation (USA).  
\end{acknowledgments}

\bibliography{auto_generated}

\cleardoublepage \appendix\section{The CMS Collaboration \label{app:collab}}\begin{sloppypar}\hyphenpenalty=5000\widowpenalty=500\clubpenalty=5000\input{PRO-21-001-public-authorlist.tex}\end{sloppypar}
\end{document}

%% file: PRO-21-001-public-authorlist.tex
\cmsinstitute{Yerevan Physics Institute, Yerevan, Armenia}
{\tolerance=6000
A.~Tumasyan\cmsAuthorMark{1}\cmsorcid{0009-0000-0684-6742}
\par}
\cmsinstitute{Institut f\"{u}r Hochenergiephysik, Vienna, Austria}
{\tolerance=6000
W.~Adam\cmsorcid{0000-0001-9099-4341}, J.W.~Andrejkovic, T.~Bergauer\cmsorcid{0000-0002-5786-0293}, S.~Chatterjee\cmsorcid{0000-0003-2660-0349}, K.~Damanakis\cmsorcid{0000-0001-5389-2872}, M.~Dragicevic\cmsorcid{0000-0003-1967-6783}, A.~Escalante~Del~Valle\cmsorcid{0000-0002-9702-6359}, P.S.~Hussain\cmsorcid{0000-0002-4825-5278}, M.~Jeitler\cmsAuthorMark{2}\cmsorcid{0000-0002-5141-9560}, N.~Krammer\cmsorcid{0000-0002-0548-0985}, L.~Lechner\cmsorcid{0000-0002-3065-1141}, D.~Liko\cmsorcid{0000-0002-3380-473X}, I.~Mikulec\cmsorcid{0000-0003-0385-2746}, P.~Paulitsch, F.M.~Pitters, J.~Schieck\cmsAuthorMark{2}\cmsorcid{0000-0002-1058-8093}, R.~Sch\"{o}fbeck\cmsorcid{0000-0002-2332-8784}, D.~Schwarz\cmsorcid{0000-0002-3821-7331}, S.~Templ\cmsorcid{0000-0003-3137-5692}, W.~Waltenberger\cmsorcid{0000-0002-6215-7228}, C.-E.~Wulz\cmsAuthorMark{2}\cmsorcid{0000-0001-9226-5812}
% ...
\par}
\vskip\cmsinstskip
\dag:~Deceased\\
$^{1}$Also at Yerevan State University, Yerevan, Armenia\\
$^{2}$Also at TU Wien, Vienna, Austria\\
% ...

\section{The TOTEM Collaboration\label{app:totem}}
\newcommand{\AddAuthor}[2]{#1$^{#2}$,\ }
\newcommand\AddAuthorLast[2]{#1$^{#2}$}
\noindent
\AddAuthor{G.~Antchev\cmsorcid{0000-0003-3210-5037}}{a}
\AddAuthor{P.~Aspell}{8}
\AddAuthor{I.~Atanassov\cmsorcid{0000-0002-5728-9103}}{a}
\AddAuthor{V.~Avati}{7,8}
\AddAuthor{J.~Baechler}{8}
\AddAuthor{C.~Baldenegro~Barrera\cmsorcid{0000-0002-6033-8885}}{10}
\AddAuthor{V.~Berardi\cmsorcid{0000-0002-8387-4568}}{4a,4b}
\AddAuthor{M.~Berretti\cmsorcid{0000-0003-4122-8282}}{2a}
\AddAuthor{V.~Borshch\cmsorcid{0000-0002-5479-1982}}{11}
\AddAuthor{E.~Bossini\cmsorcid{0000-0002-2303-2588}}{6a}
\AddAuthor{U.~Bottigli\cmsorcid{0000-0002-0666-3433}}{6c}
\AddAuthor{M.~Bozzo\cmsorcid{0000-0002-1715-0457}}{5a,5b}
\AddAuthor{H.~Burkhardt}{8}
\AddAuthor{F.S.~Cafagna\cmsorcid{0000-0002-7450-4784}}{4a}
\AddAuthor{M.G.~Catanesi\cmsorcid{0000-0002-2987-7688}}{4a}
\AddAuthor{M.~Csan\'{a}d\cmsorcid{0000-0002-3154-6925}}{3a,b}
\AddAuthor{T.~Cs\"{o}rg\H{o}\cmsorcid{0000-0002-9110-9663}}{3a,3b}
\AddAuthor{M.~Deile\cmsorcid{0000-0001-5085-7270}}{8}
\AddAuthor{F.~De~Leonardis}{4a,4c}
\AddAuthor{M.~Doubek\cmsorcid{0000-0002-0026-9558}}{1c}
\AddAuthor{D.~Druzhkin}{8,11}
\AddAuthor{K.~Eggert}{9}
\AddAuthor{V.~Eremin}{e}
\AddAuthor{A.~Fiergolski}{8}
\AddAuthor{F.~Garcia\cmsorcid{0000-0002-4023-7964}}{2a}
\AddAuthor{V.~Georgiev}{1a}
\AddAuthor{S.~Giani}{8}
\AddAuthor{L.~Grzanka\cmsorcid{0000-0002-3599-854X}}{7}
\AddAuthor{J.~Hammerbauer}{1a}
\AddAuthor{T.~Isidori\cmsorcid{0000-0002-7934-4038}}{10}
\AddAuthor{V.~Ivanchenko\cmsorcid{0000-0002-1844-5433}}{11}
\AddAuthor{M.~Janda\cmsorcid{0000-0001-5736-6183}}{1c}
\AddAuthor{A.~Karev}{8}
\AddAuthor{J.~Ka\v{s}par\cmsorcid{0000-0001-5639-2267}}{1b,8}
\AddAuthor{B.~Kaynak\cmsorcid{0000-0003-3857-2496}}{c}
\AddAuthor{J.~Kopal\cmsorcid{0000-0001-9751-7409}}{8}
\AddAuthor{V.~Kundr\'{a}t\cmsorcid{0000-0003-2868-5550}}{1b}
\AddAuthor{S.~Lami\cmsorcid{0000-0001-9492-0147}}{6a}
\AddAuthor{R.~Linhart}{1a}
\AddAuthor{C.~Lindsey}{10}
\AddAuthor{M.V.~Lokaj\'{i}\v{c}ek\cmsorcid{0000-0002-2052-1220}$^{\textrm{\dag,}}$}{1b}
\AddAuthor{L.~Losurdo\cmsorcid{0000-0002-4964-7951}}{6c}
\AddAuthor{F.~Lucas~Rodr\'{i}guez}{8}
\AddAuthor{M.~Macr\'{i}}{\dag,5a}
\AddAuthor{M.~Malawski\cmsorcid{0000-0001-6005-0243}}{7}
\AddAuthor{N.~Minafra\cmsorcid{0000-0003-4002-1888}}{10}
\AddAuthor{S.~Minutoli}{5a}
\AddAuthor{K.~Misan\cmsorcid{0009-0007-2416-042X}}{7}
\AddAuthor{T.~Naaranoja\cmsorcid{0000-0001-5797-7929}}{2a,2b}
\AddAuthor{F.~Nemes\cmsorcid{0000-0002-1451-6484}}{3a,3b,8}
\AddAuthor{H.~Niewiadomski}{9}
\AddAuthor{T.~Nov\'{a}k\cmsorcid{0000-0001-6253-4356}}{3b}
\AddAuthor{E.~Oliveri\cmsorcid{0000-0002-0832-6975}}{8}
\AddAuthor{F.~Oljemark\cmsorcid{0000-0003-0121-2761}}{2a,2b}
\AddAuthor{M.~Oriunno}{d}
\AddAuthor{K.~\"{O}sterberg\cmsorcid{0000-0003-4807-0414}}{2a,2b}
\AddAuthor{P.~Palazzi\cmsorcid{0000-0002-4861-391X}}{8}
\AddAuthor{V.~Passaro\cmsorcid{0000-0003-0802-4464}}{4a,4c}
\AddAuthor{Z.~Peroutka}{1a}
\AddAuthor{J.~Proch\'{a}zka\cmsorcid{0000-0002-2774-2245}}{1b}
\AddAuthor{M.~Quinto\cmsorcid{0000-0002-6363-6132}}{4a,4b}
\AddAuthor{E.~Radermacher}{8}
\AddAuthor{E.~Radicioni\cmsorcid{0000-0002-2231-1067}}{4a}
\AddAuthor{F.~Ravotti\cmsorcid{0000-0002-5709-6934}}{8}
\AddAuthor{C.~Royon\cmsorcid{0000-0002-7672-9709}}{10}
\AddAuthor{G.~Ruggiero}{8}
\AddAuthor{H.~Saarikko}{2a,2b}
\AddAuthor{V.D.~Samoylenko}{e}
\AddAuthor{A.~Scribano\cmsorcid{0000-0002-4338-6332}}{6a,10}
\AddAuthor{J.~\v{S}irok\'{y}}{1a}
\AddAuthor{J.~Smajek}{8}
\AddAuthor{W.~Snoeys\cmsorcid{0000-0003-3541-9066}}{8}
\AddAuthor{R.~Stefanovitch}{8}
\AddAuthor{J.~Sziklai}{3a}
\AddAuthor{C.~Taylor\cmsorcid{0000-0001-6816-8051}}{9}
\AddAuthor{E.~Tcherniaev\cmsorcid{0000-0002-3685-0635}}{11}
\AddAuthor{N.~Turini\cmsorcid{0000-0002-9395-5230}}{6c}
\AddAuthor{R.~Turpeinen}{2a}
\AddAuthor{O.~Urban}{1a}
\AddAuthor{V.~Vacek\cmsorcid{0000-0001-9584-0392}}{1c}
\AddAuthor{O.~Vavroch}{1a}
\AddAuthor{J.~Welti}{2a,2b}
\AddAuthor{J.~Williams\cmsorcid{0000-0002-9810-7097}}{10}
\AddAuthorLast{J.~Zich}{1a}

\vskip 4pt plus 4pt
\let\thefootnote\relax
\newcommand{\AddInstitute}[2]{${}^{#1}$#2\\}
\newcommand{\AddExternalInstitute}[2]{\footnote{${}^{#1}$ #2}}
\noindent
\dag Deceased\\
\AddInstitute{1a}{University of West Bohemia, Pilsen, Czech Republic}
\AddInstitute{1b}{Institute of Physics of the Academy of Sciences of the Czech Republic, Prague, Czech Republic}
\AddInstitute{1c}{Czech Technical University, Prague, Czech Republic}
\AddInstitute{2a}{Helsinki Institute of Physics, University of Helsinki, Helsinki, Finland}
\AddInstitute{2b}{Department of Physics, University of Helsinki, Helsinki, Finland}
\AddInstitute{3a}{Wigner Research Centre for Physics, RMKI, Budapest, Hungary}
\AddInstitute{3b}{MATE Institute of Technology KRC, Gy\"{o}ngy\"{o}s, Hungary}
\AddInstitute{4a}{INFN Sezione di Bari, Bari, Italy}
\AddInstitute{4b}{Dipartimento Interateneo di Fisica di Bari, University of Bari, Bari, Italy}
\AddInstitute{4c}{Dipartimento di Ingegneria Elettrica e dell'Informazione --- Politecnico di Bari, Bari, Italy}
\AddInstitute{5a}{INFN Sezione di Genova, Genova, Italy}
\AddInstitute{5b}{Universit\`{a} degli Studi di Genova, Genova, Italy}
\AddInstitute{6a}{INFN Sezione di Pisa, Pisa, Italy}
\AddInstitute{6c}{Universit\`{a} degli Studi di Siena and Gruppo Collegato INFN di Siena, Siena, Italy}
\AddInstitute{7}{Akademia G\'{o}rniczo-Hutnicza (AGH) University of Science and Technology, Krakow, Poland}
\AddInstitute{8}{CERN, Geneva, Switzerland}
\AddInstitute{9}{Case Western Reserve University, Department of Physics, Cleveland, Ohio, USA}
\AddInstitute{10}{The University of Kansas, Lawrence, Kansas, USA}
\AddInstitute{11}{Authors affiliated with an institute or an international laboratory covered by a cooperation agreement with CERN}
\AddExternalInstitute{a}{INRNE-BAS, Institute for Nuclear Research and Nuclear Energy, Bulgarian Academy of Sciences, Sofia, Bulgaria}
\AddExternalInstitute{b}{Department of Atomic Physics, E\"{o}tv\"{o}s Lor\'{a}nd University, Budapest, Hungary}
\AddExternalInstitute{c}{Istanbul University, Istanbul, Turkey}
\AddExternalInstitute{d}{SLAC, Stanford University, California, USA}
\AddExternalInstitute{e}{Authors affiliated with an institute or an international laboratory covered by a cooperation agreement with CERN}